\documentclass[longauth]{aa} 

\usepackage{graphicx}
\usepackage{txfonts}
\usepackage{kotex}
\usepackage{systeme}
\usepackage{caption}
\usepackage{fnpct} 

\usepackage{pdflscape} 
\usepackage[section]{placeins}

\usepackage{gensymb}

\usepackage{color}
\definecolor{color1}{rgb}{0, 0.5, 1.0}
\definecolor{trans}{rgb}{1, 0.5, 0}
\definecolor{grey}{rgb}{0.7, 0.7, 0.7}
\definecolor{ro}{rgb}{0.945, 0.467, 0.667}

\usepackage{natbib}
\bibpunct{(}{)}{;}{a}{}{,} 
\usepackage[colorlinks=true,allcolors=blue]{hyperref}

\usepackage{amsmath}

\usepackage{makeidx}
\makeindex

\begin{document}

\title{Spectral analysis of a parsec-scale jet in M87: Observational constraint on the magnetic field strengths in the jet}

\titlerunning{Spectral analysis of parsec-scale jet in M87}

   \author{Hyunwook Ro \inst{1, 2}
            \and
            Motoki Kino \inst{3, 4}
            \and
            Bong Won Sohn \inst{2, 5, 1}
            \and
            Kazuhiro Hada \inst{6, 7}
            \and
            Jongho Park \inst{2, 8, 9}
            \and
            Masanori Nakamura \inst{10, 8}
            \and
            Yuzhu Cui \inst{11, 6, 7}
            \and
            Kunwoo Yi \inst{9}
            \and
            Aeree Chung \inst{1}
            \and
            Jeffrey Hodgson \inst{12}
            \and
            Tomohisa Kawashima \inst{13}
            \and
            Tao An \inst{14}
            \and
            Sascha Trippe \inst{9, 15}
            \and
            Juan-Carlos Algaba \inst{16}
            \and
            Jae-Young Kim \inst{17, 2, 18}
            \and
            Satoko Sawada-Satoh \inst{19, 20}
            \and
            Kiyoaki Wajima \inst{2, 5}
            \and
            Zhiqiang Shen \inst{14, 21}
            \and
            Xiaopeng Cheng \inst{2}
            \and
            Ilje Cho \inst{22, 2, 5}
            \and
            Wu Jiang \inst{14, 21}
            \and
            Taehyun Jung \inst{2, 5}
            \and
            Jee-Won Lee \inst{2}
            \and
            Kotaro Niinuma \inst{23, 20}
            \and
            Junghwan Oh \inst{24, 12, 2}
            \and
            Fumie Tazaki \inst{25, 6}
            \and
            Guang-Yao Zhao \inst{22, 2}
            \and
            Kazunori Akiyama \inst{26, 27, 3}
            \and            
            Mareki Honma \inst{6, 7}
            \and
            Jeong Ae Lee \inst{2}
            \and
            Rusen Lu \inst{14, 21, 18}
            \and
            Yingkang Zhang \inst{14}
            \and
            Keiichi Asada \inst{8}
            \and
            Lang Cui \inst{28, 21}
            \and
            Yoshiaki Hagiwara \inst{29}
            \and
            Tomoya Hirota \inst{3}
            \and
            Noriyuki Kawaguchi \inst{3}
            \and
            Shoko Koyama \inst{30, 8}
            \and
            Sang-Sung Lee \inst{2, 5}
            \and
            Se-Jin Oh \inst{2}
            \and
            Koichiro Sugiyama \inst{31}
            \and
            Mieko Takamura \inst{3, 32}
            \and
            Xuezheng Wang \inst{14}
            \and
            Ju-Yeon Hwang \inst{2}
            \and
            Dong-Kyu Jung \inst{2}
            \and
            Hyo-Ryoung Kim \inst{2}
            \and
            Jeong-Sook Kim \inst{2, 33, 34}
            \and
            Hideyuki Kobayashi \inst{6}
            \and
            Chung-Sik Oh \inst{2}
            \and
            Tomoaki Oyama \inst{6}
            \and
            Duk-Gyoo Roh \inst{2}
            \and
            Jae-Hwan Yeom \inst{2}
          }
    \institute{Department of Astronomy, Yonsei University, Yonsei-ro 50, Seodaemun-gu, Seoul 03722, Republic of Korea\\
             \email{hwro@yonsei.ac.kr}
    \and
    Korea Astronomy \& Space Science Institute, Daedeokdae-ro 776, Yuseong-gu, Daejeon 34055, Republic of Korea
    \and
    National Astronomical Observatory of Japan, 2-21-1 Osawa, Mitaka, Tokyo 181-8588, Japan
    \and
    Kogakuin University of Technology \& Engineering, Academic Support Center, 2665-1 Nakano, Hachioji, Tokyo 192-0015, Japan
    \and
    University of Science and Technology, Gajeong-ro 217, Yuseong-gu, Daejeon 34113, Republic of Korea
    \and
    Mizusawa VLBI Observatory, National Astronomical Observatory of Japan, 2-12 Hoshigaoka, Mizusawa, Oshu, Iwate 023-0861, Japan
    \and
    Department of Astronomical Science, The Graduate University for Advanced Studies (SOKENDAI), 2-21-1 Osawa, Mitaka, Tokyo 181-8588, Japan
    \and
    Institute of Astronomy and Astrophysics, Academia Sinica, 11F of Astronomy-Mathematics Building, AS/NTU No. 1, Sec. 4, Roosevelt Rd, Taipei 10617, Taiwan, R.O.C.
    \and
    Department of Physics and Astronomy, Seoul National University, 1 Gwanak-ro Gwanak-gu, Seoul 08826, Republic of Korea
    \and
    National Institute of Technology, Hachinohe College, Yubinbango Aomori Prefecture Hachinohe Oaza Tamonoki character Ueno flat 16-1, 039-1192, Japan
    \and
    Tsung-Dao Lee Institute, Shanghai Jiao Tong University, Shanghai 201210, China
    \and
    Department of Physics and Astronomy, Sejong University, 209 Neungdong-ro, Gwangjin-gu, Seoul 05006, Republic of Korea
    \and
    Institute for Cosmic Ray Research, The University of Tokyo, 5-1-5 Kashiwanoha, Kashiwa, Chiba 277-8582, Japan
    \and
    Shanghai Astronomical Observatory, Chinese Academy of Sciences, 80 Nandan Road, Shanghai 200030, China
    \and
    SNU Astronomy Research Center (SNUARC), Seoul National University, 1 Gwanak-ro Gwanak-gu, Seoul 08826, Republic of Korea
    \and
    Department of Physics, Faculty of Science, University of Malaya, 50603 Kuala Lumpur, Malaysia
    \and
    Department of Astronomy and Atmospheric Sciences, Kyungpook National University, Daegu 702-701, Republic of Korea
    \and
    Max-Planck-Institut f\"{u}r Radioastronomie, Auf dem H\"{u}gel 69, D-53121 Bonn, Germany
    \and
    Graduate School of Science, Osaka Metropolitan University, Osaka 599-8531, Japan
    \and
    The Research Institute of Time Studies, Yamaguchi University, Yoshida 1677-1, Yamaguchi-city, Yamaguchi 753-8511, Japan
    \and
    Key Laboratory of Radio Astronomy, Chinese Academy of Sciences, Nanjing 210008, China 
    \and
    Instituto de Astrof\'{\i}sica de Andaluc\'{\i}a - CSIC, Glorieta de la Astronom\'{\i}a s/n, E-18008 Granada, Spain
    \and
    Graduate School of Sciences and Technology for Innovation, Yamaguchi University, 1677-1 Yoshida, Yamaguchi, Yamaguchi 753-8511, Japan
    \and
    Joint Institute for VLBI ERIC, 7991 PD Dwingeloo, The Netherlands
    \and
    Tokyo Electron Technology Solutions Limited, Iwate 023-1101, Japan
    \and
    Massachusetts Institute of Technology Haystack Observatory, 99 Millstone Road, Westford, MA 01886, USA
    \and
    Black Hole Initiative at Harvard University, 20 Garden Street, Cambridge, MA 02138, USA
    \and
    Xinjiang Astronomical Observatory, Chinese Academy of Sciences, Urumqi 830011, China
    \and
    Toyo University, 5-28-20 Hakusan, Bunkyo-ku, Tokyo 112-8606, Japan
    \and
    Niigata University, 8050 Ikarashi 2-no-cho, Nishi-ku, Niigata 950-2181, Japan
    \and
    National Astronomical Research Institute of Thailand (Public Organization), 260 Moo 4, T. Donkaew, A. Maerim, Chiangmai, 50180 Thailand
    \and
    Department of Astronomy, The University of Tokyo, 7-3-1 Hongo, Bunkyo, Tokyo 113-0033, Japan
    \and
    {Department of Physics, UNIST, Ulsan 44919, Korea}
    \and
    {Basic Science Research Institute, Chungbuk National University, Chungdae-ro 1, Seowon-Gu, Cheongju, Chungbuk 28644, Korea}
    }

  \date{Received December 31, 2021; accepted November 22, 2022}

 
  \abstract
  {{Because of its proximity and the large size of its black hole, M87 is one of the best targets for studying the launching mechanism of active galactic nucleus jets. Currently, magnetic fields are considered to be an essential factor in the launching and accelerating of the jet. However, current observational estimates of the magnetic field strength of the M87 jet are limited to the innermost part of the jet ($\lesssim100$ $r_s$) or to HST-1 ($\sim10^5$ $r_s$). No attempt has yet been made to measure the magnetic field strength in between.}}
  {{We aim to infer the magnetic field strength of the M87 jet out to a distance of several thousand $r_s$ by tracking the distance-dependent changes in the synchrotron spectrum of the jet from high-resolution very long baseline interferometry observations.}}
  {{In order to obtain high-quality spectral index maps, quasi-simultaneous observations at 22 and 43 GHz were conducted using the KVN and VERA Array (KaVA) and the Very Long Baseline Array (VLBA). We compared the spectral index distributions obtained from the observations with a model and placed limits on the magnetic field strengths as a function of distance.}}
  {The overall spectral morphology is broadly consistent over the course of these observations. The observed synchrotron spectrum rapidly steepens from $\alpha_{\text{22-43 GHz}}\sim-0.7$ at $\sim$ 2 mas to $\alpha_{\text{22-43 GHz}}\sim-2.5$ at $\sim$ 6 mas. In the KaVA observations, the spectral index remains unchanged until $\sim$ 10 mas, but this trend is unclear in the VLBA observations.
  A spectral index model in which nonthermal electron injections inside the jet decrease with distance can adequately reproduce the observed trend. This suggests the magnetic field strength of the jet at a distance of 2 $-$ 10 mas ($\sim$ 900 $r_s-\sim$ 4500 $r_s$ in the deprojected distance) has a range of $B=\left(0.3-1.0\text{ G}\right)\left(\text{z}/\text{2 mas}\right)^{-0.73}$. 
  Extrapolating to the Event Horizon Telescope scale yields consistent results, suggesting that the majority of the magnetic flux of the jet near the black hole is preserved out to $\sim$ 4500 $r_s$ without significant dissipation.}
  {}

   \keywords{<galaxies: active - galaxies: individual: M87 - galaxies: jets - radio continuum: galaxies - relativistic processes - techniques: interferometric>}
   \maketitle
   
%

\section{Introduction}\label{section1}

M87 is known as the best target for investigating the active galactic nucleus (AGN) jet formation mechanism due to its proximity 
(distance = 16.7 Mpc; \citealt{blakeslee09}) and 
its large central supermassive black hole \citep[SMBH;][]{macchetto1997, gebhardt2009, gebhardt2011, walsh2013}.
Recent progress in very long baseline interferometry (VLBI)
studies has brought us further key knowledge about the M87 jet.
First, the Event Horizon Telescope (EHT) successfully imaged the first ever black hole shadow \citep{eht2019, eht2021} 
and firmly determined the mass of the central SMBH
in M87 as
$M_{\bullet}=6.5 \pm 0.7 \times10^9$ $M_{\odot}$; \citealt{eht2019}). 
At the distance of 16.7 Mpc, 1 milli-arcsecond (mas) corresponds to $\approx$ 130 $r_{s}$ $\approx$ 0.08 parsec, where $r_{s}$ is the Schwarzschild radius of the SMBH.
M87 hosts a prominent radio jet that is detected from the radio 
to TeV $\gamma$-rays \citep[e.g.,][]{abramowski2012, ehtMWL2021}. 

The current leading model for jet launching describes a jet that is driven by magnetic force and then accelerated by the relatively slow conversion of magnetic energy into kinetic energy \citep[e.g.,][]{blandford77, mckinney2006, komissarov2007, nakamura2018, chatterjee19}.
One way to test this scenario is to determine the
radial profiles of the velocity field and 
the magnetic field along the jets.
Over the last 15 years or so, the velocity field
of the M87 jet has been intensively explored via multi-epoch VLBI observations that have allowed us to directly probe the kinematics of the jet components
\citep[e.g.,][]{kovalev2007, ly2007, asada2014, mertens2016}.
Furthermore, in order to avoid errors in the identification of knot motions due to low cadence, high-cadence observations were made and the global profile of the M87 jet velocity field from $10^{2}$ $r_{s}$ to  $10^{7}$ $r_{s}$ was revealed
in great detail by \citet{park2019_kine}.
Interestingly, the obtained velocity field data showed an apparent discrepancy with those predicted by general relativistic magnetohydrodynamics simulations
\citep[e.g.,][]{mckinney2006, nakamura2018},
posing a new challenge to the jet formation mechanism. Future VLBI observations with higher angular resolution and higher cadence may allow us to address this issue.

{{Meanwhile, there has been recent progress in constraining the magnetic field strength of the M87 jet \citep[for a review, see][]{hawley15}. In the bright knot HST-1, which is located in the $\sim3.8\times10^5$ $r_{s}$ downstream of the SMBH, the magnetic field strength is constrained by the synchrotron cooling time in the X-ray energy band \citep{harris2003, harris2009}.
In the region within $\sim100$ $r_{s}$, there has been major progress in millimeter and submillimeter VLBI observations, which successfully measured the size and the flux of the jet base (the radio core), where the optical depth for synchrotron self-absorption (SSA) becomes unity at the observed frequency \citep[e.g.,][]{doeleman2012, hada2013, hada2016}. 
From this, the magnetic field strength of the radio core was estimated. 
The estimated strength suggests that the energy density of the magnetic field dominates the nonthermal electron energy density near the SMBH, which in turn suggests that magnetic fields are important for jet formation \citep{kino2014, kino2015}. The successful measurement of the frequency-dependent position of the core (the so-called core-shift) provides another way of constraining the magnetic field strength of the jet \citep{hada2011, zamaninasab2014, zdziarski2015}.
}
{Most recently, from images of the central black hole's photon ring and its polarization distribution obtained from EHT observations \citep{eht2019, eht2021}, the magnetic field strength of the surrounding plasma was constrained.}
}

{However, our understanding of the magnetic field profile along the jet is currently limited. This is because only   the magnetic field strengths of notable individual components have been estimated so far. The magnetic field strength between the radio core and HST-1 remains unknown. In particular, there has been no attempt to date to measure the magnetic field strength of the extended jet.}

{In this study we explore the magnetic field properties of the M87 jet using spectral index maps.}
{Although there have been many previous studies that have published spectral index maps of M87 \citep{zavala2003, dodson2006, ly2007, hada2011, hada2012, pushkarev2012, niinuma2014, hovatta2014, asada2016, kravchenko20}, there has been no dedicated study trying to limit physical quantities using spectral index maps.}
{In order to thoroughly constrain and characterize the 
radial profiles of physical quantities along the M87 jet,
we set up a dedicated observation program using the KVN and VERA Array (KaVA).
After commissioning observations \citep{niinuma2014}, 
M87 has been observed as part of the KaVA AGN Large Program (hereafter, LP) since 2016. 
}

This paper is organized as follows. 
In Sect. \ref{section2}
we describe the multifrequency observations and data reduction of the KaVA LP and other archival data. 
In Sect. \ref{section3} we present the spectral index maps of M87 between 22 and 43GHz.
In Sect. \ref{discussion} 
we discuss a model of the spectral index distribution that can explain the observed spectral index profile of M87 and constrain the magnetic field distribution of the jet. Furthermore, we compare our measurements to those from previous studies. Throughout the paper, we define the sign of the spectral index as $S\propto \nu^{+\alpha}$.

\section{Observations and data reduction}\label{section2}

\subsection{KaVA observations}\label{subsection2.1}

M87 observations using KaVA were performed nine times from February to June 2016. Each epoch consists of two sessions, at 22 and 43 GHz, which were observed within 1 to 2 days of each other. Each observation was allocated every 2$-$3 weeks. Among them, the data on June 1 were excluded due to antenna problems at Mizusawa and Ishigaki. As a result, eight out of the nine data sets are used for the spectral analysis. At each frequency, the total observation time at each epoch is approximately 7 hours, and M87's on-source time is about 4 hours and 30 minutes. More information about the observations is described in \cite{park2019_kine}. Data reduction was performed following the process described in \cite{park2019_kine}. {The amplitude is calibrated via an a priori method using an opacity-corrected system temperature and the elevation-dependent gain curve for each telescope.} The amplitudes were scaled up by a factor of 1.3 to correct for amplitude losses that occurred during correlation \citep{lee2015, hada2017}. However, since the same factor is applied at both frequencies, the spectral index maps do not change after the correction. Imaging with CLEAN and self-calibration was performed using the Difmap software package \citep{shepherd1994}. When natural weighting is applied, the typical size of the full width at half maximum of the synthesized beam at 22 and 43 GHz is close to a circular shape with radii of 1.2 mas and 0.6 mas, respectively.

\subsection{VLBA archival data}\label{subsection2.2}
{We also included archival quasi-simultaneous data from the Very Long Baseline Array (VLBA). A total of five observations, with two in 2010 and three in 2014 were obtained. Details of the observations for the data in 2010 and 2014 are described in \cite{hada2011} and \cite{hada2016}, respectively. We re-performed the data reduction for these data. Initial data calibrations (a priori amplitude correction, atmospheric opacity correction, fringe-fitting, and bandpass) were performed using the~Astronomical Image Processing System (AIPS; \citealt{greisen2003}). For the 2014 observations using the RDBE digital backend system, the auto-correlation amplitude was further corrected after bandpass correction, as suggested in VLBA Memo \#37. The subsequent image reconstruction was performed in Difmap based on the usual CLEAN/self-calibration procedure. The typical beam size is $0.8\times0.4$ mas at 22 GHz, and $0.45\times0.22$ mas at 43 GHz when natural weighing is applied.}


\section{Results}\label{section3}

\subsection{Spectral index maps}\label{subsection3.1}

\begin{figure*}[t]
\centering
\includegraphics[width=9cm]{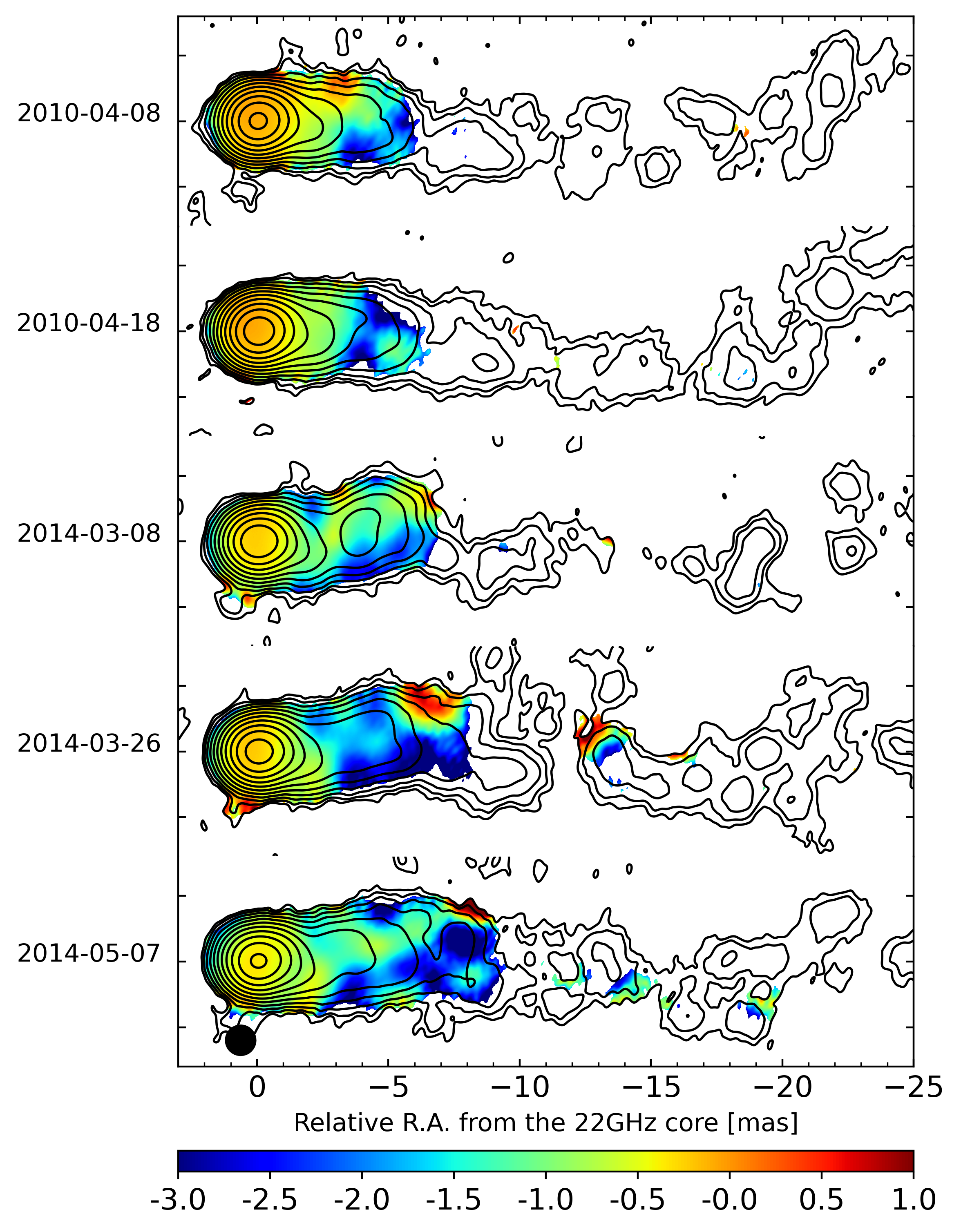}
\includegraphics[width=9cm]{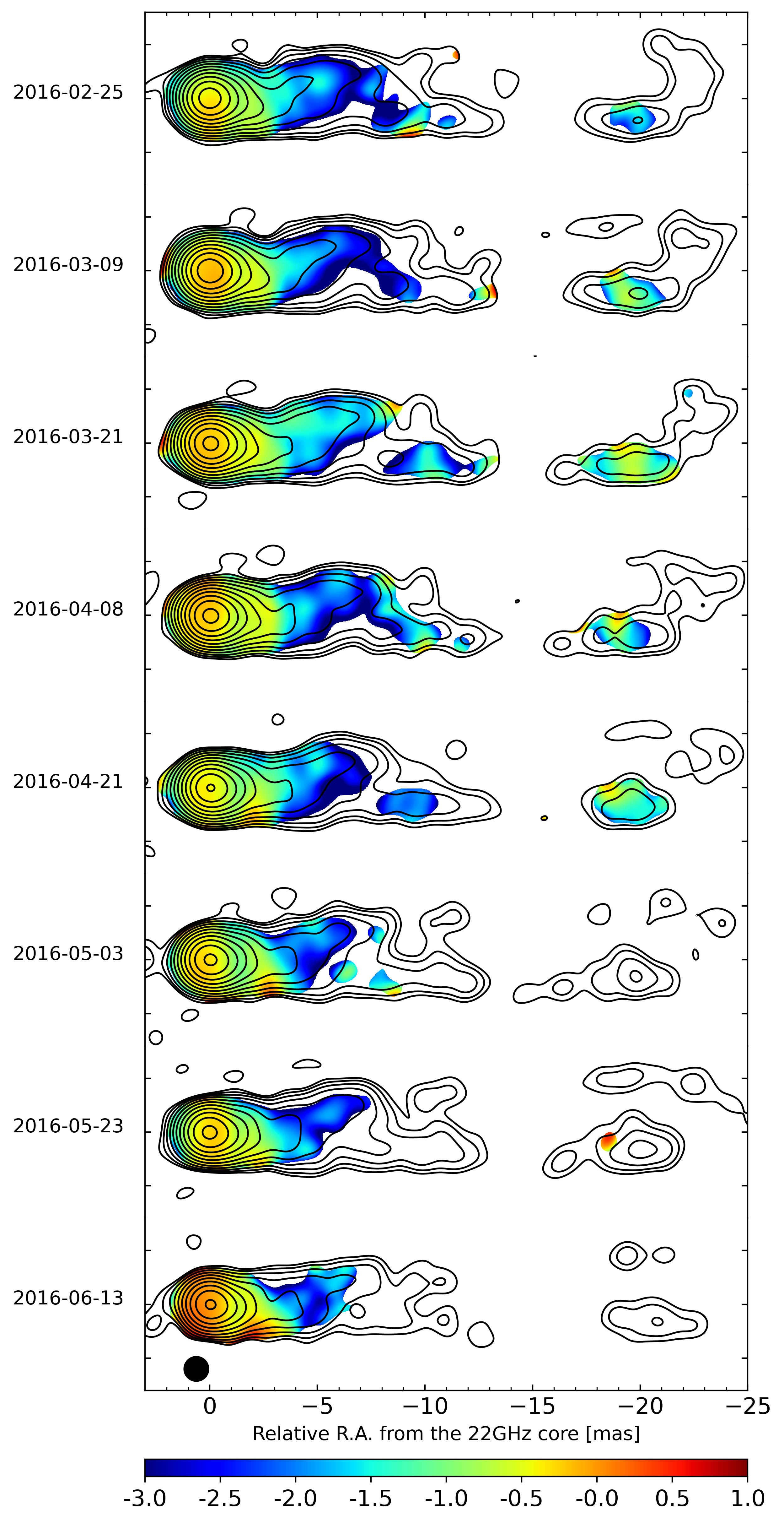}
    \caption{Spectral index maps between 22 and 43 GHz obtained from VLBA and KaVA observations. All images have been rotated by -18$\degree$. The restoring beam size is $1.2$ mas $\times1.2$ mas, drawn as a black circle in the bottom-left corner. Observing dates are shown to the left of each map. The contours represent the total intensity at 22 GHz. Contours start at 3$\sigma_{\text{rms}}$, increasing in steps of 2.}
    \label{fig:spixmap_all}
\end{figure*}

{When observations are made at two frequencies in the same array, the low-frequency data are distributed over a shorter distance in the $(u, v)-$plane than the high-frequency data (e.g., see Fig. 1 in \cite{park2019_kine} for the $(u, v)-$coverage of KaVA 22 and 43 GHz).}

{Creating spectral index maps using data with a nonidentical $(u, v)$ range can cause the spectral index to become artificially steep in the extended region, where the sensitivity of the lower frequency is higher than that of higher frequency.}
{Therefore, we excluded data on the long baseline of 43 GHz and the short baseline of 22 GHz, respectively. The resulting $(u, v)$ range of the data is 33$-$170 M$\lambda$ for KaVA and 25$-$685 M$\lambda$ for VLBA.}
{More details on the $(u, v)$ range matching and its effects are described in Appendix \ref{section.a.1}.}

Then, we restored all the maps to the same circular beam of $1.2$ mas $\times1.2$ mas ($\sim$160 $r_s$), which is the comparable size of the synthesized beam of KaVA at 22 GHz. This beam size is 1.5 times larger than the original size of the VLBA's synthesized beam. The selection of the size of the restored beam may affect the final spectral index map and further analysis. We explored this in Appendix \ref{section.a.2} and found that the main conclusions are not changed.

During the self-calibration process, the absolute coordinate position of the source is lost and the brightest component in the image (the radio core) is shifted to the center of the map. Therefore, an additional step is required to align images of different frequencies. For the M87 jet, the positions of the radio core at different frequencies have been measured using the phase-referencing technique \citep{hada2011}. According to their asymptotic relation, $r_{\text{RA}}=A\nu^{-\alpha}+B$ ($\alpha=0.94\pm0.09$, $A=1.40\pm0.16$ and $B=-0.041\pm0.012$),
the position difference between the 22 GHz core and the 43 GHz core is $0.036^{+0.013}_{-0.010}$ mas in right ascension. Assuming the jet position angle is $-72\degree$ \citep{walker2018}, then the difference in declination is $0.012^{+0.004}_{-0.003}$ mas. We took into account this shift when aligning the maps. {Recently, it has been shown that the core-shift of AGN jets could change with time \citep{plavin19}.}
{However, the phase-referencing observations of the M87 jet showed little change in 22$-$43GHz core-shifts from 2010 to 2019 \citep[$\lesssim$ 10 micro-arcsecond;][]{hada2012, hada2014, jiang21}, and thus we use a constant core-shift in our analysis.}\footnote{{The core-shift can be estimated by performing two-dimensional cross-correlation on optically thin jet regions. However, \citet[][]{pushkarev2012b} showed that this method has a large systematic alignment error in smooth, straight jets that exhibit significant spectral index gradients along the jet, such as the M87 jet. Indeed, in all epochs, we have failed to estimate reliable values of core-shift of the M87 jet using this method.}}

After aligning the images of the two frequencies, the spectral index map is created by calculating the spectral index of each pixel using the following relation:

\begin{equation}\label{eq1}
        \alpha = \frac{\log(S_{\nu_2}/S_{\nu_1})}{\log(\nu_2/\nu_1)}
,\end{equation}where $\nu_1$ and $\nu_2$ are the two observed frequencies, and $S_{\nu_1}$ and $S_{\nu_2}$ are the flux densities at each frequency. The error of the spectral index at each pixel was estimated according to the discussion in \citet{kim2014}; The error of the intensity of each pixel ($i$, $j$) is considered to be the sum of the systematic error and the thermal random noise (i.e., $\sigma_{I_{\nu,ij}} = \delta_{\nu} I_{\nu,ij} + \sigma_{\text{rms}_{\nu}}$). The factor of systematic amplitude error is empirically assumed as {$\delta \sim 10\%$} \citep[e.g.,][]{hada2012, niinuma2014, hovatta2014, cho2017}. The thermal random noise ($\sigma_{\text{rms}_{\nu}}$) is obtained from the residual map after the CLEAN process for each individual epoch. Then, the error of the spectral index at each pixel is calculated as follows: 

\begin{equation}\label{eq2}
    \sigma_{\alpha, ij}=\frac{1}{\log(\nu_{2}/\nu_{1})}\times\sqrt   { \left(\frac {\sigma_{I_{\nu_{1}},ij}} {I_{\nu_{1},ij}}\right)^{2} + \left(\frac{\sigma_{I_{\nu_{2}},ij}} {I_{\nu_{2},ij}}\right)^{2} }
.\end{equation}The spectral index error is typically $\sigma_{\alpha}\sim0.5$ near the core and increases to $\sigma_{\alpha}\gtrsim1$ in the downstream jet.
Because of the uncertainty in the core-shift, additional errors can occur during image alignment. We measured the amount of error due to image alignment for some epochs. As a result, $\sigma_{\alpha}\sim0.1$ was found in the core region but it is almost negligible in the downstream jet ($\sigma_{\alpha}\lesssim0.02$). Since the spectral index error is mainly resulting from the error of the intensity, the uncertainty of image alignment is not considered in this work.

{The left and right panels in Fig. \ref{fig:spixmap_all} are spectral index maps of the M87 jet using 22 and 43 GHz images overlaid with 22 GHz contours observed by the VLBA and KaVA, respectively. All images have been rotated by -18$\degree$ in order to align the jet central axis with the horizontal axis \citep[e.g.,][]{walker2018}. The color indicates the spectral index between 22 and 43 GHz at each pixel. We blank pixels where the total intensity is less than 3$\sigma_{\text{rms}_{\nu}}$ at one of the observed frequencies. In most maps, spectral indices can be obtained at distances up to 10 mas from the core, which is more than double than had been previously observed \citep{ly2007, niinuma2014, kravchenko20}. In particular, for some KaVA observations, spectral indices can also be obtained from structures located at $\approx$ 20 mas. However, we are not concentrating on this structure in this paper.}

It is notable that in general, the spectral index morphology appears to be quite similar over the course of the observations. While there are differences from epoch to epoch, the general structures remain remarkably consistent.

In the spectral index maps, the M87 jet shows a flatter spectrum at the core, while the extended jet has a steeper spectrum. These spectral distributions are commonly found in many AGN jets \citep[e.g.,][]{o'sullivan2009, pushkarev2012, hovatta2014, BoccardiThesis}. {We determined the weighted average spectral indices of the core from a 0.5 $\times$ 0.5 mas box centered on the location of the peak flux density. We find $\alpha_{\text{core, KaVA}}$ = $-$0.25 $\pm$ 0.50 and $\alpha_{\text{core, VLBA}}$ = $-$0.20 $\pm$ 0.55, respectively.
The spectral index values obtained from both facilities are identical within the error bars, giving an almost flat or slightly steep spectrum. 
The spectral index of the M87 jet core at 22$-$43 GHz is more or less consistent with what has been previously reported \citep{hada2012, niinuma2014, kim2018b, kravchenko20}, although it is not identical \footnote{The spectrum at the peak of the M87 jet reported by \cite{hada2012} gives a slightly inverted value of $\alpha\sim0.1$. The reason why different spectral indices were obtained despite their data being included in our study is that the size of the convolving beam used in our study is more than twice the size they used. The larger the beam size, the steeper the core spectral index is expected because the jet emission from the optically thin region is included as a structure blending effect. We confirmed that the result of \cite{hada2012} is recovered when a convolving beam of the same size as used in their study.}.

{We note that there is a sudden spectral index increase at the edge of the jet for a few epochs (March 26, 2014, and May 7, 2014). However, these are possibly spurious features because they only appear intermittently in a few epochs and then disappear immediately. Such features are frequently found in many spectral index maps \citep[e.g.,][]{o'sullivan2009, muller2011, fromm2013, hovatta2014, BoccardiThesis}. One possible reason is the sparse sampling of the $(u,v)-$coverage, especially on short ranges, leading to strong variations in the outermost structure. In the following analysis, we exclude these high spectral index features at the edge of the jet.
}}

\subsection{Spectral index distribution along the M87 jet}\label{subsection3.2}

\begin{figure}
\centering
\includegraphics[width=\hsize]{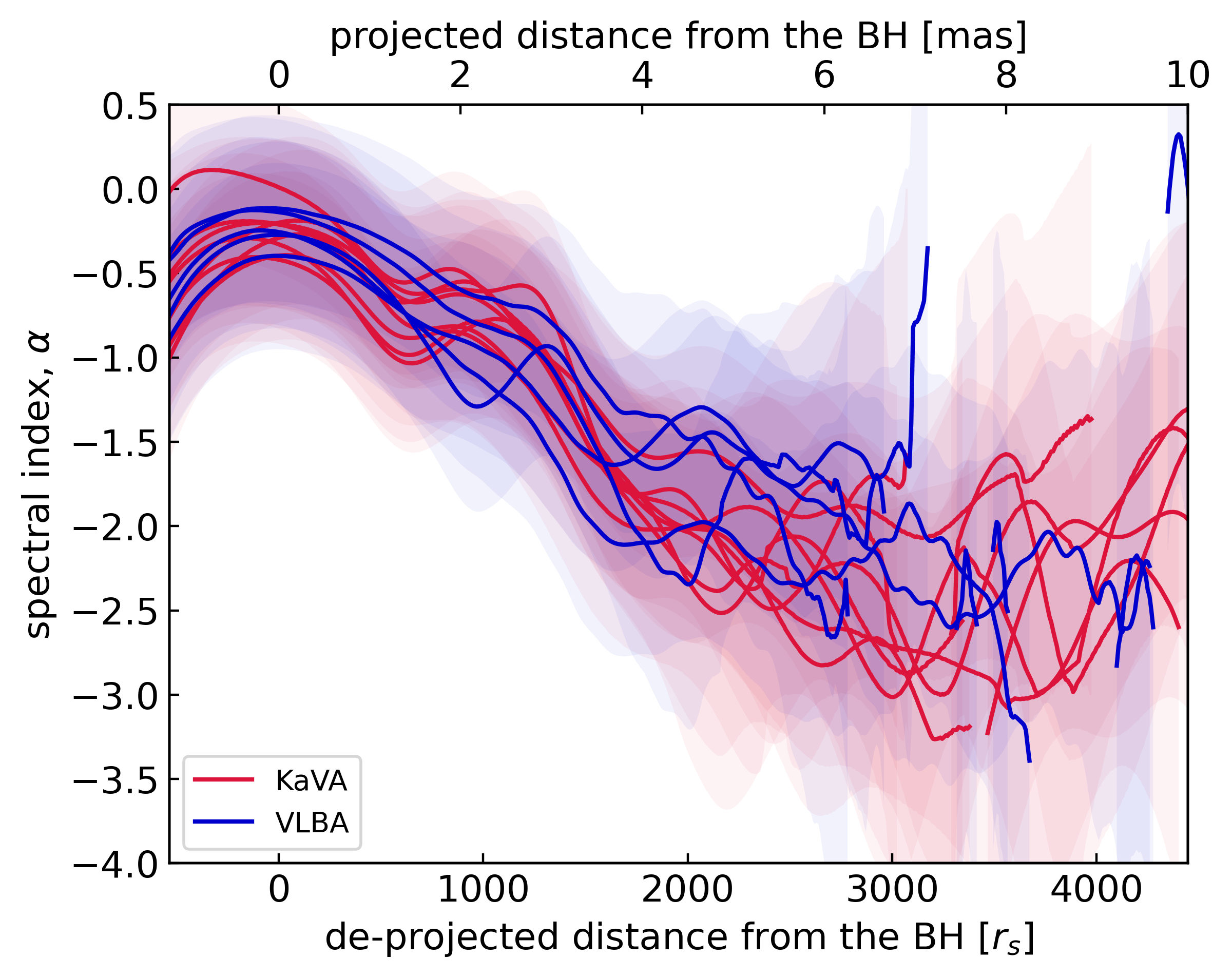}
    \caption{Evolution of the spectral index between 22 and 43 GHz along the deprojected distance from the SMBH. Weighted mean values across the jet are used in this distribution (see Sect. \ref{subsection3.2}). The projected distance is displayed on the upper axis in mas, and the corresponding deprojected distance is displayed on the lower axis in $r_s$. Red lines represent the spectral index distribution from 2016 KaVA observations and blue lines the spectral index distribution from VLBA observations from 2010 and 2014. The shaded area is the 1$\sigma$ error of the spectral index.} 
    \label{fig:radial_distribution_spectral_index}
\end{figure}

The spectral distribution along the jet can be studied in several ways. One method is to extract the spectral index values along the ridge line of the jet \citep[e.g.,][]{fromm2013, hovatta2014}. However, it is not straightforward to apply this method in our study because the M87 jet is known to consist of double ridges \citep[e.g.,][]{hada2013, hada2016}. Instead, we constructed the spectral index distribution of the jet by taking the weighted average of the spectral indices in the direction perpendicular to the jet as it moves down the jet pixel by pixel:

\begin{equation}
    \bar{\alpha} = \frac{\sum_{i=1}^{n}(\alpha_{i}\sigma_{\alpha,i}^{-2})}{\sum_{i=1}^{n}{\sigma_{\alpha,i}^{-2}}} \label{eq:weighted_mean}
,\end{equation} where $\alpha_i$ is the pixel value of the spectral index map, $\sigma_{\alpha, i}$ is the value obtained from the spectral index errors derived from Eq. \ref{eq2}, and $n$ is the number of the pixels used for averaging. When calculating the weighted mean, we used pixel values located within one beam size from the jet axis ($\pm$ 1.2 mas). The reason is that it is wide enough to cover the spectral indices of the two ridges, while at the same time excluding edge structures with spurious features at some epochs (see Sect. \ref{subsection3.1}). The error is obtained by multiplying the formal error of the weighted mean by $\sqrt{n}$ to compensate for the biases due to correlations between pixel values \citep{park2019_faraday}.

{Figure \ref{fig:radial_distribution_spectral_index} shows the radial distributions of the spectral index as a function of deprojected distance from the SMBH in units of $r_s$.} Distributions obtained from KaVA and the VLBA are indicated by red and blue lines, respectively. The positional difference between the map center and the SMBH is corrected using the results of \citet{hada2011}. We assumed a viewing angle of 17$\degree$ \citep{walker2018} to convert the observed projected jet distance to a deprojected distance, where 10 mas corresponds to $\sim$4500 $r_s$. {The spectrum of the innermost jet is relatively flat as was previously described in Sect. \ref{subsection3.1}. Downstream of the jet base the spectrum becomes steeper. At a distance of about 2 mas ($\sim$900 $r_s$), the spectral index becomes $\alpha\sim-0.7$, which is similar to the average spectral index of the M87 jet between 22$-$43 GHz reported in previous studies using the same VLBA data as used in this study \citep{hada2012, hada2016}. This value is also similar to the typical spectral index of jets observed in many AGNs \citep[e.g.,][]{pushkarev2012, hovatta2014}.}

Two interesting features are found in Fig. \ref{fig:radial_distribution_spectral_index} in the region farther than $\sim$2 mas. 
First, the spectral indices obtained from both VLBA and KaVA decrease rapidly with distance, reaching a steep spectrum of $\alpha\sim-$2.5. This high rate of decline in the spectral index with distance and the resulting very steep spectrum have not been reported in previous works on the M87 jet. 
Second, there are hints that the spectral index stops declining at $\sim$6 mas in the KaVA data.
Interestingly, a constant spectral index distribution after a certain distance is often found in many other AGN jets \citep[e.g.,][]{Kovalev2008, hovatta2014, haga2015, BoccardiThesis, Lisakov2017, pushkarev2019, baczko2019, Park2021}.
However, it is not clear that this is also seen in the VLBA data.
We cannot completely rule out the possibility that the spectral index may continue to decline in this region. There is a time difference of about $\sim$2 years between the KaVA and the VLBA observations. Therefore, the spectral distribution of the jet in this region could vary with the observation time.

\section{Discussion}\label{discussion}

\subsection{Comparison with previous works}\label{subsection4.1}

{In Sect. \ref{section3} we were able to obtain spectral index distributions of the M87 jet up to 10 mas ($\sim$ 4500 $r_s$) from the core for the first time thanks to the good image quality of KaVA and the VLBA at 22 and 43 GHz. An important characteristic of the distribution is that the spectral index decreases between 2$-$6 mas ($\sim$ 900 $r_s$ – $\sim$ 2500 $r_s$) from $\alpha\sim-$0.7 to $\alpha\sim-$2.5. In the KaVA data, it does not decrease after $\sim$6 mas.}

{The change of the spectral index can be obtained from the change in the energy distribution of nonthermal electrons $N(\gamma)$, since the slope of the electron distribution ($p$) and the spectral index is related by $\alpha = (p+1)/2$ \citep{rybicki1979}. The evolution of $N(\gamma)$ as a function of time can be investigated by solving the transfer equation (or continuity equation) including nonthermal electron injection and energy losses.}

{One previous study discussing the synchrotron spectrum of parsec-scale AGN jets has proposed two possibilities for spectral steepening using analytical solutions of the transfer equations for two nonthermal electron injection scenarios \citep[][]{hovatta2014}. (1) In the case of a constant and continuous injection of nonthermal electrons, the synchrotron spectrum has a break at $\gamma_{\text{break}}$, and the spectral index steepens by 0.5 before and after the break. (2) If there is no injection after the initial injection, the cutoff energy $\gamma_{\text{max}}$ decreases with time. At the cutoff energy, the synchrotron spectrum decreases exponentially, and thus the spectral index decreases to an arbitrarily low value.}

{However, both scenarios cannot explain the properties observed in the spectral index distribution of the M87 jet (Fig. \ref{fig:radial_distribution_spectral_index}). Firstly, the observed decrease in the spectral index is much greater than 0.5. Also, the decrease in the spectral index may not continue and may stop at a certain value, remaining at that value after that. 

{Therefore, in this study, we attempt to explain the properties of the observed spectral index distribution by numerically solving the transfer equation. Section \ref{subsection4.2} describes the transfer equation, and assumptions for physical parameters to solve the equation numerically, including magnetic field, jet radius, bulk jet velocity, and electron injection. In particular, the importance of electron injection is discussed in order to explain key features of the observed spectral index distribution. In Sect. \ref{subsection4.4} we try to constrain the magnetic field strength and the electron injection of the M87 jet at the distance of 2$-$10 mas by comparing the observed spectral index distribution to the model. Then we compare our estimate for the magnetic field strength with previous estimations.}}

\begin{figure*}
    \centering
    \includegraphics[width=0.457\hsize]{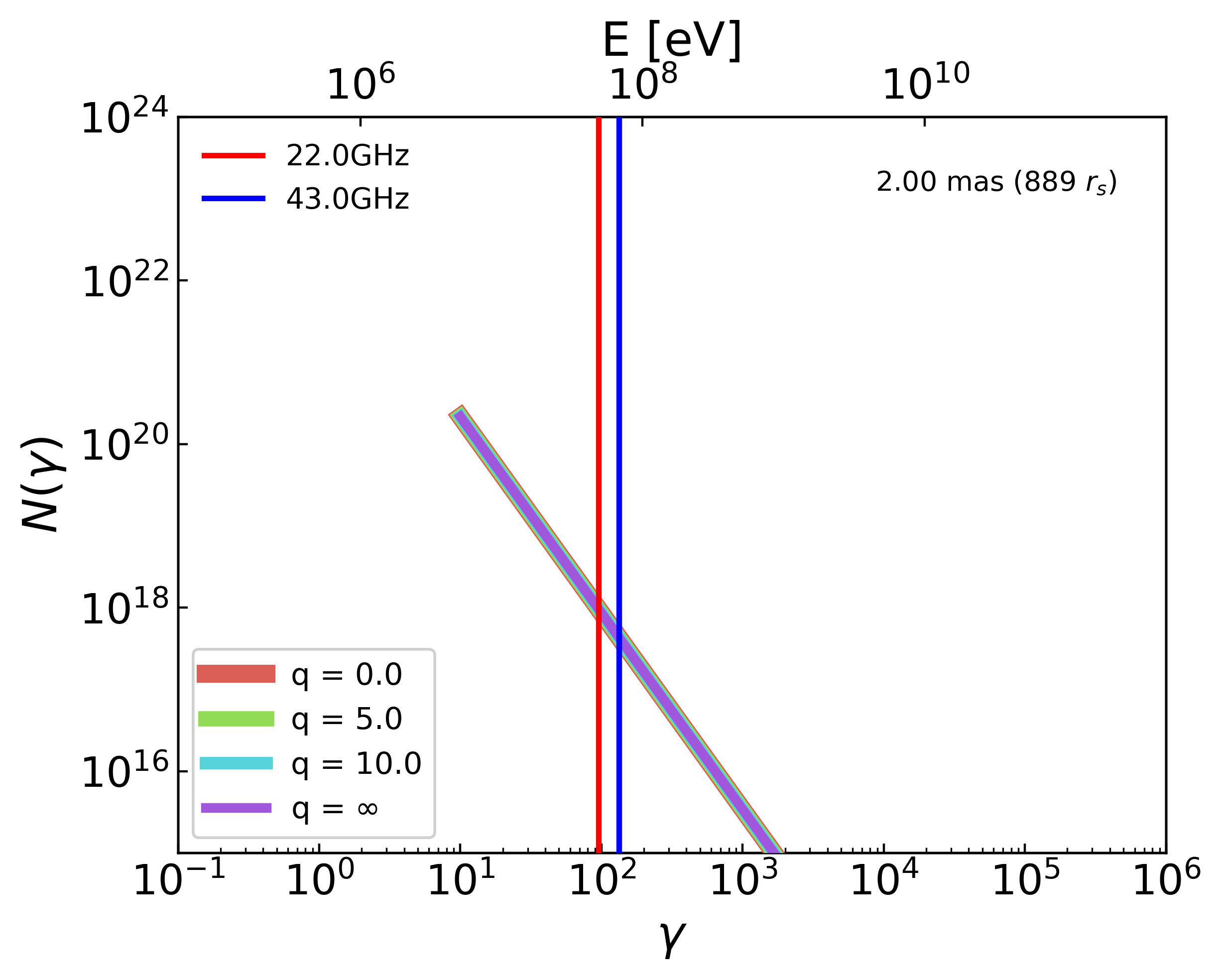}
    \includegraphics[width=0.43\hsize]{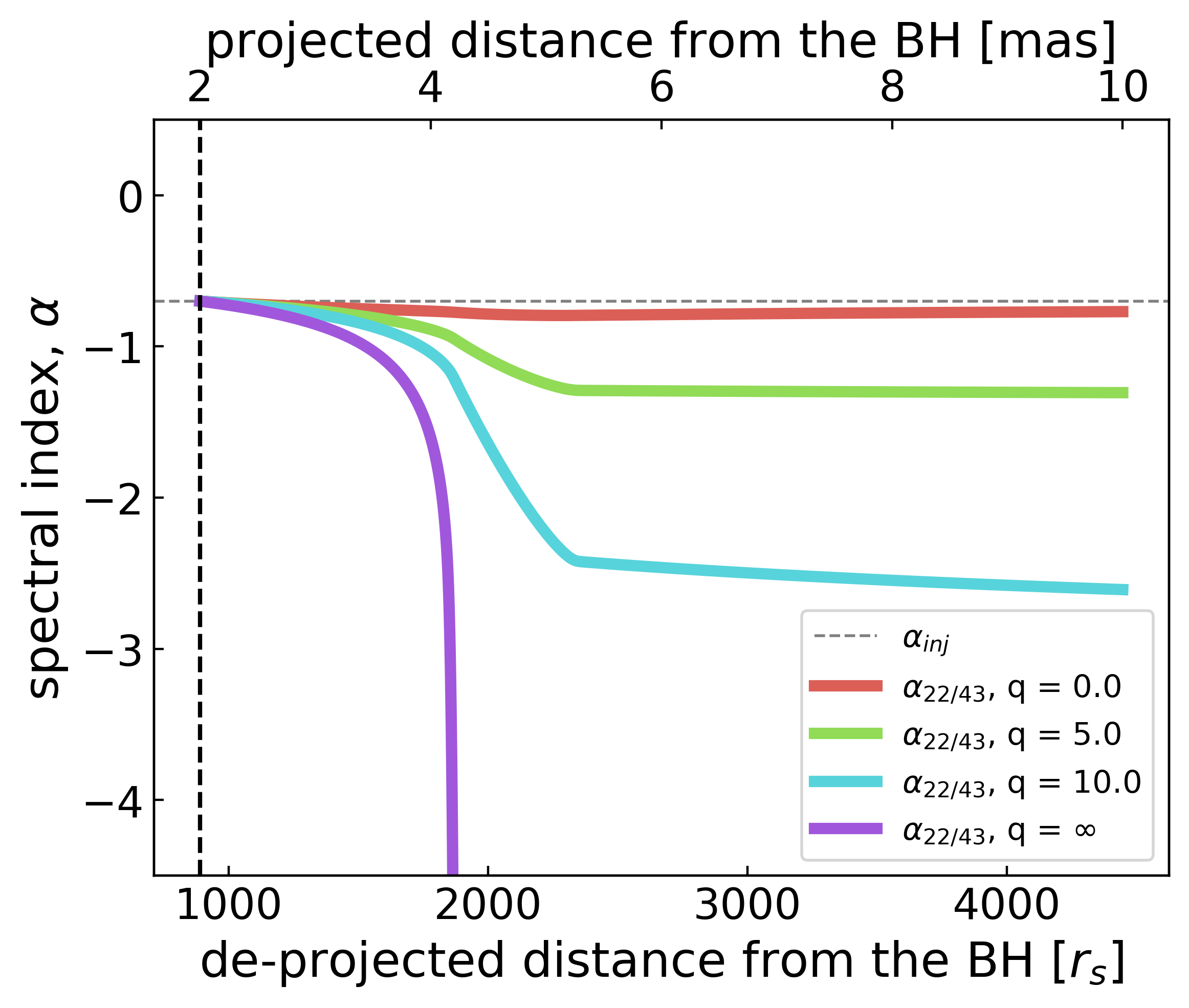}
    \includegraphics[width=0.457\hsize]{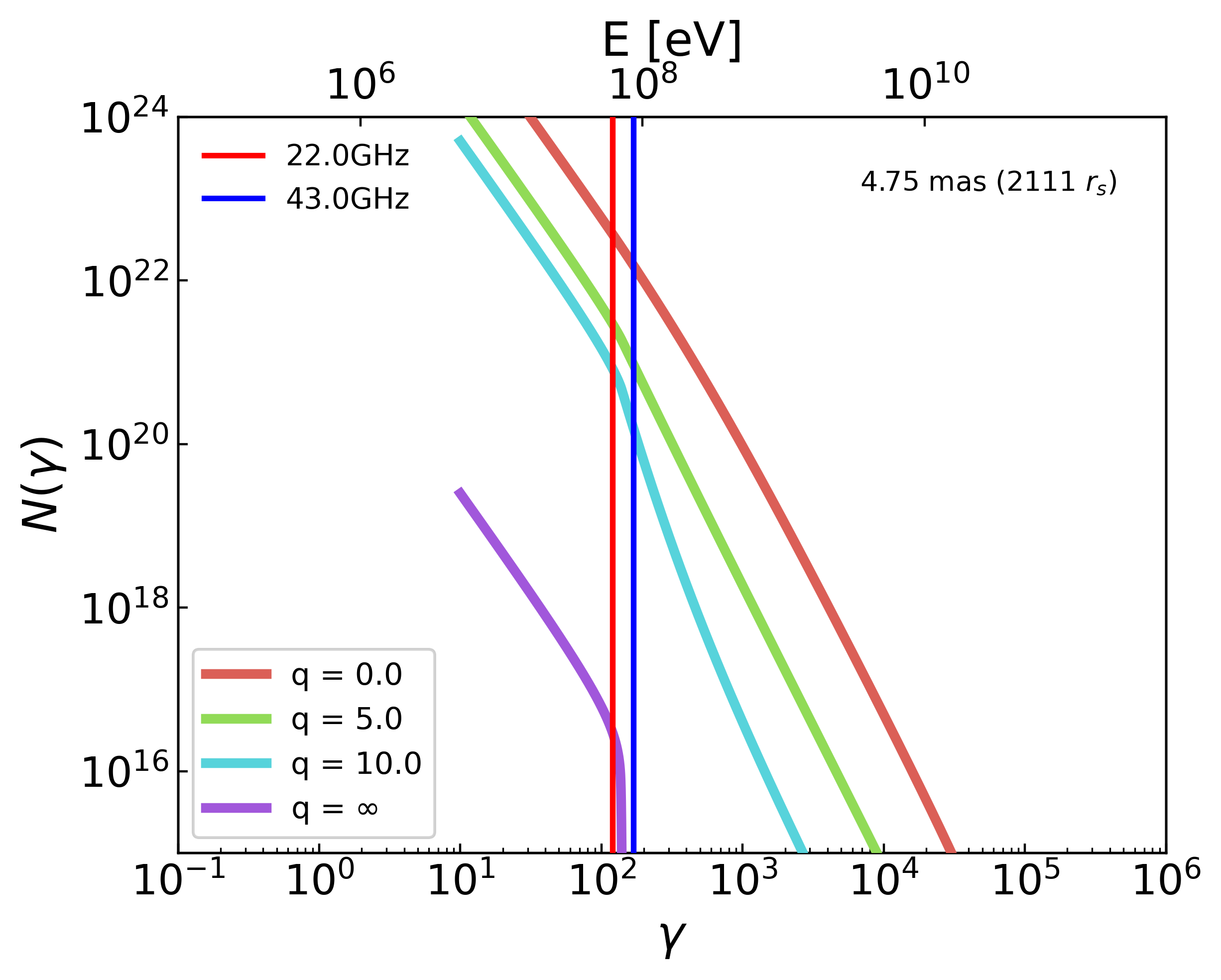}
    \includegraphics[width=0.43\hsize]{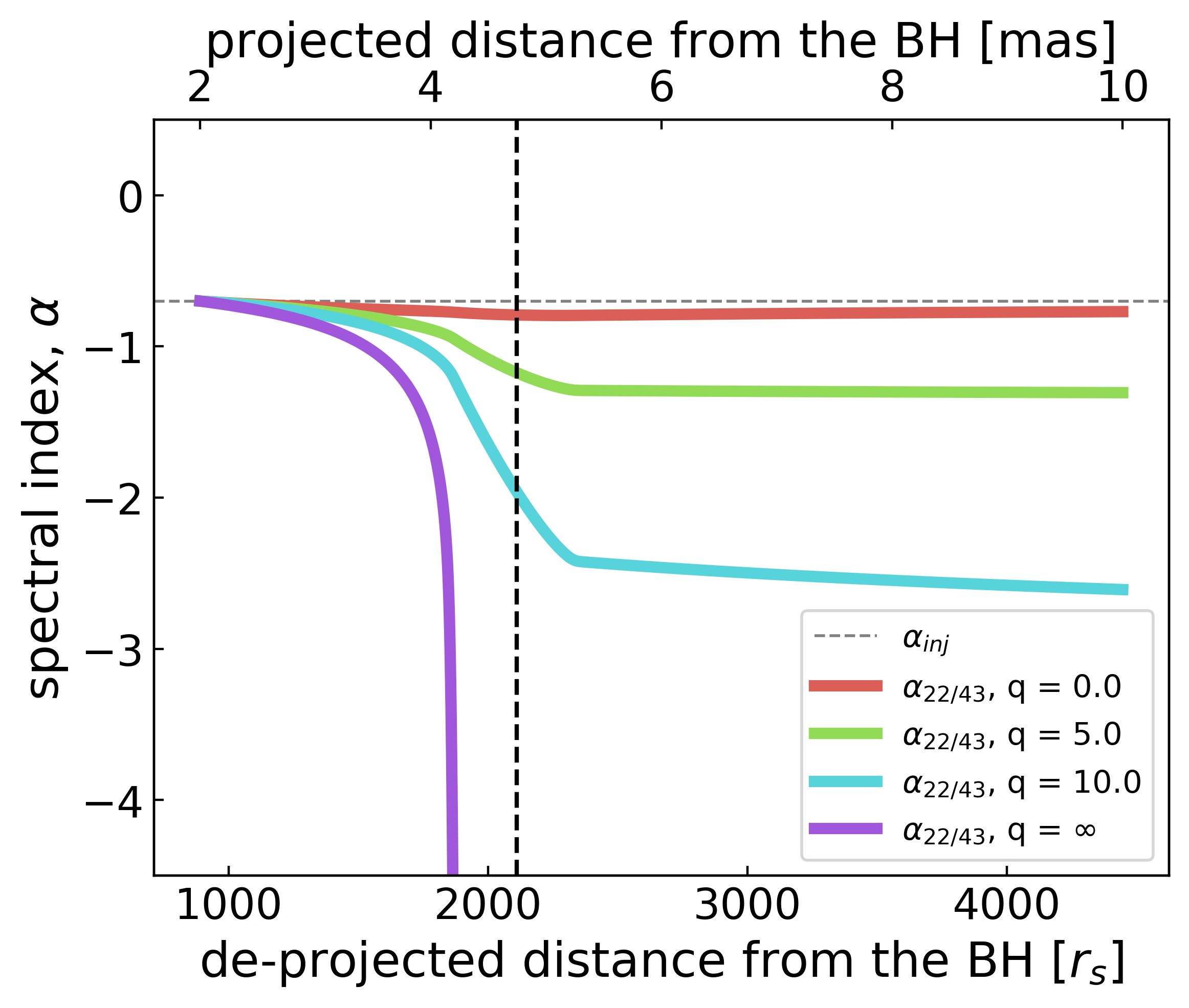}
    \includegraphics[width=0.457\hsize]{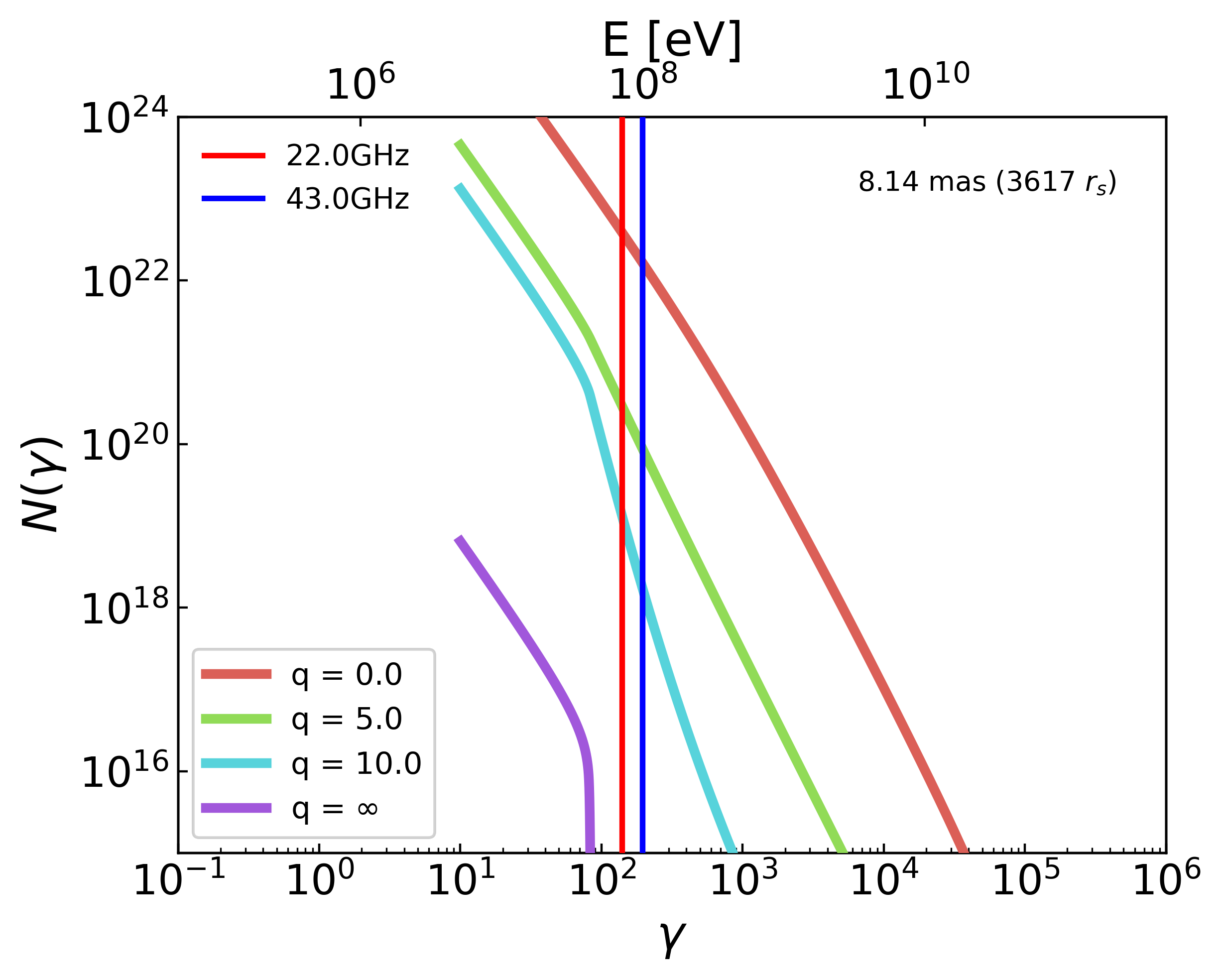}
    \includegraphics[width=0.43\hsize]{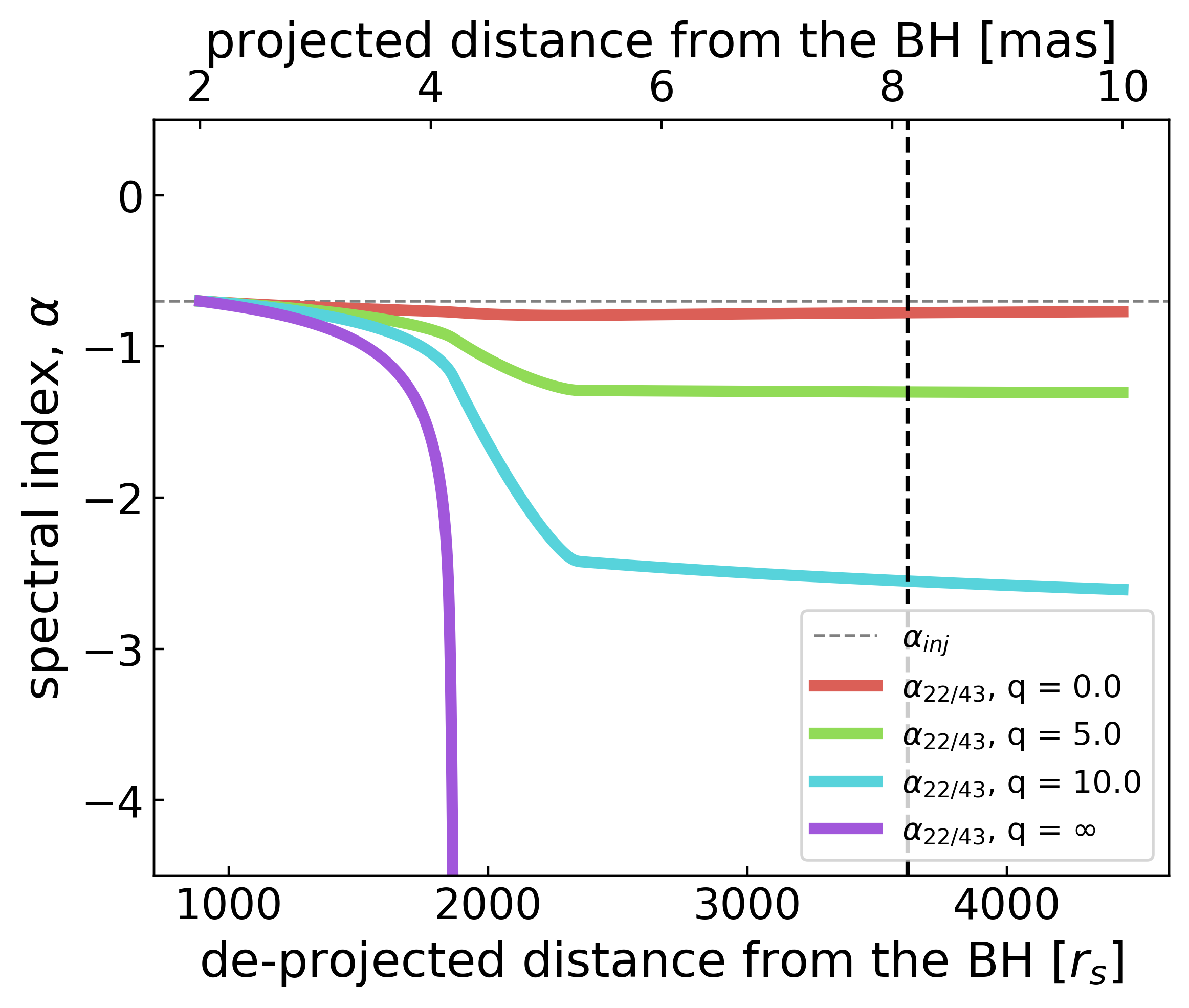}

    \caption{{Results of the modeled electron energy distributions ($N(\gamma)$) and the spectral index distributions as a function of distance along the jet. This is done to test the effect of the nonthermal electron injection rate in the jet. Left: Modeled electron number density distributions with different values of $q$ = 0, 5, 10, and $\infty$. The different colors represent different values of $q$ in the model: $q$ = 0 represents continuous new electron injections in the jet, $q$ = $\infty$  no new electron injections in the jet, and $q$ = 5 and 10 new electrons injected into the jet but the volume decreases with distance. In all cases, we set the same magnetic field distribution ($B_i$ = $0.3 \text{ G}$). The slope of the nonthermal electron injection is $p_{\text{inj}} = -2.4$. The vertical red and blue lines represent the energies of the nonthermal electrons, which emit at 22 and 43 GHz. The location down the jet where the calculations were made is written in the top right corner. Right: Modeled spectral index distributions as a function of distance, which correspond to the slope of the electron number density distributions between the vertical lines in the left panel. The vertical dashed black line is the location down the jet where the calculations were made. The horizontal dashed line is the slope of the injection function ($\alpha_{\text{inj}} = -0.7$). An animation of this figure is available}.}
    \label{fig:spectrum_index_model_different_q}
\end{figure*}

\subsection{Spectral index model}\label{subsection4.2}
{In the frame of a cross section of the jet co-moving with the jet’s flow, the transfer equation is written as \citep[e.g.,][]{ginzburg1964, longair2011, blasi2013}}

{
\begin{equation}
\frac{N(\gamma, \tau)}{\partial\tau}+(\nabla\cdot v)N(\gamma, \tau)+\frac{\partial}{\partial\gamma} [b(\gamma,\tau)N(\gamma,\tau)]=Q(\gamma,\tau)
\label{equation.5}
.\end{equation}
}
{Here, $N(\gamma, \tau)$ is the number density of the nonthermal electrons as a function of energy $\gamma$ and time $\tau$, $v$ is the velocity of the system, $Q$ is the nonthermal electron injection rate, and $b$ is the energy loss rate. In the co-moving frame, the bulk velocity along the jet direction is zero and only the expansion (or contraction) rate along the radial direction remains. In our discussion, we assume that adiabatic losses and synchrotron losses are the dominant energy loss processes. Then, $b$ can be written as \citep[e.g.,][]{rybicki1979, longair2011}} 

{\begin{equation}
    b(\gamma, \tau) = \frac{d\gamma}{d\tau} = -b_{\text{adi}}\gamma - b_{\text{sync}}\gamma^2\label{equation.6}
,\end{equation}
where $b_{\text{adi}}=\frac{1}{R} \frac{dR}{d\tau}$ and $b_{\text{sync}}=\frac{4}{3}\frac{\sigma_{T}}{m_{\text{e}} c}\frac{B^2}{8\pi}$ is the coefficient of the adiabatic losses and the synchrotron losses, respectively. Here, $R$ is the radius of the cross section, and $B$ is the magnetic field strength in the co-moving frame. We assumed that the pitch angle distribution is rapidly isotropized during synchrotron cooling\footnote{See Appendix \ref{section.a.3} for a discussion of when the pitch angle is not randomized.}.}

{In order to numerically solve Eqs. \ref{equation.5} and \ref{equation.6} along the jet, several physical quantities should be known as a function of distance: the bulk jet velocity profile ($\Gamma(z)$, where $\Gamma$ is the Lorentz factor of the bulk jet velocity), the jet radius profile ($R(z)$), the magnetic field strength profile ($B(z)$) and the nonthermal electron injection function ($Q(z)$).} Among them, we take the bulk jet velocity and the jet radius as a function of jet distance from previous observations; $\Gamma(z)\propto z^{0.16}$ \citep{park2019_kine}, $R(z)\propto z^{0.56}$ \citep{asada2012, hada2013, hada2016, nakamura2018}. The magnetic field at this distance is assumed to be dominated by the toroidal component, in which case the magnetic field strength is inversely proportional to the radius of the jet as the magnetic flux is conserved \citep[][]{Marscher2010}. Then, the magnetic field strength in the co-moving frame is \citep{lyutikov2005, zamaninasab2014}
\begin{equation}
    B(z) \propto \frac{1}{R\Gamma} = B_i\left(\frac{z}{z_i}\right)^{-0.72}
,\end{equation} where $B_i$ is the strength of the magnetic field at $z_i$.

{The energy spectrum of the nonthermal electron injection function, $Q(\gamma)$, is generally assumed to be a power law with a slope $p_{\text{inj}}$ limited to the minimum and maximum energies, $\gamma_{\text{inj min}}$ and $\gamma_{\text{inj max}}$. In addition to this, we assume that the number of nonthermal electron injections varies with distance with a power of $-q$:}
\begin{equation}
    Q(\gamma, z) = Q_0 \gamma^{p_{\text{inj}}} \left(\frac{z}{z_i}\right)^{-q},
    \\
    \gamma_{\text{inj min}} < \gamma < \gamma_{\text{inj max}}
,\end{equation}
{where $Q_0$ is the nonthermal electron injection rate at $z_i$ for electrons of $\gamma = 1$. For example, if $q = 0$, the same amount of nonthermal electrons as $Q(z_i)$ is injected along the jet continuously. For $q = \infty$, there is no additional injection after the initial injection at $z_i$. The above two cases resemble the two scenarios of the spectral steepening suggested by \cite{hovatta2014}. If $0< q< \infty$, nonthermal electrons are injected continuously, but the injection volume decreases with distance. {There are indeed theoretical models that suggest that electron acceleration along the jet is possible; one is the boundary shear acceleration between 
the jet and the surrounding matter \citep{ostrowski1998}, while the other is  
Kelvin-Helmholtz instability driven magnetic reconnection
\citep[e.g.,][]{sironi2021}.}
In Sect. \ref{subsection4.3}, we discuss the effect of the electron injections on the spectral index distribution.}

{Given $B_i$ and $q$, Eqs. \ref{equation.5} and \ref{equation.6} are numerically solved and $N(\gamma, \tau)$ is obtained, and the spectral index distribution $\alpha(z)$ between the two observed frequencies is obtained from $N(\gamma, \tau)$. Throughout the paper, the initial and final distance $z_i$ and $z_f$ are set as $\sim900$ $r_s$ and $\sim4500$ $r_s$, which is the deprojected distance of 2 mas and 10 mas of the M87 jet, respectively. The minimum and maximum energy of the injection function, $\gamma_{\text{inj min}}$ and $\gamma_{\text{inj max}}$, is $10$ and $10^5$, and the energy slope of the injection function is $p_{\text{inj}}=-2.4$ $ (\alpha_{\text{inj}}=-0.7),$ which is similar to the average value of the observed spectral index at 2 mas (Fig. \ref{fig:radial_distribution_spectral_index}). As $Q_0$ does not affect the spectral index we arbitrarily selected $Q_0$ as $1.00\times10^{20}$. Details of the model are described in Appendix \ref{section.a.4}. }

\begin{figure*}
    \centering
    \includegraphics[width=0.46\hsize]{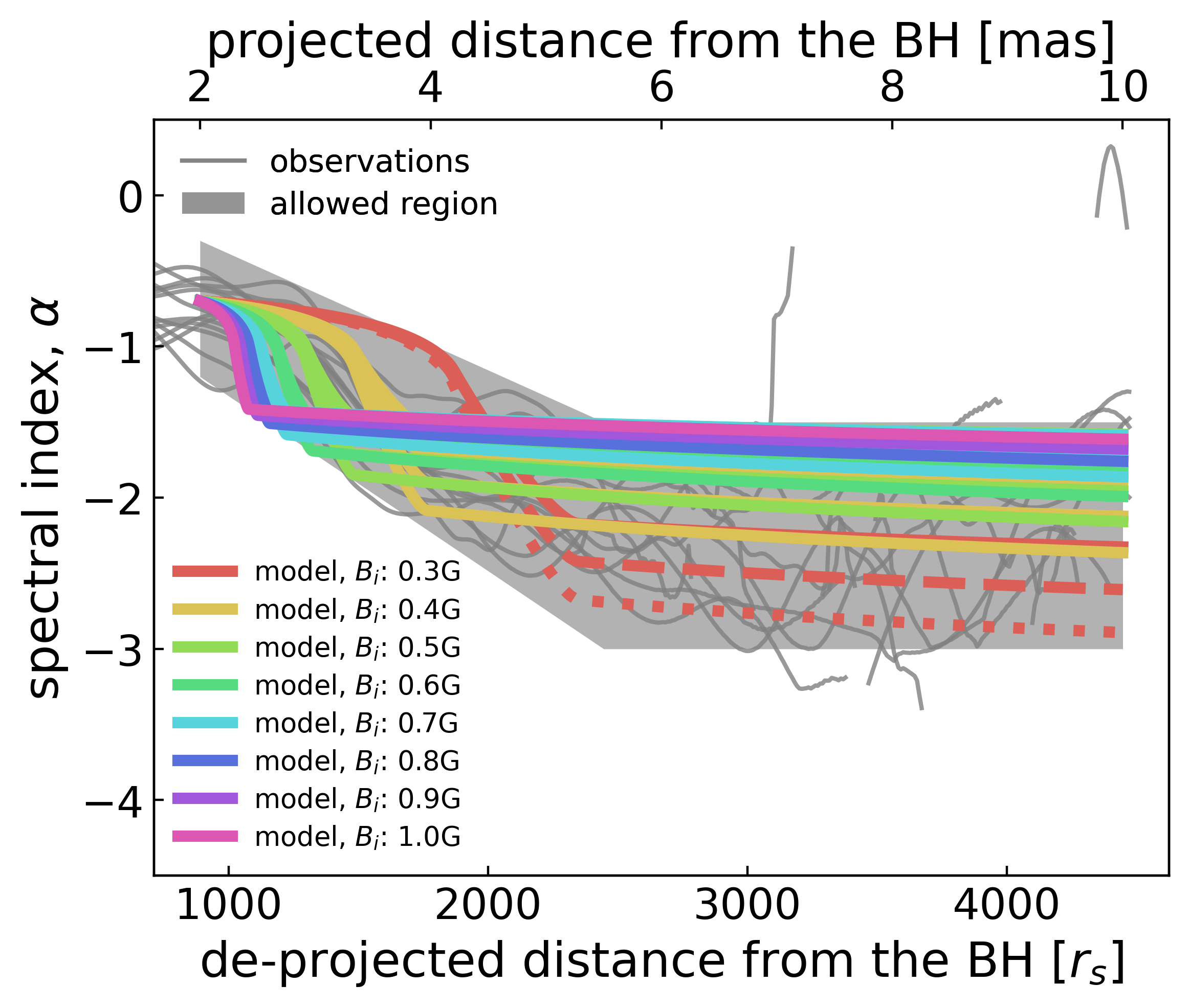}
    \includegraphics[width=0.46\hsize]{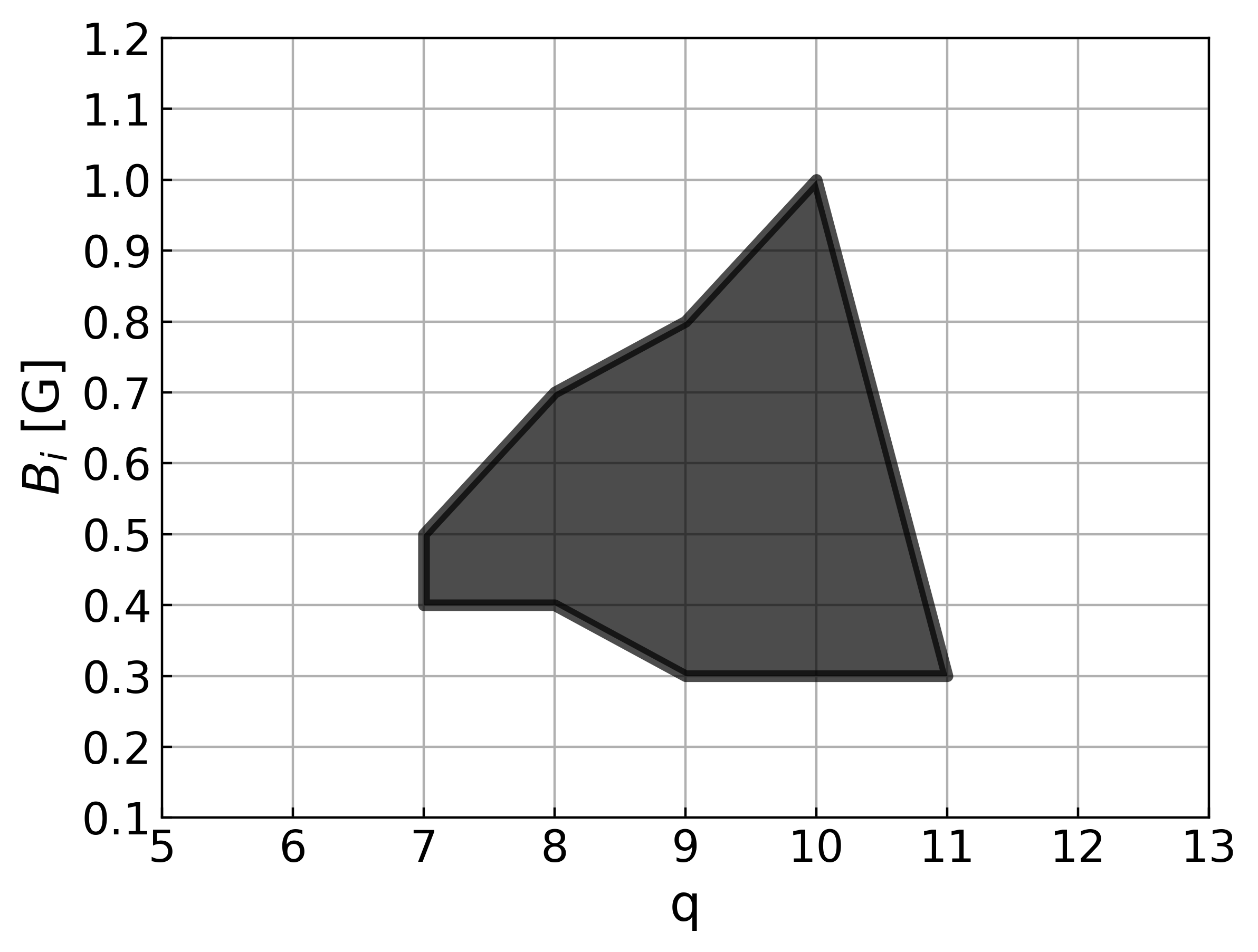}
    \caption{Constraining the magnetic field strength and nonthermal electron injection of the M87 jet. Left: Modeled spectral index distributions constrained by the observations. The distributions must lie in the gray shaded allowed region (see Sect. \ref{subsection4.4} for details). Colored lines are the modeled spectral index distributions. Different colors represent different initial magnetic field strengths, $B_i$. Right: Parameter space ($B_i$, $q$). The black region is the allowed values of $B_i$ and $q$ constrained by the observations. The dashed and dotted red line is a modeled spectrum index distribution with $q$ = 10 and $q$ = 11, respectively, with $B_i=0.3\text{ G}$.}
    \label{Fig:parameter_space}
\end{figure*}

\subsection{The effect of the nonthermal electron injections on the spectral index distribution}\label{subsection4.3}
If we assume that the flattening of the spectral index seen in Fig. \ref{fig:radial_distribution_spectral_index} is real, we can test the effect of the nonthermal electron injections on the spectral index distribution. If this assumption is wrong, we discuss the implications of this in Sect. \ref{subsection4.4}. We made a number of spectral index distribution models by adopting different values of $q$. Here, we set the four different cases of the nonthermal electron injection function in the jet to be $q$ = 0, 5, 10, and $\infty$, while the magnetic field distribution is assumed to be the same $B_i$ with a value of $0.3$ G.

{Figure \ref{fig:spectrum_index_model_different_q} summarizes the results of the test models. The left panel shows the evolution of electron energy distribution $N(\gamma)$ as a function of the jet distance drawn as a vertical dashed line in the right panel. The color of the distribution represents the value of $q$ used in the model. A blue and a red vertical line represent the electron energies emitting synchrotron radiation at 22 and 43 GHz in the observer's frame. The slope of the nonthermal electron injection is $p_{\text{inj}}=-2.4$. The right panel of Fig. \ref{fig:spectrum_index_model_different_q} shows the spectral index distribution between 22$-$43 GHz versus jet distance, $\alpha_{\text{22-43 GHz}}(z)$, obtained from $N(\gamma, z)$. The horizontal dashed line represents the spectral index of the injection function ($\alpha_{\text{inj}}=-0.7$).
}

In the absence of electron injection ($q = \infty$), cutoff energy is shown in $N(\gamma, z)$ where no higher energy electrons can be found. The cutoff energy moves to lower energy over distance. At $z \sim2000$ $r_s$ (middle left panel), the cutoff energy is located in between $\gamma(\nu_{\text{obs}}=43\text{ GHz})$ and $\gamma(\nu_{\text {obs}}=22\text{ GHz})$. Therefore, $\alpha_{\text{22-43 GHz}}(z)$ diverges to $-\infty$ around this distance. On the other hand, in the constant injection case ($q = 0$), the electrons still exist beyond the cutoff energy due to the newly injected one. Therefore, the spectral index changes only slightly with distance.\footnote{Note that $\left|\Delta \alpha\right|$ in our calculations is much less than 0.5, which was suggested by \cite{hovatta2014}. However, $\left|\Delta \alpha\right| = 0.5$ is achieved only if the magnetic field strength is constant over jet distance and no adiabatic losses exist. When the magnetic field strength decreases with distance and adiabatic losses exist, $\left|\Delta \alpha\right|$ is reduced.}

In the case of $q = 5$ and $10$, the amount of newly injected electrons decreases with distance, thus the slope of $N(\gamma, z)$ at energies higher than the cutoff energy, is steeper than in the case of $q = 0$. As a result, the energy distribution becomes a broken power law, where the position of the break ($\gamma_{\text{break}}$) is the same as the cutoff energy in the $q = \infty$ case. Interestingly, $\alpha_{\text{22-43 GHz}}(z)$ for $q = 5$ and $10$ show similar trends to the observed distributions, that is, $\alpha$ decreases until $\sim5$ mas and no longer decreasing or slowly decreases after that. This is because $\gamma_{\text{break}}$ first passes through $\gamma(\nu_{\text{obs}}=43\text{ GHz})$ and then $\gamma(\nu_{\text {obs}}=22\text{ GHz})$, and accordingly, the slope between these two frequencies shifts from the first slope of the broken power-law to the second. The larger $q$, the steeper the second slope of $N(\gamma)$, so the change in $\alpha$ will be greater. For example, if $q = 10$ then $\left|\Delta \alpha\right| \sim 2$, which is far greater than the expectation by \citet{hovatta2014} ($\left|\Delta \alpha\right| = 0.5$), and similar to our observations.

{From this experiment, we found that our model can reproduce the characteristic trend of the spectral index distribution seen in the KaVA observations for cases where the amount of nonthermal electron injections have a distance dependence. The combination of nonthermal electron injections and energy losses creates a broken power-law energy distribution in which the slope beyond $\gamma_{\text{break}}$ is determined by the change in the amount of injection with distance (i.e., the parameter $q$ of the electron injection function). The larger the $q$, the greater the slope beyond $\gamma_{\text{break}}$ and the change in $\alpha$. The continuous injection and non-injection cases proposed by \cite{hovatta2014} correspond to two extreme cases in our model ($q = 0$ and $\infty$, respectively). In the former case, the change in $\alpha$ is the smallest, and in the latter, the break energy becomes the cutoff energy with no electrons above it, making the spectral index diverge.}

\subsection{Constraining the magnetic field strength and nonthermal electron injection along the M87 jet}\label{subsection4.4}

{Now we constrain the physical quantities of the M87 jet using our numerical model. In order to compare the modeled spectral index distributions with the observations quantitatively, we define an ``allowed region'' based on the observational spectral index distributions with three parameters: (1) The spectral index range at $z_i$ ($\alpha_i$), (2) the location of the end of the steepening region ($z_{\text{s}, f}$), and (3) the spectral index range after the steepening region ($\alpha_f$). The gray shaded region in Fig. \ref{Fig:parameter_space} is the allowed region of the spectral index of the M87 jet drawn with parameters of -1.2 $<\alpha_i<$ -0.3, $z_{\text{s},f}=$ 5.5 mas, -3.0 $<\alpha_f<$ -1.5. More than 90\% of the observed profiles lie within the allowed region.}

{We then create many modeled spectral index distributions for different values of the parameters $B_i$ and $q$. The parameters $B_i$ were set from $0.1$ G to $1.2$ G in $0.1$ G increments, and the parameter $q$ was set from 5 to 13. We limited the parameters by excluding models that exist outside the allowed range. The colored lines in the left panel of Fig. \ref{Fig:parameter_space} are the modeled spectral index distributions that lie completely inside the allowed region. The color indicates the initial magnetic field strength used in the model. In a strong magnetic field, the steepening region is located closer to the core, and the spectral index decreases less. As the magnetic field strength becomes weaker, the spectral index decreases over a longer distance, and the change in the spectral index are greater. For the same magnetic field strength, $q$ essentially determines the final spectral index as discussed in Sect. \ref{subsection4.3}. For example, for the same $B_i = 0.3 \text{ G}$, the model with $q = 11$ (red dashed line) have steeper spectral index distribution than the $q = 10$ case (red dotted line). By constraining the modeled distributions to the allowed region, $B_i$ is constrained to $0.3 \text{ G}<B_i<1.0 \text{ G}$ and $q$ is constrained to $7<q<11$. The right panel of Fig. \ref{Fig:parameter_space} shows the allowed parameter space of $B_i$ and $q$.}

{We note that our method is highly sensitive to the choice of the allowed region. In Sect. \ref{subsection3.2} we found that a constant spectral index distribution from the VLBA data is less clear, so it is possible that the spectral index distribution could continue to decay farther down the jet (blue lines in Fig. \ref{fig:radial_distribution_spectral_index}). {This would suggest two possibilities:}
{(1) Spectral index flattening still occurs but farther downstream. In this case, the model indicates a weaker $B_i$.}
{(2) The spectral index declines indefinitely. In this case, this would suggest that no electrons are injected beyond $\sim6$ mas.}
{Considering that there are several years between the VLBA and KaVA observations, there is a possibility that the magnetic field strength and the electron injection rate could change with time.}}
{Since 2017, we have been monitoring the M87 jet with the East Asia VLBI Network \citep[EAVN;][]{cui2021} at 22 and 43GHz quasi-simultaneously with better sensitivity and $(u,v)-$coverage than KaVA. These observations could help resolve these possibilities.}

{\subsection{Radial magnetic field distribution of the M87 jet}}\label{subsection4.5}

\begin{figure*}
\centering
\includegraphics[width=0.9\hsize]{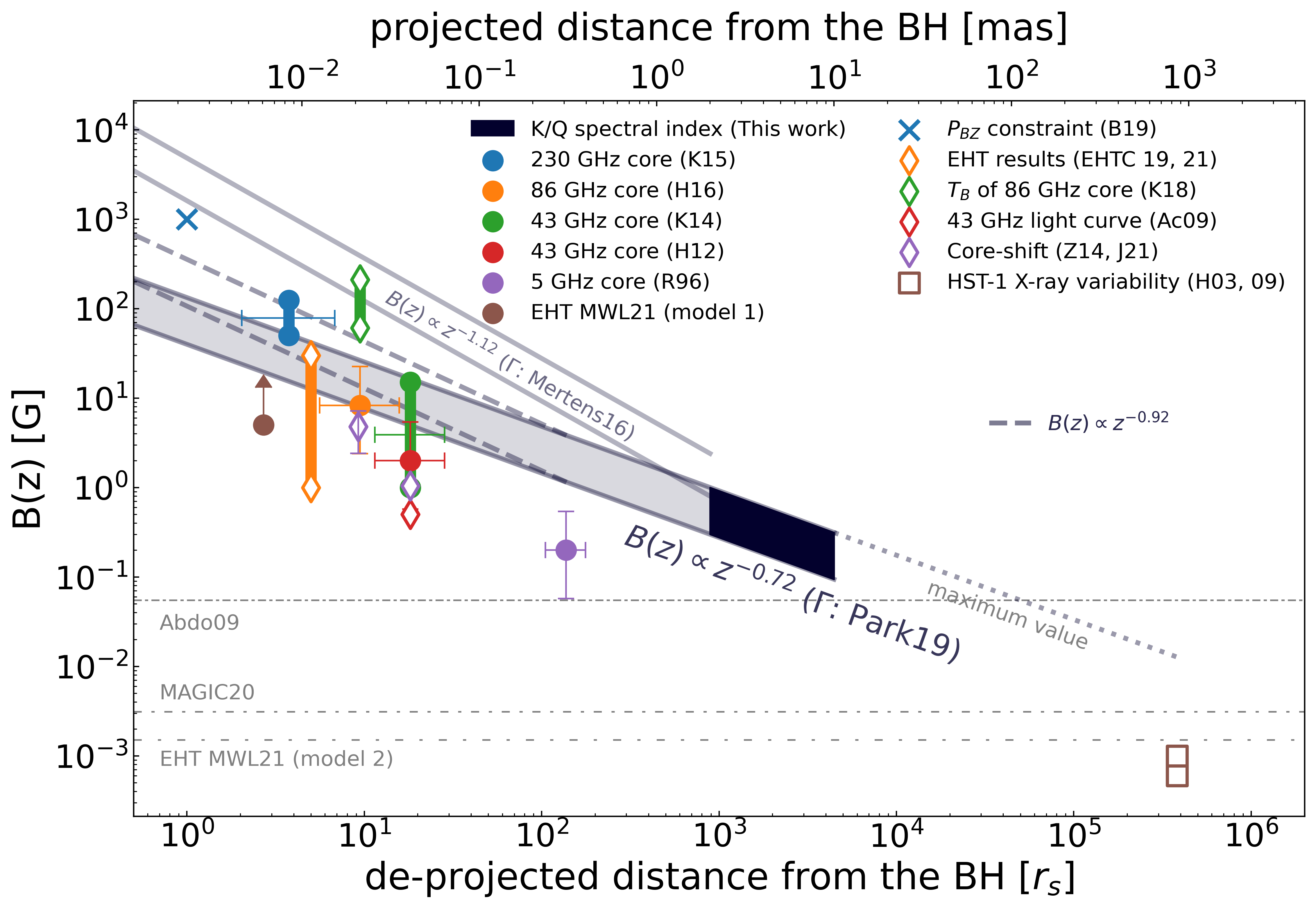}
    \caption{Magnetic field distribution of the M87 jet. The black area is the distribution of the magnetic field strength estimated in this study. The solid lines with the gray shaded region are the extrapolation of the upper and lower limits of the result in the upstream direction ($B(z)\propto z^{-0.72}$). The dashed lines are the magnetic field distribution which assumes a steeper slope of the jet radius profile in the region closer to the SMBH ($B(z)\propto z^{-0.92}$). The solid lines without a shaded region are the upper and lower bounds of the magnetic field distribution, assuming fast acceleration in the region closer within 2 mas ($B(z)\propto z^{-1.12}$). The dotted line is the extrapolation of the maximum values of the result to the downstream region. Markers are estimated magnetic field strength from previous studies. The blue cross is the theoretically expected magnetic field strength in a Blandford-Znajek jet \citep{blandford19}. The circles are magnetic field strength values obtained in VLBI cores derived using SSA theory \citep{kino2014, kino2015, hada2012, hada2016, reynolds1996}. The diamonds are values obtained from VLBI observations using various methods other than SSA theory \citep{acciari2009, kim2018, zamaninasab2014, eht2019, eht2021, ehtMWL2021, jiang21}. The squares are the magnetic field strengths at HST-1 \citep{harris2003, harris2009}. The horizontal dotted-dashed lines are the magnetic field strengths estimated by fitting the broadband SED of the jet using a single-zone SSC emission model \citep{abdo2009, magic2020, ehtMWL2021}.}
\label{Fig:Bdistribution}
\end{figure*}

{In Sect. \ref{subsection4.4} we successfully inferred the magnetic field strength from the modeled spectral index distributions in the M87 jet from 2$-$10 mas ($\sim900$ $r_s-\sim4500$ $r_s$ in the deprojected distance). Therefore, the strength of the magnetic field evolves a function of distance as\begin{equation}
B(z) = \left(0.3-1.0\text{ G}\right)\left(\frac{z}{\text{2 mas}}\right)^{-0.72},
\\
2 \text{ mas}<z<10 \text{ mas},\label{equation.8}\end{equation}}where exponent -0.72 comes from Eq. (\ref{equation.6}).

{Figure \ref{Fig:Bdistribution} shows the magnetic field strength of the M87 jet as a function of the distance from the SMBH. The magnetic field distribution obtained by our models and the observations is drawn as a black area. The figure also summarizes the magnetic field strengths at various locations of the M87 jet using observations over a wide range of wavelengths. The cross represents the theoretically expected magnetic field strength at the black hole surface based on Blandford-Znajek power estimation \citep{blandford19}. \footnote{{This is somewhat larger than those estimated from nonthermal emission properties.
This difference is probably due to the difference in the location of the jet of interest between radiation-based and energetic-based magnetic field estimation.
The limb-brightened structure of the M87 jet \citep[e.g.,][]{junor1999, kovalev2007, hada2016, kim2018, walker2018} suggests that the majority of the radio emission is generated  
near the boundary of the jet, where the magnetization parameter 
$\sigma \left(=\frac{B^2}{4\pi \rho_i c^2}\right) \lesssim 1$ holds \citep[e.g.,][]{nakamura2018, chatterjee19, narayan2022}.
On the other hand, the estimation by \citet{blandford19} 
seems to focus on the high $\sigma$ region.}
}
A recent estimation by \citet{kino22} also indicates a similar magnetic field strength in this region.
The circles represent magnetic field strengths in radio cores at various VLBI frequencies, obtained based on SSA theory \citep{kino2014, kino2015, hada2012, hada2016, reynolds1996}.\footnote{{The upper limit of the magnetic field strength of \citet{kino2015} $\sim 124$ G critically depends on their assumption of $P_{\text{jet}} \sim 5\times10^{44}$ erg/s. If smaller a jet power is assumed ($P_{\text{jet}} \sim 1\times10^{44}$ erg/s), then the maximum magnetic field strength is reduced to $\sim65$ G.}}\footnote{\citet{ehtMWL2021} used two SED fitting models oriented at different wavelengths. One is oriented at radio wavelengths (model 1) and the other is at VHE emissions (model 2). Since model 1 describes the radio core flux using the conventional SSA model, we plot the value as a circle. We also note that the magnetic field strength of model 1 is a lower limit since they ``maximally tuned'' their parameters in order to maximize the gamma-ray flux.}
Here, the distance from the core to the SMBH and its error are determined using the result of core-shift \citep{hada2011}.} 
{The error of the magnetic field strength is calculated by applying 25\% error to the radio core size and 5\% error to the flux of the core \citep{hada2013}. For the data shown as a ranged value \citep[][]{kino2014, kino2015}, the error bars of the field strength is not displayed.}
{The diamonds indicate field strengths obtained with VLBI observations using the core-shift method \citep{zamaninasab2014, jiang21}\footnote{\citet{zamaninasab2014} calculated the magnetic field strength at 1 parsec based on the assumption of a conical jet, which is incorrect for the M87 jet. Since this causes a larger error in the magnetic field strength when farther away from the core, we have shown the magnetic field strength at the 43 GHz core instead. \citet{jiang21} measured the magnetic field strength at the 88 GHz core.}, light curve modeling \citep{acciari2009}, brightness temperature \citep{kim2018}, and EHT observations \citep{eht2019, eht2021}.} {The field strengths at HST-1 estimated by applying the synchrotron cooling time to the variability of the X-ray band are drawn as squares \citep{harris2003, harris2009}.}

{The solid lines with gray shaded region are the extrapolation of the upper and lower bounds of the magnetic field distributions estimated in this study ($B(z)\propto z^{-0.72}$; Eq. (\ref{equation.8})).}
{The dashed lines represent the distribution that reflects a steeper slope of the jet radius profile in the region close to the SMBH as $R(z)\propto z^{-0.76}$ (accordingly, $B(z)\propto z^{-0.92}$) \citep{hada2013}.}
{Interestingly, 
our estimation of $B(z)$ on $900-4500~r_{s}$ scale
smoothly connects to values based on synchrotron emission down to the EHT observing scale,
keeping $B(z)\propto z^{-0.72}$.}
{The consistency of the slope of the magnetic field distributions with distance may imply that the majority of the magnetic flux of the jet near the SMBH is preserved without serious dissipation out to $\sim4500~r_s$.}

\citet{mertens2016} reported that the M87 jet shows linear acceleration in the region closer than 2 mas, meaning that the bulk jet velocity profile in this region is proportional to the jet radius ($\Gamma(z) \propto R(z) \propto z^{0.56}$). The solid lines without shaded regions are the upper and the lower bounds of the magnetic field distribution estimated by applying it. In this case, due to the fast acceleration, the magnetic field strength at 2 mas should increase to $0.8 - 2.4 \text{ G}$ in order to achieve a similar amount of spectral index steepening. Also, the slope of the magnetic field distribution becomes steeper ($B(z)\propto z^{-1.12}$) in the region closer than 2 mas. As a result, a very strong magnetic field distribution is inferred that deviates from the magnetic field strengths estimated by observations and theory.

Extending our scaling relation downstream (dotted line) appears to be considerably higher than the magnetic field strength of HST-1. HST-1 is a complex structure in which re-collimation occurs, and the magnetic field strength of the upstream jet before HST-1 is expected to be smaller than that of HST-1. Therefore, the magnetic field strength measured in HST-1 should be considered as an upper limit.
Given that our extrapolated maximum value to the location of HST-1 is roughly an order of magnitude higher than the upper limits set by X-ray variability studies in HST-1, this suggests that assumptions such as the conservation of the magnetic flux along the jet may be wrong.
For example, a recent large-scale jet simulation showed that pinch instability can lead to magnetic flux dissipation along the jet \citep[][]{chatterjee19}.

{The horizontal dotted-dashed lines are the magnetic field strengths estimated by fitting a single-zone synchrotron self-Compton (SSC) emission model to the broadband spectral energy distribution (SED) of the jet, specifically focusing on the very high energy (VHE) gamma-rays \citep{abdo2009, magic2020, ehtMWL2021}.}
{Comparing these values directly with the magnetic field distributions in our study, the estimated location of the VHE emission is in the downstream jet between $z\sim4500$ $r_s$ and $\sim 10^5$ $r_s$, which differs significantly from the most favored location of VHE flare activities \citep[a few tens of $r_s$ from the SMBH; see, e.g.,][]{acciari2009, abramowski2012, hada2014, akiyama2015}.}
{One possible explanation is that VHE gamma rays are emitted from the outer layer of the radio-emitting region \citep[e.g.,][]{tavecchio2008}.}
{Indeed, a recent simultaneous broadband SED study found that the VHE emitting region is not overlapping with the radio-emitting region, and requires a larger area \citep{ehtMWL2021}.}
{However, there are various physical processes that suggest the possibility of the VHE emission not only originating in the innermost jet base but also in the downstream jet (e.g., the HST-1 knot) \citep[see][and references therein]{abramowski2012, rieger2018}. The origin and location of the VHE emission in the M87 jet are not yet conclusive.}



\section{Summary and conclusions}\label{section5}

{In this work we have investigated the spectral index distribution of the M87 jet using KaVA LP observations from 2016 and archival VLBA observations from 2010 and 2014. This is the first systematic study of the spectral properties in the innermost regions of the M87 jet at 22 and 43 GHz, and it allows us to investigate the spectral index out to $\sim$ 10 mas from the jet base. Our study can be summarized as follows:}

\begin{enumerate}\itemsep3pt
\item The overall spectral index morphology of the M87 jet does not appear to have changed significantly over the course of these observations.

\item The observed spectral index distributions {of the M87 jet from the KaVA observations} show a rapid steepening from $\sim$ 2 mas until $\sim$ 6 mas from $\alpha\sim-$0.7 to $-$2.5. Beyond $\sim$ 6 mas, the spectral steepening stops and remains at the same value until $\sim$ 10 mas. {The trend of the spectral index distribution obtained from the VLBA data showed similar rapid spectral index decay, but the constant spectral index beyond $\sim$ 6 mas is less clear. This suggests that the spectral index distribution of the M87 jet may change with time.}

\item To understand the details of the spectral evolution in the M87 jet, we modeled the synchrotron spectral index by solving the transfer equation and calculating the evolution of the energy distribution of the nonthermal electron densities in the jet. We find that if the amount of the nonthermal electron injections decreases with distance, the spectral index distribution reproduces the characteristics of the observed distributions.

\item We compare the model of the spectral index distributions with the allowed region, which is defined based on the observations. As a result, we find that the initial magnetic field strength at 2 mas of $0.3 \text{ G}<B_i<1.0 \text{ G}$ and electron injection function with $7<q<11$ reproduce the observed spectral index distribution appropriately (see Sect. \ref{subsection4.2} for details). We conclude that the distribution of the magnetic field strengths as inferred from the model at 2$-$10 mas ($\sim$ 900 $r_s-\sim$ 4500 $r_s$ in the deprojected distance) in the M87 jet is $B=\left(0.3-1.0\text{ G}\right)\left(\text{z}/\text{2 mas}\right)^{-0.72}$.

\item {We compared the magnetic field strengths of the M87 jet with those estimated in previous observational studies. The extrapolation of the magnetic field distribution is in good agreement with the estimates by VLBI in the inner jet region, {indicating that the majority of the magnetic flux of the jet near the SMBH is preserved without serious dissipation up to $\lesssim 4500 r_s$}. However, when extrapolating to the HST-1 region, our maximum magnetic field strength is roughly an order of magnitude higher than the upper limits found from X-ray variability studies.}

\end{enumerate}


In this work, the magnetic field strength of the M87 jet at a distance of several thousands $r_s$ has been estimated for the first time. This study showed that, if the bulk jet velocity profile and radius profile are known, the magnetic field strength and nonthermal electron injection of the extended jet can be inferred from the spectral index profile. Consequently, it provides the magnetic field profile of the jet, which is an important physical quantity to validate the launching model of the jet.

\begin{acknowledgements}
We thank the anonymous referee for constructive comments to improve the quality of this paper.
This work is based on observations made with the KaVA, which is operated by the Korea Astronomy and Space Science Institute and the National Astronomical Observatory of Japan. 
{This research was supported by the Korea Astronomy and Space Science Institute under the R\&D program supervised by the Ministry of Science and ICT.}
{H.R., B.W.S., and A.C. acknowledge support from the KASI-Yonsei DRC program of the Korea Research Council of Fundamental Science and Technology (DRC-12-2-KASI).}
A.C. acknowledges support from the National Research Foundation of Korea (NRF), grant No. 2018R1D1A1B07048314. 
J.P. and I.C. acknowledge financial support from the Korean National Research Foundation (NRF) via Global Ph.D. Fellowship Grant 2014H1A2A1018695 and 2015H1A2A1033752, respectively. 
J.P. is supported by an EACOA Fellowship awarded by the East Asia Core Observatories Association, which consists of the Academia Sinica Institute of Astronomy and Astrophysics, the National Astronomical Observatory of Japan, the Center for Astronomical Mega-Science, the Chinese Academy of Sciences, and the Korea Astronomy and Space Science Institute. 
I.C. acknowledges financial support by the Spanish Ministerio de Econom\'{\i}a y Competitividad (grants PID2019-108995GB-C21).
K.H. acknowledges support from the Mitsubishi Foundation grant No. 201911019. 
This work is partially supported by JSPS/MEXT KAKENHI (grants 18K03656, 18H03721, 19H01943, 18KK0090, 21H01137, 21H04488). 
S.S.-S. is supported by JSPS KAKENHI grant No. JP21372953. 
K.L. and S.T. acknowledge financial support from the National Research Foundation (NRF) of Korea through grant 2022R1F1A1075115. 
J.O. was supported by Basic Science Research Program through the National Research Foundation of Korea (NRF) funded by the Ministry of Education (NRF-2021R1A6A3A01086420). J.-S.K. has been supported by the Basic Science Research Program through the National Research Foundation of Korea (NRF) funded by the Ministry of Education (2016R1A5A1013277 and 2020R1A2C1007219).
J.-Y.K. acknowledges support from the National Research Foundation (NRF) of Korea (grant no. 2022R1C1C1005255).

\end{acknowledgements}

\bibliographystyle{aa} 
\bibliography{aa.bib} 

\begin{thebibliography}{93}
\expandafter\ifx\csname natexlab\endcsname\relax\def\natexlab#1{#1}\fi

\bibitem[{{Abdo} {et~al.}(2009){Abdo}, {Ackermann}, {Ajello}, {Atwood},
  {Axelsson}, {Baldini}, {Ballet}, {Barbiellini}, {Bastieri}, {Bechtol},
  {Bellazzini}, {Berenji}, {Blandford}, {Bloom}, {Bonamente}, {Borgland},
  {Bregeon}, {Brez}, {Brigida}, {Bruel}, {Burnett}, {Caliandro}, {Cameron},
  {Cannon}, {Caraveo}, {Casandjian}, {Cavazzuti}, {Cecchi}, {{\c{C}}elik},
  {Charles}, {Cheung}, {Chiang}, {Ciprini}, {Claus}, {Cohen-Tanugi},
  {Colafrancesco}, {Conrad}, {Costamante}, {Cutini}, {Davis}, {Dermer}, {de
  Angelis}, {de Palma}, {Digel}, {Donato}, {Silva}, {Drell}, {Dubois},
  {Dumora}, {Edmonds}, {Farnier}, {Favuzzi}, {Fegan}, {Finke}, {Focke},
  {Fortin}, {Frailis}, {Fukazawa}, {Funk}, {Fusco}, {Gargano}, {Gasparrini},
  {Gehrels}, {Georganopoulos}, {Germani}, {Giebels}, {Giglietto}, {Giommi},
  {Giordano}, {Giroletti}, {Glanzman}, {Godfrey}, {Grenier}, {Grondin},
  {Grove}, {Guillemot}, {Guiriec}, {Hanabata}, {Harding}, {Hayashida}, {Hays},
  {Horan}, {J{\'o}hannesson}, {Johnson}, {Johnson}, {Johnson}, {Johnson},
  {Kamae}, {Katagiri}, {Kataoka}, {Kawai}, {Kerr}, {Kn{\"o}dlseder}, {Kocian},
  {Kuss}, {Lande}, {Latronico}, {Lemoine-Goumard}, {Longo}, {Loparco}, {Lott},
  {Lovellette}, {Lubrano}, {Madejski}, {Makeev}, {Mazziotta}, {McConville},
  {McEnery}, {Meurer}, {Michelson}, {Mitthumsiri}, {Mizuno}, {Moiseev},
  {Monte}, {Monzani}, {Morselli}, {Moskalenko}, {Murgia}, {Nolan}, {Norris},
  {Nuss}, {Ohsugi}, {Omodei}, {Orlando}, {Ormes}, {Ozaki}, {Paneque},
  {Panetta}, {Parent}, {Pelassa}, {Pepe}, {Pesce-Rollins}, {Piron}, {Porter},
  {Rain{\`o}}, {Rando}, {Razzano}, {Reimer}, {Reimer}, {Reposeur}, {Ritz},
  {Rochester}, {Rodriguez}, {Romani}, {Roth}, {Ryde}, {Sadrozinski},
  {Sambruna}, {Sanchez}, {Sander}, {Saz Parkinson}, {Scargle}, {Sgr{\`o}},
  {Shaw}, {Smith}, {Smith}, {Spandre}, {Spinelli}, {Strickman}, {Suson},
  {Tajima}, {Takahashi}, {Tanaka}, {Taylor}, {Thayer}, {Thompson}, {Tibaldo},
  {Torres}, {Tosti}, {Tramacere}, {Uchiyama}, {Usher}, {Vasileiou}, {Vilchez},
  {Waite}, {Wang}, {Winer}, {Wood}, {Ylinen}, {Ziegler}, {Harris}, {Massaro},
  \& {Stawarz}}]{abdo2009}
{Abdo}, A.~A., {Ackermann}, M., {Ajello}, M., {et~al.} 2009, \apj, 707, 55

\bibitem[{{Abramowski} {et~al.}(2012){Abramowski}, {Acero}, {Aharonian},
  {Akhperjanian}, {Anton}, {Balzer}, {Barnacka}, {Barres de Almeida},
  {Becherini}, {Becker}, {Behera}, {Bernl{\"o}hr}, {Birsin}, {Biteau},
  {Bochow}, {Boisson}, {Bolmont}, {Bordas}, {Brucker}, {Brun}, {Brun}, {Bulik},
  {B{\"u}sching}, {Carrigan}, {Casanova}, {Cerruti}, {Chadwick}, {Charbonnier},
  {Chaves}, {Cheesebrough}, {Clapson}, {Coignet}, {Cologna}, {Conrad},
  {Dalton}, {Daniel}, {Davids}, {Degrange}, {Deil}, {Dickinson},
  {Djannati-Ata{\"\i}}, {Domainko}, {Drury}, {Dubus}, {Dutson}, {Dyks},
  {Dyrda}, {Egberts}, {Eger}, {Espigat}, {Fallon}, {Farnier}, {Fegan},
  {Feinstein}, {Fernandes}, {Fiasson}, {Fontaine}, {F{\"o}rster},
  {F{\"u}{\ss}ling}, {Gallant}, {Gast}, {G{\'e}rard}, {Gerbig}, {Giebels},
  {Glicenstein}, {Gl{\"u}ck}, {Goret}, {G{\"o}ring}, {H{\"a}ffner}, {Hague},
  {Hampf}, {Hauser}, {Heinz}, {Heinzelmann}, {Henri}, {Hermann}, {Hinton},
  {Hoffmann}, {Hofmann}, {Hofverberg}, {Holler}, {Horns}, {Jacholkowska}, {de
  Jager}, {Jahn}, {Jamrozy}, {Jung}, {Kastendieck}, {Katarzy{\'n}ski}, {Katz},
  {Kaufmann}, {Keogh}, {Khangulyan}, {Kh{\'e}lifi}, {Klochkov}, {Klu{\'z}niak},
  {Kneiske}, {Komin}, {Kosack}, {Kossakowski}, {Laffon}, {Lamanna}, {Lennarz},
  {Lohse}, {Lopatin}, {Lu}, {Marandon}, {Marcowith}, {Masbou}, {Maurin},
  {Maxted}, {Mayer}, {McComb}, {Medina}, {M{\'e}hault}, {Moderski}, {Moulin},
  {Naumann}, {Naumann-Godo}, {de Naurois}, {Nedbal}, {Nekrassov}, {Nguyen},
  {Nicholas}, {Niemiec}, {Nolan}, {Ohm}, {de O{\~n}a Wilhelmi}, {Opitz},
  {Ostrowski}, {Oya}, {Panter}, {Paz Arribas}, {Pedaletti}, {Pelletier},
  {Petrucci}, {Pita}, {P{\"u}hlhofer}, {Punch}, {Quirrenbach}, {Raue},
  {Rayner}, {Reimer}, {Reimer}, {Renaud}, {de los Reyes}, {Rieger}, {Ripken},
  {Rob}, {Rosier-Lees}, {Rowell}, {Rudak}, {Rulten}, {Ruppel}, {Sahakian},
  {Sanchez}, {Santangelo}, {Schlickeiser}, {Sch{\"o}ck}, {Schulz}, {Schwanke},
  {Schwarzburg}, {Schwemmer}, {Sheidaei}, {Skilton}, {Sol}, {Spengler},
  {Stawarz}, {Steenkamp}, {Stegmann}, {Stinzing}, {Stycz}, {Sushch}, {Szostek},
  {Tavernet}, {Terrier}, {Tluczykont}, {Valerius}, {van Eldik}, {Vasileiadis},
  {Venter}, {Vialle}, {Viana}, {Vincent}, {V{\"o}lk}, {Volpe}, {Vorobiov},
  {Vorster}, {Wagner}, {Ward}, {White}, {Wierzcholska}, {Zacharias}, {Zajczyk},
  {Zdziarski}, {Zech}, {Zechlin}, {H.~E.~S.~S. Collaboration}, {Aleksi{\'c}},
  {Antonelli}, {Antoranz}, {Backes}, {Barrio}, {Bastieri}, {Becerra
  Gonz{\'a}lez}, {Bednarek}, {Berdyugin}, {Berger}, {Bernardini}, {Biland},
  {Blanch}, {Bock}, {Boller}, {Bonnoli}, {Borla Tridon}, {Braun}, {Bretz},
  {Ca{\~n}ellas}, {Carmona}, {Carosi}, {Colin}, {Colombo}, {Contreras},
  {Cortina}, {Cossio}, {Covino}, {Dazzi}, {De Angelis}, {De Cea del Pozo}, {De
  Lotto}, {Delgado Mendez}, {Diago Ortega}, {Doert}, {Dom{\'\i}nguez}, {Dominis
  Prester}, {Dorner}, {Doro}, {Elsaesser}, {Ferenc}, {Fonseca}, {Font},
  {Fruck}, {Garc{\'\i}a L{\'o}pez}, {Garczarczyk}, {Garrido}, {Giavitto},
  {Godinovi{\'c}}, {Hadasch}, {H{\"a}fner}, {Herrero}, {Hildebrand},
  {H{\"o}hne-M{\"o}nch}, {Hose}, {Hrupec}, {Huber}, {Jogler}, {Klepser},
  {Kr{\"a}henb{\"u}hl}, {Krause}, {La Barbera}, {Lelas}, {Leonardo},
  {Lindfors}, {Lombardi}, {L{\'o}pez}, {Lorenz}, {Makariev}, {Maneva},
  {Mankuzhiyil}, {Mannheim}, {Maraschi}, {Mariotti}, {Mart{\'\i}nez}, {Mazin},
  {Meucci}, {Miranda}, {Mirzoyan}, {Miyamoto}, {Mold{\'o}n}, {Moralejo},
  {Munar}, {Nieto}, {Nilsson}, {Orito}, {Oya}, {Paneque}, {Paoletti}, {Pardo},
  {Paredes}, {Partini}, {Pasanen}, {Pauss}, {Perez-Torres}, {Persic},
  {Peruzzo}, {Pilia}, {Pochon}, {Prada}, {Prada Moroni}, {Prandini}, {Puljak},
  {Reichardt}, {Reinthal}, {Rhode}, {Rib{\'o}}, {Rico}, {R{\"u}gamer},
  {Saggion}, {Saito}, {Saito}, {Salvati}, {Satalecka}, {Scalzotto}, {Scapin},
  {Schultz}, {Schweizer}, {Shayduk}, {Shore}, {Sillanp{\"a}{\"a}}, {Sitarek},
  {Sobczynska}, {Spanier}, {Spiro}, {Stamerra}, {Steinke}, {Storz}, {Strah},
  {Suri{\'c}}, {Takalo}, {Takami}, {Tavecchio}, {Temnikov}, {Terzi{\'c}},
  {Tescaro}, {Teshima}, {Thom}, {Tibolla}, {Torres}, {Treves}, {Vankov},
  {Vogler}, {Wagner}, {Weitzel}, {Zabalza}, {Zandanel}, {Zanin}, {MAGIC
  Collaboration}, {Arlen}, {Aune}, {Beilicke}, {Benbow}, {Bouvier}, {Bradbury},
  {Buckley}, {Bugaev}, {Byrum}, {Cannon}, {Cesarini}, {Ciupik}, {Connolly},
  {Cui}, {Dickherber}, {Duke}, {Errando}, {Falcone}, {Finley}, {Finnegan},
  {Fortson}, {Furniss}, {Galante}, {Gall}, {Godambe}, {Griffin}, {Grube},
  {Gyuk}, {Hanna}, {Holder}, {Huan}, {Hui}, {Kaaret}, {Karlsson}, {Kertzman},
  {Khassen}, {Kieda}, {Krawczynski}, {Krennrich}, {Lang}, {LeBohec}, {Maier},
  {McArthur}, {McCann}, {Moriarty}, {Mukherjee}, {Nu{\~n}ez}, {Ong}, {Orr},
  {Otte}, {Park}, {Perkins}, {Pichel}, {Pohl}, {Prokoph}, {Ragan}, {Reyes},
  {Reynolds}, {Roache}, {Rose}, {Ruppel}, {Schroedter}, {Sembroski},
  {{\c{S}}ent{\"u}rk}, {Telezhinsky}, {Te{\v{s}}i{\'c}}, {Theiling},
  {Thibadeau}, {Varlotta}, {Vassiliev}, {Vivier}, {Wakely}, {Weekes},
  {Williams}, {Zitzer}, {VERITAS Collaboration}, {Barres de Almeida}, {Cara},
  {Casadio}, {Cheung}, {McConville}, {Davies}, {Doi}, {Giovannini},
  {Giroletti}, {Hada}, {Hardee}, {Harris}, {Junor}, {Kino}, {Lee}, {Ly},
  {Madrid}, {Massaro}, {Mundell}, {Nagai}, {Perlman}, {Steele}, {Walker}, \&
  {Wood}}]{abramowski2012}
{Abramowski}, A., {Acero}, F., {Aharonian}, F., {et~al.} 2012, \apj, 746, 151

\bibitem[{{Acciari} {et~al.}(2009){Acciari}, {Aliu}, {Arlen}, {Bautista},
  {Beilicke}, {Benbow}, {Bradbury}, {Buckley}, {Bugaev}, {Butt}, \&
  et~al.}]{acciari2009}
{Acciari}, V.~A., {Aliu}, E., {Arlen}, T., {et~al.} 2009, Science, 325, 444

\bibitem[{{Akiyama} {et~al.}(2015){Akiyama}, {Lu}, {Fish}, {Doeleman},
  {Broderick}, {Dexter}, {Hada}, {Kino}, {Nagai}, {Honma}, {Johnson}, {Algaba},
  {Asada}, {Brinkerink}, {Blundell}, {Bower}, {Cappallo}, {Crew}, {Dexter},
  {Dzib}, {Freund}, {Friberg}, {Gurwell}, {Ho}, {Inoue}, {Krichbaum},
  {Loinard}, {MacMahon}, {Marrone}, {Moran}, {Nakamura}, {Nagar}, {Ortiz-Leon},
  {Plambeck}, {Pradel}, {Primiani}, {Rogers}, {Roy}, {SooHoo}, {Tavares},
  {Tilanus}, {Titus}, {Wagner}, {Weintroub}, {Yamaguchi}, {Young}, {Zensus}, \&
  {Ziurys}}]{akiyama2015}
{Akiyama}, K., {Lu}, R.-S., {Fish}, V.~L., {et~al.} 2015, \apj, 807, 150

\bibitem[{{Arfken} \& {Weber}(2005)}]{arfken}
{Arfken}, G.~B. \& {Weber}, H.~J. 2005, {Mathematical methods for physicists
  6th ed.}

\bibitem[{{Asada} \& {Nakamura}(2012)}]{asada2012}
{Asada}, K. \& {Nakamura}, M. 2012, \apjl, 745, L28

\bibitem[{{Asada} {et~al.}(2014){Asada}, {Nakamura}, {Doi}, {Nagai}, \&
  {Inoue}}]{asada2014}
{Asada}, K., {Nakamura}, M., {Doi}, A., {Nagai}, H., \& {Inoue}, M. 2014,
  \apjl, 781, L2

\bibitem[{{Asada} {et~al.}(2016){Asada}, {Nakamura}, \& {Pu}}]{asada2016}
{Asada}, K., {Nakamura}, M., \& {Pu}, H.-Y. 2016, \apj, 833, 56

\bibitem[{{Baczko} {et~al.}(2019){Baczko}, {Schulz}, {Kadler}, {Ros},
  {Perucho}, {Fromm}, \& {Wilms}}]{baczko2019}
{Baczko}, A.~K., {Schulz}, R., {Kadler}, M., {et~al.} 2019, \aap, 623, A27

\bibitem[{{Blakeslee} {et~al.}(2009){Blakeslee}, {Jord{\'a}n}, {Mei},
  {C{\^o}t{\'e}}, {Ferrarese}, {Infante}, {Peng}, {Tonry}, \&
  {West}}]{blakeslee09}
{Blakeslee}, J.~P., {Jord{\'a}n}, A., {Mei}, S., {et~al.} 2009, \apj, 694, 556

\bibitem[{{Blandford} {et~al.}(2019){Blandford}, {Meier}, \&
  {Readhead}}]{blandford19}
{Blandford}, R., {Meier}, D., \& {Readhead}, A. 2019, \araa, 57, 467

\bibitem[{{Blandford} \& {Znajek}(1977)}]{blandford77}
{Blandford}, R.~D. \& {Znajek}, R.~L. 1977, \mnras, 179, 433

\bibitem[{{Blasi}(2013)}]{blasi2013}
{Blasi}, P. 2013, \aapr, 21, 70

\bibitem[{{Boccardi}(2015)}]{BoccardiThesis}
{Boccardi}. 2015, PhD thesis, Universit{\"a}t zu K{\"o}ln

\bibitem[{{Chatterjee} {et~al.}(2019){Chatterjee}, {Liska}, {Tchekhovskoy}, \&
  {Markoff}}]{chatterjee19}
{Chatterjee}, K., {Liska}, M., {Tchekhovskoy}, A., \& {Markoff}, S.~B. 2019,
  \mnras, 490, 2200

\bibitem[{{Cheung} {et~al.}(2007){Cheung}, {Harris}, \& {Stawarz}}]{cheung2007}
{Cheung}, C.~C., {Harris}, D.~E., \& {Stawarz}, {\L}. 2007, \apjl, 663, L65

\bibitem[{{Cho} {et~al.}(2017){Cho}, {Jung}, {Zhao}, {Akiyama}, {Sawada-Satoh},
  {Kino}, {Byun}, {Sohn}, {Shibata}, {Hirota}, {Niinuma}, {Yonekura},
  {Fujisawa}, \& {Oyama}}]{cho2017}
{Cho}, I., {Jung}, T., {Zhao}, G.-Y., {et~al.} 2017, \pasj, 69, 87

\bibitem[{{Cui} {et~al.}(2021){Cui}, {Hada}, {Kino}, {Sohn}, {Park}, {Ro},
  {Sawada-Satoh}, {Jiang}, {Cui}, {Honma}, {Shen}, {Tazaki}, {An}, {Cho},
  {Zhao}, {Cheng}, {Niinuma}, {Wajima}, {Zhang}, {Kawaguchi}, {Algaba},
  {Koyama}, {Hirota}, {Yonekura}, {Sakai}, {Xia}, {Jiang}, {Yu}, {Gou},
  {Hwang}, {Jiang}, {Sun}, {Jung}, {Kim}, {Kim}, {Kobayashi}, {Lee}, {Lee},
  {Zhang}, {Li}, {Xu}, {Li}, {Oh}, {Oh}, {Oh}, {Oyama}, {Roh}, {Shibata},
  {Guo}, {Zhao}, {Zhong}, {Wang}, {Yang}, {Yan}, {Yeom}, {Li}, {Li}, {Yuan},
  {Dong}, {Chen}, {Akiyama}, {Asada}, {Byun}, {Hagiwara}, {Hodgson}, {Jung},
  {Kim}, {Lee}, {Yi}, {Liu}, {Liu}, {Lu}, {Nakamura}, {Trippe}, {Wang}, {Wang},
  \& {Zhang}}]{cui2021}
{Cui}, Y.-Z., {Hada}, K., {Kino}, M., {et~al.} 2021, Research in Astronomy and
  Astrophysics, 21, 205

\bibitem[{{de Gasperin} {et~al.}(2012){de Gasperin}, {Orr{\'u}}, {Murgia},
  {Merloni}, {Falcke}, {Beck}, {Beswick}, {B{\^\i}rzan}, {Bonafede},
  {Br{\"u}ggen}, {Brunetti}, {Chy{\.z}y}, {Conway}, {Croston}, {En{\ss}lin},
  {Ferrari}, {Heald}, {Heidenreich}, {Jackson}, {Macario}, {McKean}, {Miley},
  {Morganti}, {Offringa}, {Pizzo}, {Rafferty}, {R{\"o}ttgering}, {Shulevski},
  {Steinmetz}, {Tasse}, {van der Tol}, {van Driel}, {van Weeren}, {van
  Zwieten}, {Alexov}, {Anderson}, {Asgekar}, {Avruch}, {Bell}, {Bell},
  {Bentum}, {Bernardi}, {Best}, {Breitling}, {Broderick}, {Butcher}, {Ciardi},
  {Dettmar}, {Eisloeffel}, {Frieswijk}, {Gankema}, {Garrett}, {Gerbers},
  {Griessmeier}, {Gunst}, {Hassall}, {Hessels}, {Hoeft}, {Horneffer},
  {Karastergiou}, {K{\"o}hler}, {Koopman}, {Kuniyoshi}, {Kuper}, {Maat},
  {Mann}, {Mevius}, {Mulcahy}, {Munk}, {Nijboer}, {Noordam}, {Paas}, {Pandey},
  {Pandey}, {Polatidis}, {Reich}, {Schoenmakers}, {Sluman}, {Smirnov}, {Sobey},
  {Stappers}, {Swinbank}, {Tagger}, {Tang}, {van Bemmel}, {van Cappellen}, {van
  Duin}, {van Haarlem}, {van Leeuwen}, {Vermeulen}, {Vocks}, {White}, {Wise},
  {Wucknitz}, \& {Zarka}}]{de_Gasperin2012}
{de Gasperin}, F., {Orr{\'u}}, E., {Murgia}, M., {et~al.} 2012, \aap, 547, A56

\bibitem[{{Dodson} {et~al.}(2006){Dodson}, {Edwards}, \&
  {Hirabayashi}}]{dodson2006}
{Dodson}, R., {Edwards}, P.~G., \& {Hirabayashi}, H. 2006, \pasj, 58, 243

\bibitem[{{Doeleman} {et~al.}(2012){Doeleman}, {Fish}, {Schenck}, {Beaudoin},
  {Blundell}, {Bower}, {Broderick}, {Chamberlin}, {Freund}, {Friberg},
  {Gurwell}, {Ho}, {Honma}, {Inoue}, {Krichbaum}, {Lamb}, {Loeb}, {Lonsdale},
  {Marrone}, {Moran}, {Oyama}, {Plambeck}, {Primiani}, {Rogers}, {Smythe},
  {SooHoo}, {Strittmatter}, {Tilanus}, {Titus}, {Weintroub}, {Wright}, {Young},
  \& {Ziurys}}]{doeleman2012}
{Doeleman}, S.~S., {Fish}, V.~L., {Schenck}, D.~E., {et~al.} 2012, Science,
  338, 355

\bibitem[{{EHT MWL Science Working Group} {et~al.}(2021){EHT MWL Science
  Working Group}, {Algaba}, {Anczarski}, {Asada}, {Balokovi{\'c}}, {Chandra},
  {Cui}, {Falcone}, {Giroletti}, {Goddi}, {Hada}, {Haggard}, {Jorstad}, {Kaur},
  {Kawashima}, {Keating}, {Kim}, {Kino}, {Komossa}, {Kravchenko}, {Krichbaum},
  {Lee}, {Lu}, {Lucchini}, {Markoff}, {Neilsen}, {Nowak}, {Park}, {Principe},
  {Ramakrishnan}, {Reynolds}, {Sasada}, {Savchenko}, {Williamson}, {Event
  Horizon Telescope Collaboration}, {Akiyama}, {Alberdi}, {Alef}, {Anantua},
  {Azulay}, {Baczko}, {Ball}, {Barrett}, {Bintley}, {Benson}, {Blackburn},
  {Blundell}, {Boland}, {Bouman}, {Bower}, {Boyce}, {Bremer}, {Brinkerink},
  {Brissenden}, {Britzen}, {Broderick}, {Broguiere}, {Bronzwaer}, {Byun},
  {Carlstrom}, {Chael}, {Chan}, {Chatterjee}, {Chatterjee}, {Chen}, {Chen},
  {Chesler}, {Cho}, {Christian}, {Conway}, {Cordes}, {Crawford}, {Crew},
  {Cruz-Osorio}, {Davelaar}, {de Laurentis}, {Deane}, {Dempsey}, {Desvignes},
  {Dexter}, {Doeleman}, {Eatough}, {Falcke}, {Farah}, {Fish}, {Fomalont},
  {Ford}, {Fraga-Encinas}, {Friberg}, {Fromm}, {Fuentes}, {Galison}, {Gammie},
  {Garc{\'\i}a}, {Gentaz}, {Georgiev}, {Gold}, {G{\'o}mez}, {G{\'o}mez-Ruiz},
  {Gu}, {Gurwell}, {Hecht}, {Hesper}, {Ho}, {Ho}, {Honma}, {Huang}, {Huang},
  {Hughes}, {Ikeda}, {Inoue}, {Issaoun}, {James}, {Jannuzi}, {Janssen},
  {Jeter}, {Jiang}, {Jim{\'e}nez-Rosales}, {Johnson}, {Jung}, {Karami},
  {Karuppusamy}, {Kettenis}, {Kim}, {Kim}, {Kim}, {Koay}, {Kofuji}, {Koch},
  {Koyama}, {Kramer}, {Kramer}, {Kuo}, {Lauer}, {Levis}, {Li}, {Li},
  {Lindqvist}, {Lico}, {Lindahl}, {Liu}, {Liu}, {Liuzzo}, {Lo}, {Lobanov},
  {Loinard}, {Lonsdale}, {MacDonald}, {Mao}, {Marchili}, {Marrone}, {Marscher},
  {Mart{\'\i}-Vidal}, {Matsushita}, {Matthews}, {Medeiros}, {Menten}, {Mizuno},
  {Mizuno}, {Moran}, {Moriyama}, {Moscibrodzka}, {M{\"u}ller}, {Musoke},
  {Mej{\'\i}as}, {Nagai}, {Nagar}, {Nakamura}, {Narayan}, {Narayanan},
  {Natarajan}, {Nathanail}, {Neri}, {Ni}, {Noutsos}, {Okino}, {Olivares},
  {Ortiz-Le{\'o}n}, {Oyama}, {{\"O}zel}, {Palumbo}, {Patel}, {Pen}, {Pesce},
  {Pi{\'e}tu}, {Plambeck}, {Popstefanija}, {Porth}, {P{\"o}tzl}, {Prather},
  {Preciado-L{\'o}pez}, {Psaltis}, {Pu}, {Rao}, {Rawlings}, {Raymond},
  {Rezzolla}, {Ricarte}, {Ripperda}, {Roelofs}, {Rogers}, {Ros}, {Rose},
  {Roshanineshat}, {Rottmann}, {Roy}, {Ruszczyk}, {Rygl}, {S{\'a}nchez},
  {S{\'a}nchez-Arguelles}, {Savolainen}, {Schloerb}, {Schuster}, {Shao},
  {Shen}, {Small}, {Sohn}, {Soohoo}, {Sun}, {Tazaki}, {Tetarenko}, {Tiede},
  {Tilanus}, {Titus}, {Toma}, {Torne}, {Trent}, {Traianou}, {Trippe}, {van
  Bemmel}, {van Langevelde}, {van Rossum}, {Wagner}, {Ward-Thompson}, {Wardle},
  {Weintroub}, {Wex}, {Wharton}, {Wielgus}, {Wong}, {Wu}, {Yoon}, {Young},
  {Young}, {Younsi}, {Yuan}, {Yuan}, {Zensus}, {Zhao}, {Zhao}, {Fermi Large
  Area Telescope Collaboration}, {Principe}, {Giroletti}, {D'Ammando},
  {Orienti}, {H.~E.~S.~S. Collaboration}, {Abdalla}, {Adam}, {Aharonian},
  {Benkhali}, {Ang{\"u}ner}, {Arcaro}, {Armand}, {Armstrong}, {Ashkar},
  {Backes}, {Baghmanyan}, {Barbosa Martins}, {Barnacka}, {Barnard},
  {Becherini}, {Berge}, {Bernl{\"o}hr}, {Bi}, {B{\"o}ttcher}, {Boisson},
  {Bolmont}, {de Lavergne}, {Breuhaus}, {Brun}, {Brun}, {Bryan}, {B{\"u}chele},
  {Bulik}, {Bylund}, {Caroff}, {Carosi}, {Casanova}, {Chand}, {Chen}, {Cotter},
  {Cury{\l}o}, {Damascene Mbarubucyeye}, {Davids}, {Davies}, {Deil}, {Devin},
  {Dewilt}, {Dirson}, {Djannati-Ata{\"\i}}, {Dmytriiev}, {Donath},
  {Doroshenko}, {Duffy}, {Dyks}, {Egberts}, {Eichhorn}, {Einecke}, {Emery},
  {Ernenwein}, {Feijen}, {Fegan}, {Fiasson}, {de Clairfontaine}, {Fontaine},
  {Funk}, {F{\"u}{\ss}ling}, {Gabici}, {Gallant}, {Giavitto}, {Giunti},
  {Glawion}, {Glicenstein}, {Gottschall}, {Grondin}, {Hahn}, {Haupt},
  {Hermann}, {Hinton}, {Hofmann}, {Hoischen}, {Holch}, {Holler}, {H{\"o}rbe},
  {Horns}, {Huber}, {Jamrozy}, {Jankowsky}, {Jankowsky}, {Jardin-Blicq},
  {Joshi}, {Jung-Richardt}, {Kasai}, {Kastendieck}, {Katarzy{\'n}ski}, {Katz},
  {Khangulyan}, {Kh{\'e}lifi}, {Klepser}, {Klu{\'z}niak}, {Komin}, {Konno},
  {Kosack}, {Kostunin}, {Kreter}, {Lamanna}, {Lemi{\`e}re}, {Lemoine-Goumard},
  {Lenain}, {Levy}, {Lohse}, {Lypova}, {Mackey}, {Majumdar}, {Malyshev},
  {Malyshev}, {Marandon}, {Marchegiani}, {Marcowith}, {Mares},
  {Mart{\'\i}-Devesa}, {Marx}, {Maurin}, {Meintjes}, {Meyer}, {Moderski},
  {Mohamed}, {Mohrmann}, {Montanari}, {Moore}, {Morris}, {Moulin}, {Muller},
  {Murach}, {Nakashima}, {Nayerhoda}, {de Naurois}, {Ndiyavala},
  {Niederwanger}, {Niemiec}, {Oakes}, {O'Brien}, {Odaka}, {Ohm},
  {Olivera-Nieto}, {de Ona Wilhelmi}, {Ostrowski}, {Panter}, {Panny},
  {Parsons}, {Peron}, {Peyaud}, {Piel}, {Pita}, {Poireau}, {Noel}, {Prokhorov},
  {Prokoph}, {P{\"u}hlhofer}, {Punch}, {Quirrenbach}, {Rauth}, {Reichherzer},
  {Reimer}, {Reimer}, {Remy}, {Renaud}, {Rieger}, {Rinchiuso}, {Romoli},
  {Rowell}, {Rudak}, {Ruiz-Velasco}, {Sahakian}, {Sailer}, {Sanchez},
  {Santangelo}, {Sasaki}, {Scalici}, {Schutte}, {Schwanke}, {Schwemmer},
  {Seglar-Arroyo}, {Senniappan}, {Seyffert}, {Shafi}, {Shiningayamwe},
  {Simoni}, {Sinha}, {Sol}, {Specovius}, {Spencer}, {Spir-Jacob}, {Stawarz},
  {Sun}, {Steenkamp}, {Stegmann}, {Steinmassl}, {Steppa}, {Takahashi},
  {Tavernier}, {Taylor}, {Terrier}, {Tiziani}, {Tluczykont}, {Tomankova},
  {Trichard}, {Tsirou}, {Tuffs}, {Uchiyama}, {van der Walt}, {van Eldik}, {van
  Rensburg}, {van Soelen}, {Vasileiadis}, {Veh}, {Venter}, {Vincent}, {Vink},
  {V{\"o}lk}, {Vuillaume}, {Wadiasingh}, {Wagner}, {Watson}, {Werner}, {White},
  {Wierzcholska}, {Wong}, {Yusafzai}, {Zacharias}, {Zanin}, {Zargaryan},
  {Zdziarski}, {Zech}, {Zhu}, {Zorn}, {Zouari}, {{\.Z}ywucka}, {MAGIC
  Collaboration}, {Acciari}, {Ansoldi}, {Antonelli}, {Engels}, {Artero},
  {Asano}, {Baack}, {Babi{\'c}}, {Baquero}, {de Almeida}, {Barrio}, {Becerra
  Gonz{\'a}lez}, {Bednarek}, {Bellizzi}, {Bernardini}, {Bernardos}, {Berti},
  {Besenrieder}, {Bhattacharyya}, {Bigongiari}, {Biland}, {Blanch}, {Bonnoli},
  {Bo{\v{s}}njak}, {Busetto}, {Carosi}, {Ceribella}, {Cerruti}, {Chai},
  {Chilingarian}, {Cikota}, {Colak}, {Colombo}, {Contreras}, {Cortina},
  {Covino}, {D'Amico}, {D'Elia}, {da Vela}, {Dazzi}, {de Angelis}, {de Lotto},
  {Delfino}, {Delgado}, {Delgado Mendez}, {Depaoli}, {di Pierro}, {di Venere},
  {Do Souto Espi{\~n}eira}, {Dominis Prester}, {Donini}, {Dorner}, {Doro},
  {Elsaesser}, {Ramazani}, {Fattorini}, {Ferrara}, {Fonseca}, {Font}, {Fruck},
  {Fukami}, {Garc{\'\i}a L{\'o}pez}, {Garczarczyk}, {Gasparyan}, {Gaug},
  {Giglietto}, {Giordano}, {Gliwny}, {Godinovi{\'c}}, {Green}, {Green},
  {Hadasch}, {Hahn}, {Heckmann}, {Herrera}, {Hoang}, {Hrupec}, {H{\"u}tten},
  {Inada}, {Inoue}, {Ishio}, {Iwamura}, {Jim{\'e}nez}, {Jormanainen}, {Jouvin},
  {Kajiwara}, {Karjalainen}, {Kerszberg}, {Kobayashi}, {Kubo}, {Kushida},
  {Lamastra}, {Lelas}, {Leone}, {Lindfors}, {Lombardi}, {Longo},
  {L{\'o}pez-Coto}, {L{\'o}pez-Moya}, {L{\'o}pez-Oramas}, {Loporchio}, {Machado
  de Oliveira Fraga}, {Maggio}, {Majumdar}, {Makariev}, {Mallamaci}, {Maneva},
  {Manganaro}, {Mannheim}, {Maraschi}, {Mariotti}, {Mart{\'\i}nez}, {Mazin},
  {Menchiari}, {Mender}, {Mi{\'c}anovi{\'c}}, {Miceli}, {Miener}, {Minev},
  {Miranda}, {Mirzoyan}, {Molina}, {Moralejo}, {Morcuende}, {Moreno},
  {Moretti}, {Neustroev}, {Nigro}, {Nilsson}, {Nishijima}, {Noda}, {Nozaki},
  {Ohtani}, {Oka}, {Otero-Santos}, {Paiano}, {Palatiello}, {Paneque},
  {Paoletti}, {Paredes}, {Pavleti{\'c}}, {Pe{\~n}il}, {Perennes}, {Persic},
  {Moroni}, {Prandini}, {Priyadarshi}, {Puljak}, {Rhode}, {Rib{\'o}}, {Rico},
  {Righi}, {Rugliancich}, {Saha}, {Sahakyan}, {Saito}, {Sakurai}, {Satalecka},
  {Saturni}, {Schleicher}, {Schmidt}, {Schweizer}, {Sitarek},
  {{\v{S}}nidari{\'c}}, {Sobczynska}, {Spolon}, {Stamerra}, {Strom}, {Strzys},
  {Suda}, {Suri{\'c}}, {Takahashi}, {Tavecchio}, {Temnikov}, {Terzi{\'c}},
  {Teshima}, {Tosti}, {Truzzi}, {Tutone}, {Ubach}, {van Scherpenberg}, {Vanzo},
  {Vazquez Acosta}, {Ventura}, {Verguilov}, {Vigorito}, {Vitale}, {Vovk},
  {Will}, {Wunderlich}, {Zari{\'c}}, {VERITAS Collaboration}, {Adams},
  {Benbow}, {Brill}, {Capasso}, {Christiansen}, {Chromey}, {Daniel}, {Errando},
  {Farrell}, {Feng}, {Finley}, {Fortson}, {Furniss}, {Gent}, {Giuri}, {Hassan},
  {Hervet}, {Holder}, {Hughes}, {Humensky}, {Jin}, {Kaaret}, {Kertzman},
  {Kieda}, {Kumar}, {Lang}, {Lundy}, {Maier}, {Moriarty}, {Mukherjee}, {Nieto},
  {Nievas-Rosillo}, {O'Brien}, {Ong}, {Otte}, {Patel}, {Pfrang}, {Pohl},
  {Prado}, {Pueschel}, {Quinn}, {Ragan}, {Reynolds}, {Ribeiro}, {Richards},
  {Roache}, {Rulten}, {Ryan}, {Santander}, {Sembroski}, {Shang}, {Weinstein},
  {Williams}, {Williamson}, {Eavn Collaboration}, {Hirota}, {Cui}, {Niinuma},
  {Ro}, {Sakai}, {Sawada-Satoh}, {Wajima}, {Wang}, {Liu}, \&
  {Yonekura}}]{ehtMWL2021}
{EHT MWL Science Working Group}, {Algaba}, J.~C., {Anczarski}, J., {et~al.}
  2021, \apjl, 911, L11

\bibitem[{{Event Horizon Telescope Collaboration} {et~al.}(2019){Event Horizon
  Telescope Collaboration}, {Akiyama}, {Alberdi}, {Alef}, {Asada}, {Azulay},
  {Baczko}, {Ball}, {Balokovi{\'c}}, {Barrett}, \& et~al.}]{eht2019}
{Event Horizon Telescope Collaboration}, {Akiyama}, K., {Alberdi}, A., {et~al.}
  2019, \apjl, 875, L1

\bibitem[{{Event Horizon Telescope Collaboration} {et~al.}(2021){Event Horizon
  Telescope Collaboration}, {Akiyama}, {Algaba}, {Alberdi}, {Alef}, {Anantua},
  {Asada}, {Azulay}, {Baczko}, {Ball}, {Balokovi{\'c}}, {Barrett}, {Benson},
  {Bintley}, {Blackburn}, {Blundell}, {Boland}, {Bouman}, {Bower}, {Boyce},
  {Bremer}, {Brinkerink}, {Brissenden}, {Britzen}, {Broderick}, {Broguiere},
  {Bronzwaer}, {Byun}, {Carlstrom}, {Chael}, {Chan}, {Chatterjee},
  {Chatterjee}, {Chen}, {Chen}, {Chesler}, {Cho}, {Christian}, {Conway},
  {Cordes}, {Crawford}, {Crew}, {Cruz-Osorio}, {Cui}, {Davelaar}, {De
  Laurentis}, {Deane}, {Dempsey}, {Desvignes}, {Dexter}, {Doeleman}, {Eatough},
  {Falcke}, {Farah}, {Fish}, {Fomalont}, {Ford}, {Fraga-Encinas}, {Freeman},
  {Friberg}, {Fromm}, {Fuentes}, {Galison}, {Gammie}, {Garc{\'\i}a}, {Gentaz},
  {Georgiev}, {Goddi}, {Gold}, {G{\'o}mez}, {G{\'o}mez-Ruiz}, {Gu}, {Gurwell},
  {Hada}, {Haggard}, {Hecht}, {Hesper}, {Ho}, {Ho}, {Honma}, {Huang}, {Huang},
  {Hughes}, {Ikeda}, {Inoue}, {Issaoun}, {James}, {Jannuzi}, {Janssen},
  {Jeter}, {Jiang}, {Jimenez-Rosales}, {Johnson}, {Jorstad}, {Jung}, {Karami},
  {Karuppusamy}, {Kawashima}, {Keating}, {Kettenis}, {Kim}, {Kim}, {Kim},
  {Kim}, {Kino}, {Koay}, {Kofuji}, {Koch}, {Koyama}, {Kramer}, {Kramer},
  {Krichbaum}, {Kuo}, {Lauer}, {Lee}, {Levis}, {Li}, {Li}, {Lindqvist}, {Lico},
  {Lindahl}, {Liu}, {Liu}, {Liuzzo}, {Lo}, {Lobanov}, {Loinard}, {Lonsdale},
  {Lu}, {MacDonald}, {Mao}, {Marchili}, {Markoff}, {Marrone}, {Marscher},
  {Mart{\'\i}-Vidal}, {Matsushita}, {Matthews}, {Medeiros}, {Menten}, {Mizuno},
  {Mizuno}, {Moran}, {Moriyama}, {Moscibrodzka}, {M{\"u}ller}, {Musoke},
  {Mej{\'\i}as}, {Michalik}, {Nadolski}, {Nagai}, {Nagar}, {Nakamura},
  {Narayan}, {Narayanan}, {Natarajan}, {Nathanail}, {Neilsen}, {Neri}, {Ni},
  {Noutsos}, {Nowak}, {Okino}, {Olivares}, {Ortiz-Le{\'o}n}, {Oyama},
  {{\"O}zel}, {Palumbo}, {Park}, {Patel}, {Pen}, {Pesce}, {Pi{\'e}tu},
  {Plambeck}, {PopStefanija}, {Porth}, {P{\"o}tzl}, {Prather},
  {Preciado-L{\'o}pez}, {Psaltis}, {Pu}, {Ramakrishnan}, {Rao}, {Rawlings},
  {Raymond}, {Rezzolla}, {Ricarte}, {Ripperda}, {Roelofs}, {Rogers}, {Ros},
  {Rose}, {Roshanineshat}, {Rottmann}, {Roy}, {Ruszczyk}, {Rygl},
  {S{\'a}nchez}, {S{\'a}nchez-Arguelles}, {Sasada}, {Savolainen}, {Schloerb},
  {Schuster}, {Shao}, {Shen}, {Small}, {Sohn}, {SooHoo}, {Sun}, {Tazaki},
  {Tetarenko}, {Tiede}, {Tilanus}, {Titus}, {Toma}, {Torne}, {Trent},
  {Traianou}, {Trippe}, {van Bemmel}, {van Langevelde}, {van Rossum}, {Wagner},
  {Ward-Thompson}, {Wardle}, {Weintroub}, {Wex}, {Wharton}, {Wielgus}, {Wong},
  {Wu}, {Yoon}, {Young}, {Young}, {Younsi}, {Yuan}, {Yuan}, {Zensus}, {Zhao},
  \& {Zhao}}]{eht2021}
{Event Horizon Telescope Collaboration}, {Akiyama}, K., {Algaba}, J.~C.,
  {et~al.} 2021, \apjl, 910, L12

\bibitem[{{Fromm} {et~al.}(2013){Fromm}, {Ros}, {Perucho}, {Savolainen},
  {Mimica}, {Kadler}, {Lobanov}, \& {Zensus}}]{fromm2013}
{Fromm}, C.~M., {Ros}, E., {Perucho}, M., {et~al.} 2013, \aap, 557, A105

\bibitem[{{Gebhardt} {et~al.}(2011){Gebhardt}, {Adams}, {Richstone}, {Lauer},
  {Faber}, {G{\"u}ltekin}, {Murphy}, \& {Tremaine}}]{gebhardt2011}
{Gebhardt}, K., {Adams}, J., {Richstone}, D., {et~al.} 2011, \apj, 729, 119

\bibitem[{{Gebhardt} \& {Thomas}(2009)}]{gebhardt2009}
{Gebhardt}, K. \& {Thomas}, J. 2009, \apj, 700, 1690

\bibitem[{{Ghisellini}(2013)}]{ghisellini2013}
{Ghisellini}, G. 2013, {Radiative Processes in High Energy Astrophysics}, Vol.
  873

\bibitem[{{Ginzburg} \& {Syrovatskii}(1964)}]{ginzburg1964}
{Ginzburg}, V.~L. \& {Syrovatskii}, S.~I. 1964, {The Origin of Cosmic Rays}

\bibitem[{{Greisen}(2003)}]{greisen2003}
{Greisen}, E.~W. 2003, {AIPS, the VLA, and the VLBA}, ed. A.~{Heck}, Vol. 285,
  109

\bibitem[{{Hada} {et~al.}(2011){Hada}, {Doi}, {Kino}, {Nagai}, {Hagiwara}, \&
  {Kawaguchi}}]{hada2011}
{Hada}, K., {Doi}, A., {Kino}, M., {et~al.} 2011, \nat, 477, 185

\bibitem[{{Hada} {et~al.}(2014){Hada}, {Giroletti}, {Kino}, {Giovannini},
  {D'Ammando}, {Cheung}, {Beilicke}, {Nagai}, {Doi}, {Akiyama}, {Honma},
  {Niinuma}, {Casadio}, {Orienti}, {Krawczynski}, {G{\'o}mez}, {Sawada-Satoh},
  {Koyama}, {Cesarini}, {Nakahara}, \& {Gurwell}}]{hada2014}
{Hada}, K., {Giroletti}, M., {Kino}, M., {et~al.} 2014, \apj, 788, 165

\bibitem[{{Hada} {et~al.}(2016){Hada}, {Kino}, {Doi}, {Nagai}, {Honma},
  {Akiyama}, {Tazaki}, {Lico}, {Giroletti}, {Giovannini}, {Orienti}, \&
  {Hagiwara}}]{hada2016}
{Hada}, K., {Kino}, M., {Doi}, A., {et~al.} 2016, \apj, 817, 131

\bibitem[{{Hada} {et~al.}(2013){Hada}, {Kino}, {Doi}, {Nagai}, {Honma},
  {Hagiwara}, {Giroletti}, {Giovannini}, \& {Kawaguchi}}]{hada2013}
{Hada}, K., {Kino}, M., {Doi}, A., {et~al.} 2013, \apj, 775, 70

\bibitem[{{Hada} {et~al.}(2012){Hada}, {Kino}, {Nagai}, {Doi}, {Hagiwara},
  {Honma}, {Giroletti}, {Giovannini}, \& {Kawaguchi}}]{hada2012}
{Hada}, K., {Kino}, M., {Nagai}, H., {et~al.} 2012, \apj, 760, 52

\bibitem[{{Hada} {et~al.}(2017){Hada}, {Park}, {Kino}, {Niinuma}, {Sohn}, {Ro},
  {Jung}, {Algaba}, {Zhao}, {Lee}, {Akiyama}, {Trippe}, {Wajima},
  {Sawada-Satoh}, {Tazaki}, {Cho}, {Hodgson}, {Lee}, {Hagiwara}, {Honma},
  {Koyama}, {Oh}, {Lee}, {Yoo}, {Kawaguchi}, {Roh}, {Oh}, {Yeom}, {Jung}, {Oh},
  {Kim}, {Hwang}, {Byun}, {Cho}, {Kim}, {Kobayashi}, \& {Shibata}}]{hada2017}
{Hada}, K., {Park}, J.~H., {Kino}, M., {et~al.} 2017, \pasj, 69, 71

\bibitem[{{Haga} {et~al.}(2015){Haga}, {Doi}, {Murata}, {Sudou}, {Kameno}, \&
  {Hada}}]{haga2015}
{Haga}, T., {Doi}, A., {Murata}, Y., {et~al.} 2015, \apj, 807, 15

\bibitem[{{Hardcastle} \& {Krause}(2013)}]{hardcastle2013}
{Hardcastle}, M.~J. \& {Krause}, M.~G.~H. 2013, \mnras, 430, 174

\bibitem[{{Harris} {et~al.}(2003){Harris}, {Biretta}, {Junor}, {Perlman},
  {Sparks}, \& {Wilson}}]{harris2003}
{Harris}, D.~E., {Biretta}, J.~A., {Junor}, W., {et~al.} 2003, \apjl, 586, L41

\bibitem[{{Harris} {et~al.}(2009){Harris}, {Cheung}, {Stawarz}, {Biretta}, \&
  {Perlman}}]{harris2009}
{Harris}, D.~E., {Cheung}, C.~C., {Stawarz}, {\L}., {Biretta}, J.~A., \&
  {Perlman}, E.~S. 2009, \apj, 699, 305

\bibitem[{{Hawley} {et~al.}(2015){Hawley}, {Fendt}, {Hardcastle}, {Nokhrina},
  \& {Tchekhovskoy}}]{hawley15}
{Hawley}, J.~F., {Fendt}, C., {Hardcastle}, M., {Nokhrina}, E., \&
  {Tchekhovskoy}, A. 2015, \ssr, 191, 441

\bibitem[{{Hovatta} {et~al.}(2014){Hovatta}, {Aller}, {Aller}, {Clausen-Brown},
  {Homan}, {Kovalev}, {Lister}, {Pushkarev}, \& {Savolainen}}]{hovatta2014}
{Hovatta}, T., {Aller}, M.~F., {Aller}, H.~D., {et~al.} 2014, \aj, 147, 143

\bibitem[{{Jaffe} \& {Perola}(1973)}]{jaffe1973}
{Jaffe}, W.~J. \& {Perola}, G.~C. 1973, \aap, 26, 423

\bibitem[{{Jiang} {et~al.}(2021){Jiang}, {Shen}, {Mart{\'\i}-Vidal}, {Wang},
  {Jiang}, \& {Kawaguchi}}]{jiang21}
{Jiang}, W., {Shen}, Z., {Mart{\'\i}-Vidal}, I., {et~al.} 2021, \apjl, 922, L16

\bibitem[{{Junor} {et~al.}(1999){Junor}, {Biretta}, \& {Livio}}]{junor1999}
{Junor}, W., {Biretta}, J.~A., \& {Livio}, M. 1999, \nat, 401, 891

\bibitem[{{Kardashev}(1962)}]{kardashev1962}
{Kardashev}, N.~S. 1962, \sovast, 6, 317

\bibitem[{{Kim} {et~al.}(2018{\natexlab{a}}){Kim}, {Krichbaum}, {Lu}, {Ros},
  {Bach}, {Bremer}, {de Vicente}, {Lindqvist}, \& {Zensus}}]{kim2018}
{Kim}, J.~Y., {Krichbaum}, T.~P., {Lu}, R.~S., {et~al.} 2018{\natexlab{a}},
  \aap, 616, A188

\bibitem[{{Kim} {et~al.}(2018{\natexlab{b}}){Kim}, {Lee}, {Hodgson}, {Algaba},
  {Zhao}, {Kino}, {Byun}, \& {Kang}}]{kim2018b}
{Kim}, J.-Y., {Lee}, S.-S., {Hodgson}, J.~A., {et~al.} 2018{\natexlab{b}},
  \aap, 610, L5

\bibitem[{{Kim} \& {Trippe}(2014)}]{kim2014}
{Kim}, J.-Y. \& {Trippe}, S. 2014, Journal of Korean Astronomical Society, 47,
  195

\bibitem[{{Kino} {et~al.}(2015){Kino}, {Takahara}, {Hada}, {Akiyama}, {Nagai},
  \& {Sohn}}]{kino2015}
{Kino}, M., {Takahara}, F., {Hada}, K., {et~al.} 2015, \apj, 803, 30

\bibitem[{{Kino} {et~al.}(2014){Kino}, {Takahara}, {Hada}, \& {Doi}}]{kino2014}
{Kino}, M., {Takahara}, F., {Hada}, K., \& {Doi}, A. 2014, \apj, 786, 5

\bibitem[{{Kino} {et~al.}(2022){Kino}, {Takahashi}, {Kawashima}, {Park},
  {Hada}, {Ro}, \& {Cui}}]{kino22}
{Kino}, M., {Takahashi}, M., {Kawashima}, T., {et~al.} 2022, arXiv e-prints,
  arXiv:2209.07264

\bibitem[{{Komissarov} {et~al.}(2007){Komissarov}, {Barkov}, {Vlahakis}, \&
  {K{\"o}nigl}}]{komissarov2007}
{Komissarov}, S.~S., {Barkov}, M.~V., {Vlahakis}, N., \& {K{\"o}nigl}, A. 2007,
  \mnras, 380, 51

\bibitem[{{Kovalev} {et~al.}(2007){Kovalev}, {Lister}, {Homan}, \&
  {Kellermann}}]{kovalev2007}
{Kovalev}, Y.~Y., {Lister}, M.~L., {Homan}, D.~C., \& {Kellermann}, K.~I. 2007,
  \apjl, 668, L27

\bibitem[{{Kovalev} {et~al.}(2008){Kovalev}, {Lobanov}, {Pushkarev}, \&
  {Zensus}}]{Kovalev2008}
{Kovalev}, Y.~Y., {Lobanov}, A.~P., {Pushkarev}, A.~B., \& {Zensus}, J.~A.
  2008, \aap, 483, 759

\bibitem[{{Kravchenko} {et~al.}(2020){Kravchenko}, {Giroletti}, {Hada},
  {Meier}, {Nakamura}, {Park}, \& {Walker}}]{kravchenko20}
{Kravchenko}, E., {Giroletti}, M., {Hada}, K., {et~al.} 2020, \aap, 637, L6

\bibitem[{{Lee} {et~al.}(2015){Lee}, {Byun}, {Oh}, {Kim}, {Kim}, {Jung}, {Oh},
  {Roh}, {Jung}, \& {Yeom}}]{lee2015}
{Lee}, S.-S., {Byun}, D.-Y., {Oh}, C.~S., {et~al.} 2015, Journal of Korean
  Astronomical Society, 48, 229

\bibitem[{{Lisakov} {et~al.}(2017){Lisakov}, {Kovalev}, {Savolainen},
  {Hovatta}, \& {Kutkin}}]{Lisakov2017}
{Lisakov}, M.~M., {Kovalev}, Y.~Y., {Savolainen}, T., {Hovatta}, T., \&
  {Kutkin}, A.~M. 2017, \mnras, 468, 4478

\bibitem[{{Longair}(2011)}]{longair2011}
{Longair}, M.~S. 2011, {High Energy Astrophysics}

\bibitem[{{Ly} {et~al.}(2007){Ly}, {Walker}, \& {Junor}}]{ly2007}
{Ly}, C., {Walker}, R.~C., \& {Junor}, W. 2007, \apj, 660, 200

\bibitem[{{Lyutikov} {et~al.}(2005){Lyutikov}, {Pariev}, \&
  {Gabuzda}}]{lyutikov2005}
{Lyutikov}, M., {Pariev}, V.~I., \& {Gabuzda}, D.~C. 2005, \mnras, 360, 869

\bibitem[{{Macchetto} {et~al.}(1997){Macchetto}, {Marconi}, {Axon}, {Capetti},
  {Sparks}, \& {Crane}}]{macchetto1997}
{Macchetto}, F., {Marconi}, A., {Axon}, D.~J., {et~al.} 1997, \apj, 489, 579

\bibitem[{{MAGIC Collaboration} {et~al.}(2020){MAGIC Collaboration}, {Acciari},
  {Ansoldi}, {Antonelli}, {Arbet Engels}, {Arcaro}, {Baack}, {Babi{\'c}},
  {Banerjee}, {Bangale}, {Barres de Almeida}, {Barrio}, {Becerra Gonz{\'a}lez},
  {Bednarek}, {Bellizzi}, {Bernardini}, {Berti}, {Besenrieder},
  {Bhattacharyya}, {Bigongiari}, {Biland}, {Blanch}, {Bonnoli},
  {Bo{\v{s}}njak}, {Busetto}, {Carosi}, {Ceribella}, {Chai}, {Chilingaryan},
  {Cikota}, {Colak}, {Colin}, {Colombo}, {Contreras}, {Cortina}, {Covino},
  {D'Elia}, {da Vela}, {Dazzi}, {de Angelis}, {de Lotto}, {Delfino}, {Delgado},
  {Depaoli}, {di Pierro}, {di Venere}, {Do Souto Espi{\~n}eira}, {Dominis
  Prester}, {Donini}, {Dorner}, {Doro}, {Elsaesser}, {Fallah Ramazani},
  {Fattorini}, {Fern{\'a}ndez-Barral}, {Ferrara}, {Fidalgo}, {Foffano},
  {Fonseca}, {Font}, {Fruck}, {Fukami}, {Garc{\'\i}a L{\'o}pez}, {Garczarczyk},
  {Gasparyan}, {Gaug}, {Giglietto}, {Giordano}, {Godinovi{\'c}}, {Green},
  {Guberman}, {Hadasch}, {Hahn}, {Herrera}, {Hoang}, {Hrupec}, {H{\"u}tten},
  {Inada}, {Inoue}, {Ishio}, {Iwamura}, {Jouvin}, {Kerszberg}, {Kubo},
  {Kushida}, {Lamastra}, {Lelas}, {Leone}, {Lindfors}, {Lombardi}, {Longo},
  {L{\'o}pez}, {L{\'o}pez-Coto}, {L{\'o}pez-Oramas}, {Loporchio}, {Machado de
  Oliveira Fraga}, {Maggio}, {Majumdar}, {Makariev}, {Mallamaci}, {Maneva},
  {Manganaro}, {Mannheim}, {Maraschi}, {Mariotti}, {Mart{\'\i}nez}, {Masuda},
  {Mazin}, {Mi{\'c}anovi{\'c}}, {Miceli}, {Minev}, {Miranda}, {Mirzoyan},
  {Molina}, {Moralejo}, {Morcuende}, {Moreno}, {Moretti}, {Munar-Adrover},
  {Neustroev}, {Nigro}, {Nilsson}, {Ninci}, {Nishijima}, {Noda}, {Nogu{\'e}s},
  {N{\"o}the}, {Nozaki}, {Paiano}, {Palacio}, {Palatiello}, {Paneque},
  {Paoletti}, {Paredes}, {Pe{\~n}il}, {Peresano}, {Persic}, {Prada Moroni},
  {Prandini}, {Puljak}, {Rhode}, {Rib{\'o}}, {Rico}, {Righi}, {Rugliancich},
  {Saha}, {Sahakyan}, {Saito}, {Sakurai}, {Satalecka}, {Schmidt}, {Schweizer},
  {Sitarek}, {{\v{S}}nidari{\'c}}, {Sobczynska}, {Somero}, {Stamerra}, {Strom},
  {Strzys}, {Suda}, {Suri{\'c}}, {Takahashi}, {Tavecchio}, {Temnikov},
  {Terzi{\'c}}, {Teshima}, {Torres-Alb{\`a}}, {Tosti}, {Tsujimoto}, {Vagelli},
  {van Scherpenberg}, {Vanzo}, {Acosta}, {Vigorito}, {Vitale}, {Vovk}, {Will},
  {Zari{\'c}}, {Asano}, {Hada}, {Harris}, {Giroletti}, {Jermak}, {Madrid},
  {Massaro}, {Richter}, {Spanier}, {Steele}, \& {Walker}}]{magic2020}
{MAGIC Collaboration}, {Acciari}, V.~A., {Ansoldi}, S., {et~al.} 2020, \mnras,
  492, 5354

\bibitem[{{Marscher}(2010)}]{Marscher2010}
{Marscher}, A.~P. 2010, {Jets in Active Galactic Nuclei}, ed. T.~{Belloni},
  Vol. 794, 173

\bibitem[{{McKinney}(2006)}]{mckinney2006}
{McKinney}, J.~C. 2006, \mnras, 368, 1561

\bibitem[{{Mertens} {et~al.}(2016){Mertens}, {Lobanov}, {Walker}, \&
  {Hardee}}]{mertens2016}
{Mertens}, F., {Lobanov}, A.~P., {Walker}, R.~C., \& {Hardee}, P.~E. 2016,
  \aap, 595, A54

\bibitem[{{M{\"u}ller} {et~al.}(2011){M{\"u}ller}, {Kadler}, {Ojha}, {Wilms},
  {B{\"o}ck}, {Edwards}, {Fromm}, {Hase}, {Horiuchi}, {Katz}, {Lovell},
  {Pl{\"o}tz}, {Pursimo}, {Richers}, {Ros}, {Rothschild}, {Taylor}, {Tingay},
  \& {Zensus}}]{muller2011}
{M{\"u}ller}, C., {Kadler}, M., {Ojha}, R., {et~al.} 2011, \aap, 530, L11

\bibitem[{{Nakamura} {et~al.}(2018){Nakamura}, {Asada}, {Hada}, {Pu}, {Noble},
  {Tseng}, {Toma}, {Kino}, {Nagai}, {Takahashi}, {Algaba}, {Orienti},
  {Akiyama}, {Doi}, {Giovannini}, {Giroletti}, {Honma}, {Koyama}, {Lico},
  {Niinuma}, \& {Tazaki}}]{nakamura2018}
{Nakamura}, M., {Asada}, K., {Hada}, K., {et~al.} 2018, \apj, 868, 146

\bibitem[{{Narayan} {et~al.}(2022){Narayan}, {Chael}, {Chatterjee}, {Ricarte},
  \& {Curd}}]{narayan2022}
{Narayan}, R., {Chael}, A., {Chatterjee}, K., {Ricarte}, A., \& {Curd}, B.
  2022, \mnras, 511, 3795

\bibitem[{{Niinuma} {et~al.}(2014){Niinuma}, {Lee}, {Kino}, {Sohn}, {Akiyama},
  {Zhao}, {Sawada-Satoh}, {Trippe}, {Hada}, {Jung}, {Hagiwara}, {Dodson},
  {Koyama}, {Honma}, {Nagai}, {Chung}, {Doi}, {Fujisawa}, {Han}, {Kim}, {Lee},
  {Lee}, {Miyazaki}, {Oyama}, {Sorai}, {Wajima}, {Bae}, {Byun}, {Cho}, {Choi},
  {Chung}, {Chung}, {Han}, {Hirota}, {Hwang}, {Je}, {Jike}, {Jung}, {Jung},
  {Kang}, {Kang}, {Kang}, {Kan-ya}, {Kanaguchi}, {Kawaguchi}, {Kim}, {Kim},
  {Kim}, {Kim}, {Kim}, {Kim}, {Kim}, {Kobayashi}, {Kono}, {Kurayama}, {Lee},
  {Lee}, {Lee}, {Minh}, {Matsumoto}, {Nakagawa}, {Oh}, {Oh}, {Park}, {Roh},
  {Sasao}, {Shibata}, {Song}, {Tamura}, {Wi}, {Yeom}, \& {Yun}}]{niinuma2014}
{Niinuma}, K., {Lee}, S.-S., {Kino}, M., {et~al.} 2014, \pasj, 66, 103

\bibitem[{{Ostrowski}(1998)}]{ostrowski1998}
{Ostrowski}, M. 1998, \aap, 335, 134

\bibitem[{{O'Sullivan} \& {Gabuzda}(2009)}]{o'sullivan2009}
{O'Sullivan}, S.~P. \& {Gabuzda}, D.~C. 2009, \mnras, 400, 26

\bibitem[{{Pacholczyk}(1970)}]{pacholczyk1970}
{Pacholczyk}, A.~G. 1970, {Radio astrophysics. Nonthermal processes in galactic
  and extragalactic sources}

\bibitem[{{Park} {et~al.}(2019{\natexlab{a}}){Park}, {Hada}, {Kino},
  {Nakamura}, {Hodgson}, {Ro}, {Cui}, {Asada}, {Algaba}, {Sawada-Satoh}, {Lee},
  {Cho}, {Shen}, {Jiang}, {Trippe}, {Niinuma}, {Sohn}, {Jung}, {Zhao},
  {Wajima}, {Tazaki}, {Honma}, {An}, {Akiyama}, {Byun}, {Kim}, {Zhang},
  {Cheng}, {Kobayashi}, {Shibata}, {Lee}, {Roh}, {Oh}, {Yeom}, {Jung}, {Oh},
  {Kim}, {Hwang}, \& {Hagiwara}}]{park2019_kine}
{Park}, J., {Hada}, K., {Kino}, M., {et~al.} 2019{\natexlab{a}}, \apj, 887, 147

\bibitem[{{Park} {et~al.}(2019{\natexlab{b}}){Park}, {Hada}, {Kino},
  {Nakamura}, {Ro}, \& {Trippe}}]{park2019_faraday}
{Park}, J., {Hada}, K., {Kino}, M., {et~al.} 2019{\natexlab{b}}, \apj, 871, 257

\bibitem[{{Park} {et~al.}(2021){Park}, {Hada}, {Nakamura}, {Asada}, {Zhao}, \&
  {Kino}}]{Park2021}
{Park}, J., {Hada}, K., {Nakamura}, M., {et~al.} 2021, \apj, 909, 76

\bibitem[{{Perlman} \& {Wilson}(2005)}]{perlman2005}
{Perlman}, E.~S. \& {Wilson}, A.~S. 2005, \apj, 627, 140

\bibitem[{{Plavin} {et~al.}(2019){Plavin}, {Kovalev}, {Pushkarev}, \&
  {Lobanov}}]{plavin19}
{Plavin}, A.~V., {Kovalev}, Y.~Y., {Pushkarev}, A.~B., \& {Lobanov}, A.~P.
  2019, \mnras, 485, 1822

\bibitem[{{Pushkarev} {et~al.}(2019){Pushkarev}, {Butuzova}, {Kovalev}, \&
  {Hovatta}}]{pushkarev2019}
{Pushkarev}, A.~B., {Butuzova}, M.~S., {Kovalev}, Y.~Y., \& {Hovatta}, T. 2019,
  \mnras, 482, 2336

\bibitem[{{Pushkarev} {et~al.}(2012){Pushkarev}, {Hovatta}, {Kovalev},
  {Lister}, {Lobanov}, {Savolainen}, \& {Zensus}}]{pushkarev2012b}
{Pushkarev}, A.~B., {Hovatta}, T., {Kovalev}, Y.~Y., {et~al.} 2012, \aap, 545,
  A113

\bibitem[{{Pushkarev} \& {Kovalev}(2012)}]{pushkarev2012}
{Pushkarev}, A.~B. \& {Kovalev}, Y.~Y. 2012, \aap, 544, A34

\bibitem[{{Reynolds} {et~al.}(1996){Reynolds}, {Fabian}, {Celotti}, \&
  {Rees}}]{reynolds1996}
{Reynolds}, C.~S., {Fabian}, A.~C., {Celotti}, A., \& {Rees}, M.~J. 1996,
  \mnras, 283, 873

\bibitem[{{Rieger} \& {Levinson}(2018)}]{rieger2018}
{Rieger}, F. \& {Levinson}, A. 2018, Galaxies, 6, 116

\bibitem[{{Rybicki} \& {Lightman}(1979)}]{rybicki1979}
{Rybicki}, G.~B. \& {Lightman}, A.~P. 1979, {Radiative processes in
  astrophysics}

\bibitem[{{Shepherd} {et~al.}(1994){Shepherd}, {Pearson}, \&
  {Taylor}}]{shepherd1994}
{Shepherd}, M.~C., {Pearson}, T.~J., \& {Taylor}, G.~B. 1994, in \baas,
  Vol.~26, 987--989

\bibitem[{{Sironi} {et~al.}(2021){Sironi}, {Rowan}, \& {Narayan}}]{sironi2021}
{Sironi}, L., {Rowan}, M.~E., \& {Narayan}, R. 2021, \apjl, 907, L44

\bibitem[{{Tavecchio} \& {Ghisellini}(2008)}]{tavecchio2008}
{Tavecchio}, F. \& {Ghisellini}, G. 2008, \mnras, 385, L98

\bibitem[{{Thompson} {et~al.}(2017){Thompson}, {Moran}, \&
  {Swenson}}]{thompson2017}
{Thompson}, A.~R., {Moran}, J.~M., \& {Swenson}, George~W., J. 2017,
  {Interferometry and Synthesis in Radio Astronomy, 3rd Edition}

\bibitem[{{Walker} {et~al.}(2018){Walker}, {Hardee}, {Davies}, {Ly}, \&
  {Junor}}]{walker2018}
{Walker}, R.~C., {Hardee}, P.~E., {Davies}, F.~B., {Ly}, C., \& {Junor}, W.
  2018, \apj, 855, 128

\bibitem[{{Walsh} {et~al.}(2013){Walsh}, {Barth}, {Ho}, \& {Sarzi}}]{walsh2013}
{Walsh}, J.~L., {Barth}, A.~J., {Ho}, L.~C., \& {Sarzi}, M. 2013, \apj, 770, 86

\bibitem[{{Zamaninasab} {et~al.}(2014){Zamaninasab}, {Clausen-Brown},
  {Savolainen}, \& {Tchekhovskoy}}]{zamaninasab2014}
{Zamaninasab}, M., {Clausen-Brown}, E., {Savolainen}, T., \& {Tchekhovskoy}, A.
  2014, \nat, 510, 126

\bibitem[{{Zavala} \& {Taylor}(2003)}]{zavala2003}
{Zavala}, R.~T. \& {Taylor}, G.~B. 2003, \apj, 589, 126

\bibitem[{{Zdziarski} {et~al.}(2015){Zdziarski}, {Sikora}, {Pjanka}, \&
  {Tchekhovskoy}}]{zdziarski2015}
{Zdziarski}, A.~A., {Sikora}, M., {Pjanka}, P., \& {Tchekhovskoy}, A. 2015,
  \mnras, 451, 927

\end{thebibliography}

\begin{appendix}

\section{Effect of the $(u,v)$-range matching on the spectral index distribution}\label{section.a.1}

{As described in Sect.\index{Sect} \ref{subsection3.1}, it is important to match the $(u, v)$ ranges of the two frequencies when producing spectral index maps.}
{Considering the typical $(u, v)$ ranges for KaVA and VLBA at 22 and 43 GHz given in Table \ref{tab:uvrange}, we used data within 33$-$170 M$\lambda$ for KaVA and 25$-$685 M$\lambda$ for VLBA.}
{This cuts out the data on average $\sim$ 21 \% ($\sim$ 46 \%) for KaVA 22 GHz (43 GHz), and $\sim$ 5 \% ($\sim$ 8 \%) for VLBA 22 GHz (43 GHz), respectively.}
{As a result, the image noise increases on average by $\sim$ 7 \% ($\sim$ 33 \%) for KaVA 22 GHz (43 GHz), and $\sim$ 0 \% (3 \%) for VLBA 22 and 43 GHz, respectively.}

\begin{figure}
    \centering
    \includegraphics[width=\columnwidth]{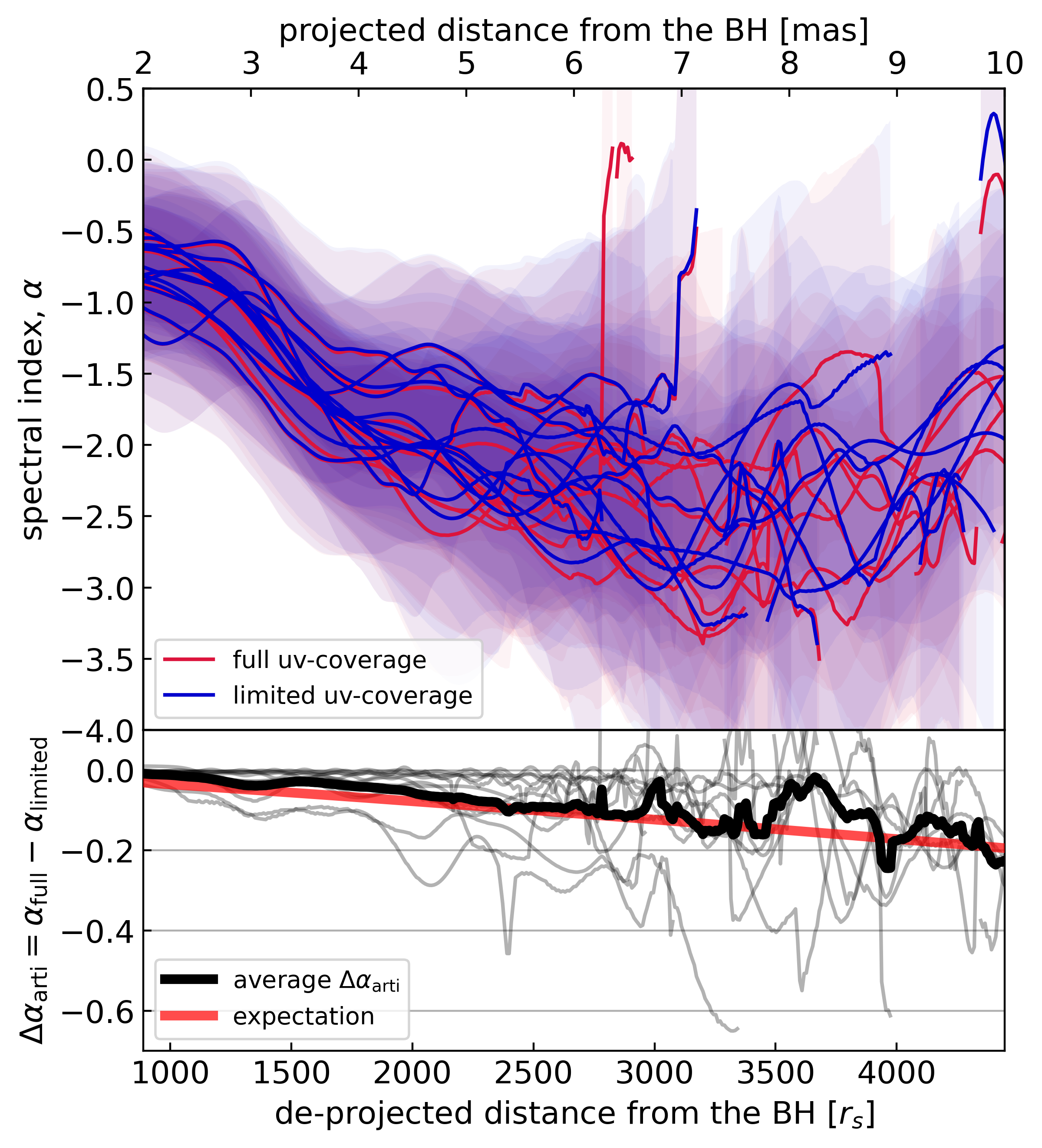}
    \caption{Effect of the $(u,v)$-range matching on the spectral index distribution. The upper panel shows spectral index distributions between 22 and 43 GHz before (red lines) and after (blue lines) matching the $(u, v)$ range.
    Distributions using the full $(u, v)$ range have steeper spectra at longer distances. Differences between the two distributions (i.e., artificial steeping due to inconsistent $(u, v)$ ranges) are indicated by gray lines in the lower panel. The thick black line is the average difference, and the red line is the expected artificial steepening along the jet. See text for details.
    }
    \label{fig:uv_coverage_comparison}
\end{figure}

{The upper panel of Fig. \ref{fig:uv_coverage_comparison} compares the spectral index distributions before (red lines) and after (blue lines) matching the $(u, v)$ ranges. This shows the distributions using the full $(u, v)$ range have steeper spectra at longer distances.}
{The difference between these two distributions -- the artificial steepening along the jet distance ($\Delta\alpha_{\text{arti}}$) -- is shown in the lower panel.}
{The gray lines are from individual epochs and the thick black line is their average.}
{We also calculated the expected $\Delta\alpha_{\text{arti}}$ under the assumption that the loss of sensitivity for large-scale structures at the shortest baseline is $\sim \sin(\pi\theta/\theta_{\text{max}})/(\pi\theta/\theta_{\text{max}})$ \citep[e.g.,][]{thompson2017}, where $\theta$ is the size of the structure and $\theta_{\text{max}}$ is the angular size corresponding to the shortest baseline. Here, we assume $\theta$ as jet widths and adopt it from \citet{hada2013}, and $\theta_{\text{max}}$ of KaVA and VLBA at 22 and 43 GHz are summarized in Table \ref{tab:uvrange}. After calculating the expected $\Delta\alpha_{\text{arti}}$ for KaVA and VLBA, they were weighted and averaged by the number of observations to calculate the final value. This value is displayed in a red line in the lower panel.}

{The averaged $\Delta\alpha_{\text{arti}}$ increases with distance, close to the expected value, and is $\sim- 0.2$ at $\sim$ 10 mas.}
{Individual cases show $\Delta\alpha_{\text{arti}}\sim-0.6$ at a distance of $\sim$ 7 mas in severe cases. Even taking this into account, the observed spectral index decreases more significantly ($\Delta\alpha \sim -2$). Therefore, we conclude that the decrease in the spectral index with the distance of the M87 jet is intrinsic.}

\begin{table}
    \caption{Typical $(u, v)$ range and the angular size corresponding to the shortest baseline of KaVA and VLBA at 22 and 43 GHz.}
    \label{tab:uvrange}
    \centering

    \begin{tabular}{c c c c c}
    \hline\hline
         & $r_{\text{22GHz}}$ & $r_{\text{43GHz}}$ & $\theta_{\text{max, 22GHz}}$ & $\theta_{\text{max, 43GHz}}$ \\
         & [$M\lambda$] & [$M\lambda$] & [mas] & [mas] \\
    \hline
    KaVA &  17$-$170  &  33$-$330 &  12.1  &  6.3 \\
    VLBA &  13$-$685  &  25$-$1230 &  16.0  &  8.3 \\
    \hline
    \end{tabular}
\end{table}

\section{Effect of beam size on the spectral index distribution}\label{section.a.2}
\FloatBarrier
\begin{figure}
    \centering
    \includegraphics[width=\columnwidth]{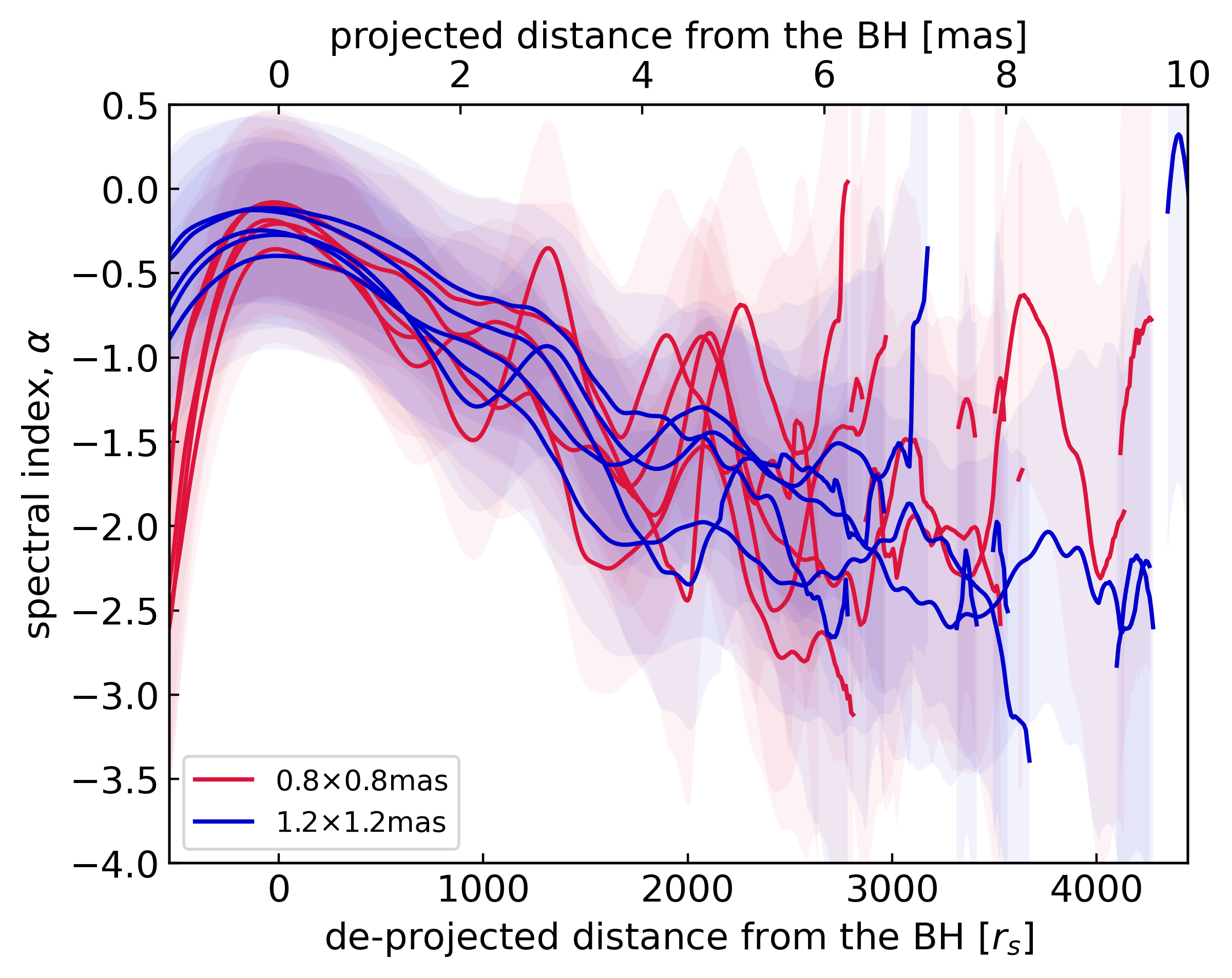}
    \caption{Spectral index distributions between VLBA 22 and 43 GHz maps convolved with beams of different sizes. The blue lines are convolved with a circular beam of $1.2\times1.2$ mas, where the radius is the same as the major axis of the KaVA 22 GHz beam. The red lines are convolved with $0.8\times0.8$ mas, which is equal to the radius of the VLBA 22 GHz beam. The distribution with the VLBA beam has a larger scattering than the distribution with the KaVA beam.}
    \label{fig:beam_comparison}
\end{figure}

{Typical beam sizes of VLBA 22 and 43 GHz are $0.8\times0.4$ mas and $0.45\times0.22$ mas, respectively, which are much smaller than the typical beam sizes of KaVA. In Fig. \ref{fig:beam_comparison} the spectral index distribution between the VLBA 22 and 43 GHz maps was convolved into two different beams: $0.8\times0.8$ mas and $1.2\times1.2$ mas. The radius of each beam is equal to the major axis of the VLBA and KaVA beams at 22 GHz, respectively. Compared to the spectral index distribution using the KaVA beam, the distribution using the VLBA beam has a greater scatter with some local fluctuations. As the beam size increases, the local structures are averaged and the overall distribution is smoother. However, it can be seen that the global trend of the spectrum does not change with the size of the beam.}

\section{Alternative explanation of the spectrum steepening}\label{section.a.3}

{In our model, we assumed a relationship of $\alpha = (p+1)/2$ between the slope of the electron distribution and the synchrotron spectrum \citep{rybicki1979}. This assumption is maintained when the electron energy corresponding to the observation frequency is within a sufficiently wide range of $\gamma$ in the optically thin region. The observation frequencies 22 and 43 GHz correspond to $\gamma\sim10^{1.6-2.0}$ and are located within the electron energy range of the injection function ($\gamma_{\text{min inj}} = 10^{1}$, $\gamma_{\text{max inj}} = 10^{5}$). However, without electron injection, $\gamma_{\text{max}}$ decreases over time and thus the above assumption may not hole. In this case, the synchrotron spectrum has a different shape depending on the synchrotron emission model we use.}

\begin{figure}
    \centering
    \includegraphics[width=\columnwidth]{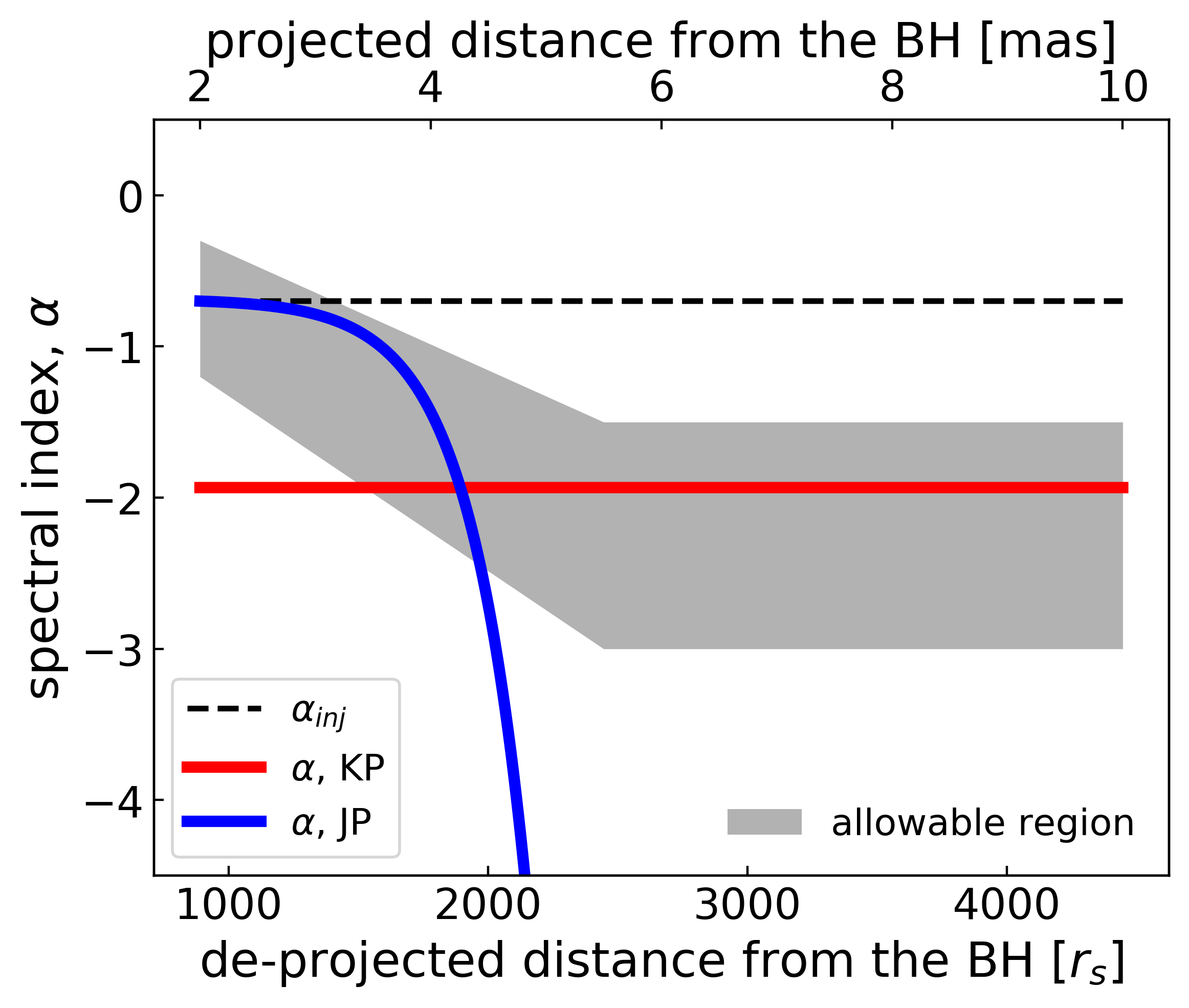}
    \caption{Schematic diagram of the spectral index distributions without electron injection, assuming different synchrotron emission models. The blue line assumes the JP model, which is used in this study. The red line shows the spectral index distribution with the KP model. Assuming that the injection spectral index is $\alpha_{\text{inj}} = -0.7$, the expected index after steepening is $4/3\alpha_{\text{inj}}-1\approx-1.933$, which is within the allowable range obtained from observation.}
    \label{fig:KP_JP_model}
\end{figure}

{The most widely used synchrotron emission models were proposed in \cite{kardashev1962} and \cite{pacholczyk1970} (KP model) and \cite{jaffe1973} (JP model). In the JP model, which is used in our spectral index model, the scattering due to turbulence in the magnetic field randomizes the electron pitch angles on shorter timescales relative to the radiative lifetime \citep[e.g.,][]{hardcastle2013}. Consequently, the cutoff energy is determined only by magnetic field strength, not by the pitch angle.}

{In the KP model, however, the pitch angle of individual electrons remains constant over the lifetime, and electrons with a pitch angle parallel to the magnetic field do not lose energy. Hence, the cutoff energy depends not only on the magnetic field strength but also on the pitch angle. Assuming an isotropic pitch angle, the synchrotron spectrum does not drop exponentially but has a break with the spectral index ranging from $\alpha_{\text{inj}}$ at low frequencies to $4/3\alpha_{\text{inj}}-1$ at high frequencies \citep{kardashev1962}\footnote{This analytical solution assumes no temporal changes in magnetic field strength and no adiabatic losses. Numerical calculations are required to accurately estimate the change of the spectral index in the M87 jet using the KP model, which is beyond the scope of the discussion.}.}

{Figure \ref{fig:KP_JP_model} shows a schematic picture of the spectral index distribution using the KP and JP models in the absence of electron injection. Applying an initial slope of $\alpha_{\text{inj}} = -0.7$, the KP type energy loss predicts the slope at energies higher than the break energy as $4/3\alpha_{\text{inj}}-1\approx-1.933$, which is within the allowed region defined from observations. Therefore, this scenario may also explain the spectral index steepening.}

{In summary, we suggest two scenarios to explain the observed spectral index distribution. (1) JP-type energy loss and particle injection with distance dependence. (2) KP-type energy loss and no electron injection. Both scenarios make an electron energy distribution a shape of a broken power law, and hence it is possible to explain the observed trend of the spectral index distributions. Further study is needed to discuss which scenario is more plausible on the M87 jet. One of the expectations is that the brightness distribution between the JP model scenario and the KP model scenario is different since the JP model scenario requires continuous electron injection and the KP model scenario does not. Another distinguishing point between the JP and the KP scenarios is the fractional polarization. The JP model needs a fully random field plus an isotropic pitch angle. This will lead to very low fractional polarization. The KP model predicts linear polarization (depending on the magnetic field geometry). Therefore, fractional polarization could be a good indicator of the synchrotron loss model.}

\section{Details of the spectral index model}\label{section.a.4}
{The general transfer equation of the electron number density distribution is written as follows \citep[e.g.,][]{ginzburg1964, longair2011, blasi2013}:}
\begin{equation}
   \frac{\partial N}{\partial\tau}-\nabla\cdot\left(D\nabla N\right)+\nabla\cdot\left(vN\right)+\frac{\partial}{\partial\gamma}\left(bN\right)=Q,
   \label{equation.a.1}
\end{equation}
{where $D$ is the diffusion coefficient, $v$ is the velocity of the system, $b$ is the energy loss rate, and $Q$ is the nonthermal electron injection rate. In our discussion, the frame of reference is a cross section of the jet co-moving with the jet’s flow, then $v \cdot \nabla N = 0$. We further neglect the diffusion of the electron. Then, Eq. (\ref{equation.a.1}) becomes Eq. (\ref{equation.5}).}

{To solve the general transfer equation numerically, we change the partial derivative equation into a series of ordinary differential equations \citep[e.g.,][]{arfken}. First, we differentiate $b(\gamma)$ (Eq. \ref{equation.6}) with respect to $\gamma$,
\begin{equation}
   \frac{\partial b}{\partial \gamma} = -b_{\text{adi}} - 2b_{\text{sync}}\gamma
   \label{equation.a.2}
.\end{equation}
Putting Eq. (\ref{equation.a.2}) into Eq. (\ref{equation.a.1}) and using the relation $\nabla\cdot v = \frac{3}{R}\frac{dR}{d\tau} =  3b_{\text{adi}}$, we get
\begin{equation}
    \frac{\partial N}{\partial \tau} +\frac{d \gamma}{d \tau} \frac{\partial N}{\partial \gamma} = Q + 2\left(b_{\text{sync}}\gamma - b_{\text{adi}}\right)N
    \label{equation.a.3}
.\end{equation}
The left-hand side of Eq. (\ref{equation.a.3}) is the same as a total derivative of $N(\gamma, \tau)$ with respect to $\tau$. Consequently, the transfer equation changes into a series of ordinary differential equations:}
\begin{flalign*}
    &\frac{dN}{d\tau} = 2(b_{\text{sync}}\gamma-b_{\text{adi}})N+Q \tag{C.4.1}\label{equation.a.4.1}\\
    &\frac{d\gamma}{d\tau} = -b_{\text{adi}}\gamma-b_{\text{sync}}\gamma^2, \tag{C.4.2}\label{equation.a.4.2}
\end{flalign*}
{which are solvable numerically. The number density of electrons $N(\gamma)$ changes with time with respect to Eq. (\ref{equation.a.4.1}), and at the same time, each electron in $N(\gamma)$ loses its energy by following Eq. (\ref{equation.a.4.2}) (same as Eq. \ref{equation.6}).}

{To solve the problem, we additionally need to know the relationship between the distance of the reference frame from the SMBH and the time of the frame (i.e., the proper time). This is because the physical quantities in the equation were obtained as a function of distance, while the transfer equation deals with the change with time. When $\tau$ is the proper time and $z$ is the distance, $\tau$ and $z$ are in the following relation:
\begin{equation}
    \frac{d\tau}{dz}=\frac{dt_{\text{obs}}}{dz}\frac{d\tau}{dt_{\text{obs}}}=\frac{1}{\Gamma\beta c}
    \tag{C.5}\label{equation.a.5}
,\end{equation}where $t_{\text{obs}}$ is the time in the observer’s frame, $\beta = \sqrt{1-\Gamma^{-2}}$ is the bulk jet speed in units of the speed of the light and $\Gamma$ is the bulk Lorentz factor.
Here, we use the equation of apparent jet speed $\beta_{\text{app}}c \equiv \frac{dz}{dt_{\text{obs}}}\sin{\theta} = \frac{\beta c\sin{\theta}}{1 - \beta\cos{\theta}}$ and Doppler factor $\delta \equiv \frac{d\tau}{dt_{\text{obs}}} = \frac{1}{\Gamma(1-\beta \cos{\theta})}$ \citep[e.g.,][]{ghisellini2013}.
The relation between $\tau$ and $z$ is then calculated by integrating Eq. (\ref{equation.a.5}):
\begin{equation}
    \int_{\tau_i}^{\tau_f} \,d\tau = \int_{z_i}^{z_f} \frac{dz}{c\sqrt{\Gamma^2-1}}
    \tag{C.6}\label{equation.a.6}
.\end{equation}The distance range of our calculation is from $z_i=2~\text{mas}$ to $z_f=10~\text{mas,}$ which correspond to $z_i\approx889~r_s$ and $z_f\approx4446~r_s$ in deprojected units assuming a viewing angle of 17$\degree$ \citep{walker2018}.
Applying the observed bulk jet velocity field of the M87 jet \citep{park2019_kine} and setting $\tau_i=0s$, the travel time is $\tau_f\sim2.2\times10^8\text{s}$ ($\sim$ 6.9 years).
If we apply the velocity field with faster acceleration in the region closer than 2 mas \citep{mertens2016}, the travel time between 2 $-$ 10 mas is $\sim$ 2.4 years.

Now we solve the Eqs. (\ref{equation.a.4.1}) and (\ref{equation.a.4.2}) to obtain the electron number density distribution as a function of time, $N(\gamma, \tau)$ \footnote{In this process, the adiabatic loss coefficient can be calculated by using the jet radius profile $R(z)$ and the bulk jet velocity profile $\Gamma(z)$ since the equation can be rewritten as $b_{\text{adi}}=\frac{1}{R}\frac{dR}{d\tau}=\frac{\Gamma\beta c}{R}\frac{dR}{dz}$.}, which is further converted to $N(\gamma, z)$ using Eq. (\ref{equation.a.6}). From $N(\gamma, z)$, the slope of the number density distribution between $\gamma(\nu_{\text{obs}} = 22\text{GHz})$ and $\gamma(\nu_{\text{obs}} = 43\text{GHz})$ along the jet distance, $p_{\text{22-43GHz}}(z)$, is obtained by using the relation of the observed frequency of synchrotron radiation and the energy of nonthermal electrons \citep{rybicki1979}:
\begin{equation}
    \nu_{\text{obs}} = \frac{3eB}{4\pi m_{\text{e}} c}\gamma^2\delta
    \tag{C.7}\label{equation.a.7}
.\end{equation}
Finally, the spectral index distribution $\alpha_{\text{22-43GHz}}(z)$ is obtained by using the relation $\alpha = (p+1)/2$ \citep{rybicki1979}.}

\section{Notes on the injection function $Q(\gamma, z)$}\label{section.a.5}

\begin{figure}
\centering
\includegraphics[width=\hsize]{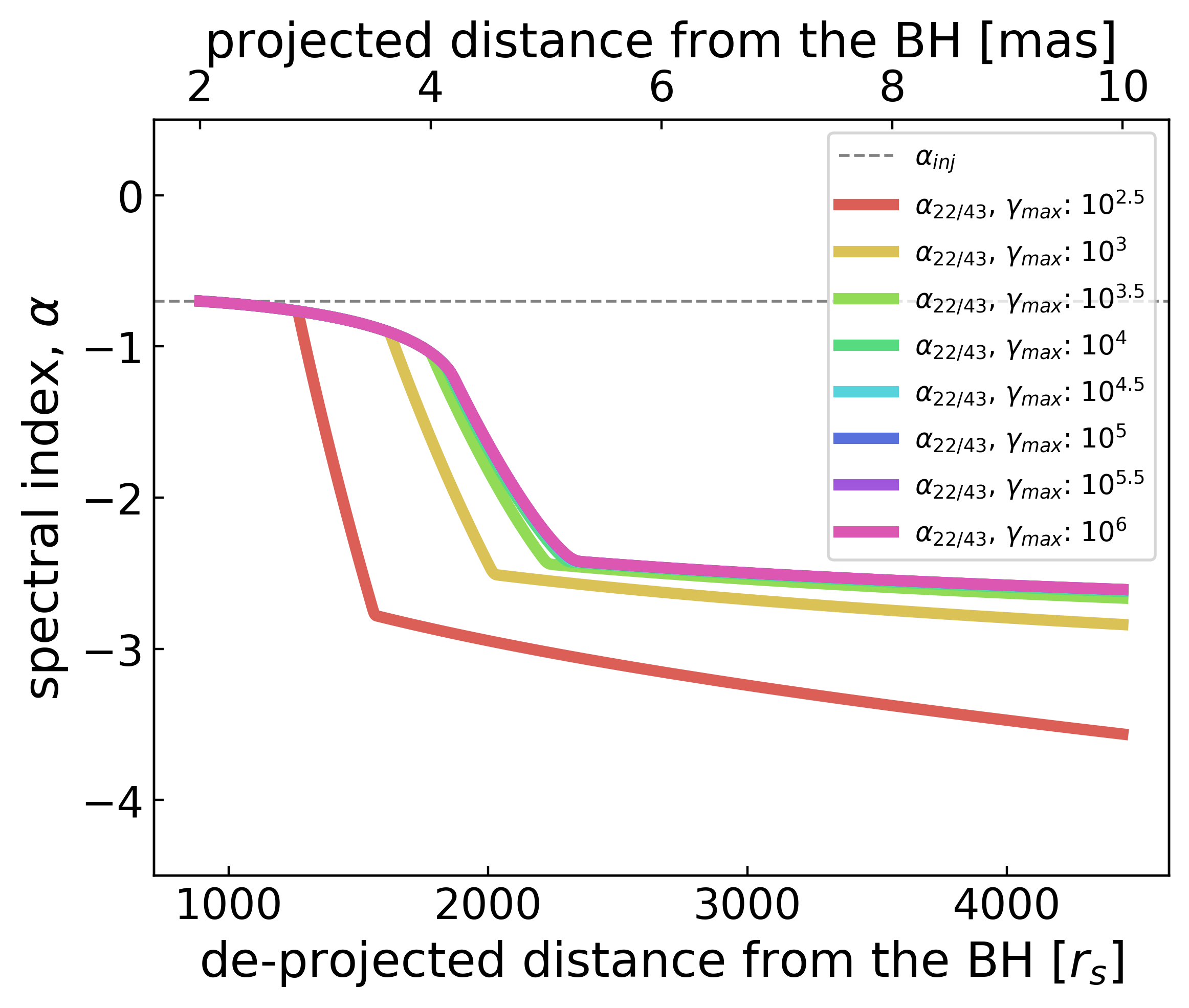}
    \caption{Comparison of model spectral index distributions using nonthermal electron injection functions with different $\gamma_{\text{max inj}}$. Distributions with different colors represent different values of $\gamma_{\text{max inj}}$. For all distributions, the magnetic field strength was assumed to be $B_i = 0.3 {\text{ G}}$. As $\gamma_{\text{max inj}}$ increases, the spectral index distributions converge and there is little change between the models with $\gamma_{\text{max inj}} = 10^{4.5}$ and $\gamma_{\text{max inj}} = 10^{6}$.}
    \label{fig:gamma_max_effect}
\end{figure}

{The electron injection function in our calculation is assumed to be a power-law distribution with a sharp cutoff at $\gamma_{\text{inj min}}=10$ and $\gamma_{\text{inj max}}=10^5$. The minimum and maximum range in the electron energy distributions are constrained by the observations. The M87 jet has been detected in VLBI observations at frequencies as low as 1.6 GHz \citep[e.g.,][]{cheung2007}, and as low as $\sim20~\text{MHz}$ using Low-Frequency Array \citep[LOFAR; e.g.,][]{de_Gasperin2012}. From Eq. (\ref{equation.a.7}), the electron energy corresponding to synchrotron radiation at 20 MHz is $\gamma(\nu_{\text{obs}} = 20\text{MHz}) \sim 10^{0.1-0.4}$ when $B_i=0.3-1.0 \text{ G}$ and $\Gamma = 2$ is assumed. Likewise, the detection of X-ray emission in the M87 jet between the core and HST-1 could constrain the maximum energy of the electron distribution \citep[e.g.,][]{perlman2005, ehtMWL2021}. From Eq. (\ref{equation.a.7}), the X-ray synchrotron emission requires $\gamma(\nu_{\text{obs}}=300\text{PHz})\sim10^{5.2-5.5}$. \citet{perlman2005} suggested that even higher energies of nonthermal electrons of $\gamma\sim10^{6-8}$ is required in order to emit X-ray synchrotron emission. We also found that electrons with energies greater than $10^5$ have little effect on the spectral index distribution at 22$-$43 GHz. Figure \ref{fig:gamma_max_effect} shows the spectral index distribution model using different $\gamma_{\text{inj max}}$ in the injection function. For all models, $q = 10$ and $B_i = 0.3 \text{ G}$ were applied. As $\gamma_{\text{inj max}}$ increases, the spectral index distributions gradually converge, and when $\gamma_{\text{inj max}}$ is greater than $ 10^{4.5}$, the spectral index distribution does not change. This is because the synchrotron cooling time of the electrons with energy $\gamma \gtrsim 10^5$ is $\tau_{\text{sync}}=\gamma\left(\frac{d\gamma}{d\tau}\right)^{-1}\lesssim8.6\times10^4\text{s}$, which is much shorter than the travel time ($\sim2.2\times10^8\text{s}$). From this experiment, we found that, for $\gamma_{\text{inj max}}>10^5$, the spectral index distribution remains unchanged regardless of the shape of the high-energy tail of the injection function (e.g., a sharp cutoff or an exponentially decreasing). Therefore, in this study, we assume the injection function as a power-law energy distribution and the sharp cutoff at $\gamma_{\text{inj max}}=10^5$.}

\end{appendix}


\end{document}